\documentclass[iop,twocolumn]{emulateapj}

\newcommand{\vect}[1]{\boldsymbol{#1}}
\newcommand{\tens}[1]{\mathbf{#1}}

\newcommand{\degree}{\ensuremath{^\circ}}
\newcommand{\msun}{\ensuremath{\mathrm{M_{\odot}}}}

\usepackage{amssymb}
\usepackage{amsmath}
\usepackage{hyperref}
\usepackage{float}
\usepackage{overpic}

\hypersetup{
 colorlinks,
 citecolor=blue,
 filecolor=black,
 linkcolor=blue,
 urlcolor=black
}

\shortauthors{Handy, Plewa, \& Odrzywo{\l}ek}
\shorttitle{Shock revival with SN 1987A energetics}

\begin{document}
\title{Toward Connecting Core-Collapse Supernova Theory with Observations: I. Shock revival in a 15 \msun\ blue supergiant progenitor with SN 1987A energetics}
\author{Timothy Handy\altaffilmark{1}}
%
\author{Tomasz Plewa\altaffilmark{1}}
\email[Corresponding author: ]{tplewa@fsu.edu}
\author{Andrzej Odrzywo{\l}ek\altaffilmark{2}}
%
\altaffiltext{1}{Department of Scientific Computing, Florida State University, Tallahassee, FL 32306, U.S.A.}
\altaffiltext{2}{Marian Smoluchowski Institute of Physics, Jagiellonian University, Reymonta 4, 30-059 Cracow, Poland}
%
%
%
\begin{abstract}
We study the evolution of the collapsing core of a 15 \msun\ blue
supergiant supernova progenitor from the core bounce until 1.5 seconds
later. We present a sample of hydrodynamic models parameterized to
match the explosion energetics of SN 1987A.

We find the spatial model dimensionality to be an important
contributing factor in the explosion process. Compared to
two-dimensional simulations, our three-dimensional models require
lower neutrino luminosities to produce equally energetic
explosions. We estimate that the convective engine in our models is
$4$\% more efficient in three dimensions than in two dimensions. We
propose that the greater efficiency of the convective engine found in
three-dimensional simulations might be due to the larger
surface-to-volume ratio of convective plumes, which aids in
distributing energy deposited by neutrinos.

We do not find evidence of the standing accretion shock instability nor
turbulence being a key factor in powering the explosion in our
models. Instead, the analysis of the energy transport in the post-shock
region reveals characteristics of penetrative convection. The
explosion energy decreases dramatically once the resolution is
inadequate to capture the morphology of convection on large
scales. This shows that the role of dimensionality is secondary to
correctly accounting for the basic physics of the explosion.

We also analyze information provided by particle tracers embedded in
the flow, and find that the unbound material has relatively long
residency times in two-dimensional models, while in three dimensions a
significant fraction of the explosion energy is carried by particles
with relatively short residency times.

\end{abstract}
%
%
%
%
%
%
%
%
%
%
\section{Introduction}\label{s:intro}
There is substantial controversy over the importance of physics processes
involved in core-collapse supernova (ccSN) explosions \citep[see, e.g., ][and
references therein]{janka+12,burrows13}. One problem is a multitude of
relevant processes which include hydrodynamics, gravity, nuclear equation of
state, and neutrino-matter interactions. Despite much effort in the last
four decades, no model accounting for these physics effects exists that
consistently produces energetic ccSNe explosions for suitable progenitor
models. Early theories of ccSNe explosions such as neutrino-driven convection
\citep[see, e.g., ][]{burrows+93, janka+96,mezzacappa+98} and the more recent
standing accretion shock instability (SASI) \citep{blondin+03} remain the
centerpiece of computational core-collapse supernova studies.

The SASI mechanism has been discovered in the course of numerical simulations
in the work by \cite{blondin+03}. Subsequent SASI studies included theoretical
analysis \citep{laming07,foglizzo09} and numerical investigations
\citep{ohnishi+06,blondin+07,scheck+08}. This shock instability is
characterized by global, low order oscillations of the supernova shock which
generate substantial non-radial flow in the post-shock region. These fluid
motions result in increased residence times of the accreted material and
possible increase the heating effect by neutrinos emitted by the nascent
proto-neutron star.

Since SASI is only one of the participating processes, it is often times
difficult to unambiguously quantify its contribution to the overall explosion
process. Only recently has careful analysis been attempted to quantify the
role of SASI and compare its contribution to the other major participating
process of neutrino-driven convection. In particular, SASI may only be present
in progenitors with higher masses when neutrino-driven convection might be
suppressed by increased accretion rates \citep[][see also \cite{bruenn+13}]{mueller+12}. Furthermore, the
situation is additionally complicated by effects due to model dimensionality
observed in some simulations
\citep{murphy+08,nordhaus+10,hanke+12,dolence+13,couch13,takiwaki+13}. Those findings
are, however, not confirmed by all studies \citep{janka+12}. (In passing, we
would like to note that is conceivable that the high frequency of SASI
observations in two-dimensional simulations can be due to the assumed symmetry
of such models. In either case, disentangling this kind of effects has not
been attempted in a systematic manner.) There is also little known about the
difference in the efficiency of convective heat transport between two and
three dimensions, especially in the context of core-collapse supernovae.
Although three-dimensional ccSNe explosion models have been available for some
time \citep[see, for example, ][]{fryer+02}, no systematic study aimed at
comparing the convection efficiency between two and three dimensions exists.

In this work, we investigate the energetic explosions of a $15$ \msun\ 
post-collapse WPE15 model of \cite{bruenn+93} in one, two, and three spatial
dimensions. This model has been used extensively in the past by Kifonidis and
his collaborators \citep[][see also
\cite{wongwathanarat+13}]{kifonidis+03,kifonidis+06,gawryszczak+10}, mainly in
application to post-explosion mixing of nucleosynthetic products. Here our
focus is solely on the explosion mechanism. To this end, we tune the
parameterized neutrino luminosity such that the explosion energy near
saturation matches the observationally constrained energetics of SN 1987A. In
the process, we have constructed a database containing more than 250
individual model realizations, differing in model dimensionality,
parameterized neutrino luminosity, and the small amplitude noise added to the
initial velocity. Obtaining such a large database of models was required given
the extreme sensitivity of models to perturbations, and one cannot draw
conclusions about the nature of the explosion process except in a statistical
manner. Furthermore, the observed differences between individual model
realizations reflect the combined model sensitivity and may provide
information about the role of, and coupling between, participating physics
processes.

In Section \ref{s:methods} we describe the physical and numerical models
adopted in our study. In Section \ref{s:expchar} we present overall properties
of the explosion models. In particular, we focus on dimensionality effects,
provide estimates of neutron star kick velocities, and analyze the flow
dynamics in the context of neutrino-driven convection and SASI. Next, in
Section \ref{s:convection}, we examine the role of convection in more detail.
We analyze the energy flow in gain region using energy flux decomposition, and
introduce and apply new methods for characterizing the morphology of the post-shock 
flow. Section \ref{s:turb} discusses the potential role of turbulence
in the explosion process, while in Section \ref{s:particle} we use the
Lagrangian view of the flow and track the energy histories of a large sample
of fluid parcels in the gain region in order to characterize conditions
leading up to shock revival. Finally, Section \ref{s:conc} contains our
discussion of results and conclusions of this work.
%
%
%
\section{Methods}\label{s:methods}
\subsection{Hydrodynamics}
We model the conservation of mass, momentum, energy, and electron fraction as
\begin{equation}
\label{e:continuity}
\frac{\partial\rho}{\partial t} + \vect{\nabla}\vect{\cdot}\left(\rho\vect{u}\right) = 0,
\end{equation}
\begin{equation}
\label{e:momentum}
\frac{\partial\rho\vect{u}}{\partial t} + \vect{\nabla}\vect{\cdot}\left(\rho\vect{u}\vect{\otimes}\vect{u}\right) + \vect{\nabla}P = -\rho\vect{\nabla}\Phi + \vect{Q}_{M},
\end{equation}
\begin{equation}
\label{e:energy}
\frac{\partial\rho E}{\partial t} + \vect{\nabla}\vect{\cdot}\left[\vect{u}\left(\rho E + P\right)\right] = -\rho\vect{u}\vect{\cdot}\vect{\nabla}\Phi + Q_{E} + \vect{u}\vect{\cdot}\vect{Q}_{M},
\end{equation}
\begin{equation}
\label{e:electronfraction}
\frac{\partial\rho Y_{e}}{\partial t} + \vect{\nabla}\vect{\cdot}\left(\rho Y_{e} \vect{u}\right) = Q_{N}.
\end{equation}
where $\rho$ is the mass density, $\vect{u}$ the fluid velocity, $P$ the
thermal pressure, $\Phi$ the gravitational potential, $E = u^2/2+e$ the
specific total energy (with $e$ being the specific internal energy), and $Y_e$
is the electron fraction. The source terms $\vect{Q}_M$, $Q_E$, and $Q_N$
account for the effects of neutrino-matter interaction (see, e.g.,
\cite{janka+96}).

Equations (\ref{e:continuity})--(\ref{e:electronfraction}) are evolved using
the neutrino-hydrodynamics code of \citep{janka+96,kifonidis+03,janka+03}. The
hydrodynamics solver of this code is built upon the Piecewise Parabolic Method
(PPM) of \cite{colella+84}, and applies Strang splitting \citep{strang+68} to
evolve multidimensional problems.

The numerical fluxes inside grid-aligned shocks are calculated using the HLLE
Riemann solver of \cite{einfeldt+88} to avoid the problem of even-odd
decoupling \citep{kifonidis+06}; otherwise, the hydrodynamic fluxes are calculated
using the Riemann solver of \cite{colella+85} is used.
\subsection{Equation of State}
We adopt the equation of state (EOS) outlined by \cite{janka+96}. This EOS
consists of contributions from free nucleons, $\alpha$-particles, and a heavy
nucleus in nuclear statistical equilibrium. Nuclei and nucleons are treated as
ideal, nonrelativistic Boltzmann gases, with electrons and positrons
contributing as arbitrarily degenerate and arbitrarily relativistic, ideal
Fermi gases. Thermal effects between photons and massive particles are taken
into account. Nuclear statistical equilibrium was assumed for $T \gtrsim 0.5$
MeV ($\sim 5.8\times10^{9}$ K). Good agreement has been reported between this
EOS and the \cite{lattimer+91} EOS for $\rho \lesssim 5\times10^{13}$
g cm$^{-3}$ \citep{janka+96}.
\subsection{Gravity}
The Newtonian gravitational potential field is treated as a composite of a point mass
field (for the excised PNS), and a distributed field (for the mass on
the mesh). The composite field is computed using a multipole
expansion of the Poisson equation for gravity,
\begin{equation}
\label{e:gravity}
\vect{\nabla}^2\Phi_{Newton} = 4\pi G\rho,
\end{equation}
where $G$ is Newton's gravitational constant. Spherically symmetric effects
due to general relativity are accounted for using the effective relativistic
potential of \cite{rampp+02} such that,
\begin{equation}
\label{e:gravitycorrection}
\Phi_{final} = \Phi_{Newton} - \Phi_{Newton}^{1D} + \Phi_{TOV}^{1D},
\end{equation}
where $\Phi_{Newton}^{1D}$ and $\Phi_{TOV}^{1D}$ are the spherically symmetric
Newtonian and Tolman-Oppenheimer-Volkoff potentials, respectively. See
\cite{kifonidis+03,marek+06,murphy+08} for the details of this method. Thus,
only in two-dimensions we account for deviations from spherical symmetry in
the Newtonian part of the potential. Otherwise, both the Newtonian and general
relativistic contributions are spherically symmetric.
\subsection{Numerical grid}
\label{s:grid}
All simulations were performed in spherical geometry. In the radial direction
we utilize 450 logarithmically spaced zones. The inner boundary is located at
a time-dependent location to mimic contraction of the collapsing proto-neutron
star core (see Section \ref{s:models} for details of this parameterization). The
outer boundary is located at $4\times10^{9}$ cm. The choice of outer boundary
radius ensures that the total mass on the mesh remains essentially constant
throughout the simulation. In multidimensional models we excise a $6$\degree\
cone around the symmetry axis in order to minimize potential numerical
artifacts associated with the symmetry axis (see \cite{scheck+07} for
justification of this approach).

Multidimensional models have angular resolution of $2$\degree\ and $3$\degree\ in
2D and 3D, respectively. Additionally, a representative 3D model was selected
to conduct a series of coarse resolution simulations at $6$\degree, $12$\degree,
and $24$\degree\ (cf.\ Section \ref{s:conc}).
\subsection{Boundary conditions}
We impose reflecting boundary conditions at the symmetry axis. In three
dimensions, we use periodic boundaries in the azimuthal direction. The outer
radial boundary is transmitting (zero gradient), while the inner boundary is
an impenetrable wall with the radial pressure gradient matching the
hydrostatic equilibrium condition.

In order to account for continued contraction of the neutron star, we
parameterize the inner boundary position, $r_{ib}$, according to
\begin{equation}
r_{ib}\left(t\right) = \frac{r_{ib}^{i}}{1 + \left[1-\exp{\left(-t/t_{ib}\right)}\right]\left(r_{ib}^{i}/r_{ib}^{f}-1\right)},
\end{equation}
as described in \cite{janka+96}. The contraction of the proto-neutron star is
parameterized by a characteristic timescale, $t_{ib}$. We make this
parameterization under the assumption that even in fully consistent models,
uncertain physics such as the neutrino-matter interaction rates and the
(stiffness of) the nuclear equation of state will inevitably lead to
uncertainties in the cooling of the neutron star, and therefore its
contraction timescale. Larger values of contraction timescale correspond to
stiffer nuclear equations of state (slow contraction), while smaller
timescales lead to softer nuclear equations of state (fast contraction). 

In the present work, we consider families of both fast and slow contracting
PNS core models, denoted as SC and FC models, respectively, to reflect the
above uncertainty. The first family are the slowly contracting proto-neutron
star models with $r_{ib}^{f} = 15$ km and $t_{ib} = 1$ s. The second family
are fast contracting proto-neutron star models with $r_{ib}^{f} = 10.5$ km and
$t_{ib} = 0.25$ s. In both cases we use $r_{ib}^{i} = 54.47$ km.

\subsection{Neutrinos}
\label{s:neutrino}
In our implementation of neutrino energy deposition we use a light-bulb
approximation \citep{janka+96} with modifications introduced later by
\cite{scheck+06} (see below). Accordingly, we impose an isotropic,
parameterized neutrino luminosity, $L_{\nu}^{tot,0}$. The initial total
neutrino luminosity is chosen such that it asymptotically represents the
gravitational binding energy released by the neutron star core
\citep{janka+96,scheck+06,scheck+07},
\begin{equation}
\Delta E_{core}^{\infty} = \int_{0}^{\infty} L_{\nu}^{tot,0}h(t)dt = 
       3L_{\nu}^{tot,0}t_{ib},
\end{equation}
The neutrino luminosity varies in time as 
\begin{equation}
L_{\nu_{x}}\left(r_{ib},t\right) = L_{\nu}^{tot,0} K_{\nu_{x}} h\left(t\right),
\end{equation}
where $\nu_{x}\equiv \nu_{e}, \overline{\nu}_{e}, \nu_{\mu},
\overline{\nu}_{\mu}, \nu_{\tau}, \overline{\nu}_{\tau}$. The $K_{\nu_{x}}$
terms represent the constant, fractional contributions from each neutrino type
to the total neutrino luminosity. The $h(t)$ function describes the temporal
evolution of the neutrino luminosity and is given by \cite{scheck+06}
\begin{equation}
h\left(t\right) = 
\begin{cases} 
\mbox{1.0,} &\mbox{if } t\le t_{ib}\mbox{,}\\
\left(t_{ib}/t\right)^{3/2} &\mbox{if } t>t_{ib}\mbox{.}
\end{cases} 
\end{equation}
\subsection{Tracer Particles}
We utilize passive tracer particles to provide additional analysis of the flow
structure. Particles are initially distributed uniformly in Lagrangian mass
coordinate. Each particle represents a fluid parcel with a mass of
$1\times10^{-6}$ \msun\ in two dimensions and $1\times10^{-7}$ \msun\ in three
dimensions (thus, the total number of particles is $7.6\times10^{5}$ and
$7.6\times10^{6}$ in two and three-dimensional models, respectively). We note
that our choice of distributing particles uniformly in Lagrangian mass
results in large numbers of particles near the inner boundary due to high
densities there. Consequently, not all particles sample the most dynamically
active parts of the flow. Therefore, we limit our analysis of particle-based
data to those inside the gain region.
\subsection{Initial conditions}
\label{s:progmodel}
For the initial conditions we use the post-collapse model WPE15 ls(180) of
\cite{bruenn+93} based on the 15 \msun\ blue supergiant progenitor model by
\cite{woosley+88}. We map the spherically symmetric collapse model to the grid
and in multidimensions we add small velocity perturbations with relative
amplitude $1\times10^{-3}$ of the radial velocity.
\subsection{Tuning explosion energies}
\label{s:models}
In the current work we are interested in explosion models with energetics
(with energy per unit mass) matching that of SN 1987A. Estimates of the SN
1987A explosion energetics are summarized in
Table \ref{t:explosionestimates}).
\begin{deluxetable*}{llll}
\tablewidth{0pt}
\tablecaption{Estimated explosion energetics of SN 1987A.\label{t:explosionestimates}}
\tablehead{
\colhead{\hspace*{15em}} & 
\colhead{E/M} & 
\colhead{$M_{ej}$} & 
\colhead{$E_{exp}$}
\\ 
\colhead{} & 
\colhead{($10^{50}$ erg/\msun)} & 
\colhead{($\msun$)} & 
\colhead{($10^{51}$ erg)}
} 
\scriptsize{4pt}
\startdata
\cite{arnett87}		&	\nodata			&	\nodata			&	1--2	   \\
\cite{chugai+88}	&	\nodata			&	\nodata			&	1.3	   \\
\cite{woosley+88}	&	0.8				&	\nodata			&	0.8--1.5	  \\
\cite{shigeyama+90}	&	1.10$\pm$0.3	&	\nodata			&	1.0$\pm$0.4	 \\
\cite{imshennik+92}	&	0.75$\pm$0.05	&	\nodata			&	1.05--1.2	 \\
\cite{utrobin+93}	&	0.85			&	15--19			&	\nodata	  \\
\cite{blinnikov+99}	&	0.75$\pm$0.17	&	14.7			&	1.05$\pm$0.25	\\
\cite{utrobin+04}	&	0.67			&	18				&	1.2	   \\
\cite{utrobin+05a}	&	0.83			&	18$\pm$1.5		&	1.5$\pm$0.12	\\
\cite{utrobin+05b}	&	0.83			&	18				&	1.5	   \\
\cite{pumo+11}		&	0.59			&	16--18			&	1.0	
\enddata
\end{deluxetable*}
Based on this summary, we adopt $E/M = 0.75\times10^{50}$ erg \msun$^{-1}$,
which for the adopted progenitor model corresponds to an explosion energy of
approximately $1.05\times10^{51}$ erg.

We define the total model positive binding energy as 
%
%
%
\begin{equation}
E_{exp} = \sum \rho\left[\frac{1}{2}u^2 + \varepsilon + \Phi\right],
\end{equation}
where the summation is taken over grid cells for which the binding energy is
positive, $1/2u^2 + \varepsilon + \Phi>0$. Here $u$ is the velocity magnitude,
$\varepsilon$ is the internal energy, and $\Phi$ is the gravitational
potential. We define the explosion time as the moment when the positive
binding energy exceeds $1\times10^{48}$ erg \citep{janka+96}. From that point
in time on the total positive binding energy is identified as the explosion
energy.

In order to match the model energetics with that of SN 1987A, we computed a
large number of trial explosion models with various neutrino luminosities
recording their explosion energies at the final time, $t = 1.5$ s. Once we
identified the range of viable neutrino luminosities, we computed additional
explosion models varying the seed velocity perturbations.

The above procedure follows from the realization that due to strong model
sensitivity to small perturbations a unique mapping between the neutrino
luminosity and the final energy does not exist. Consequently, one can only
study the model energetics in a statistical sense. Then, for a given neutrino
luminosity one can consider an explosion model characterized by average
properties with possibly large dispersion. Our numerical experiments indicate
that one can produce equally energetic explosions for neutrino luminosities
varying by as much as 3\%.
%
%
%
%
%
%
\section{General properties of explosion models}\label{s:expchar}
We obtained a large database of supernova explosion models in one-, two-, and
three-dimensions. The database contained 26 spherically symmetric models, 169
2D models, and 61 3D models. The individual model realizations differed in
mesh resolution, parameterized neutrino luminosity, and random perturbation
pattern. Table \ref{t:models} provides a summary of a subset of the database
of explosion models
\begin{deluxetable*}{lcccccc}
\tablecaption{Parameters and main properties of the explosion models. \label{t:models}}
\tablewidth{13cm}
\tablehead{
\colhead{Model} & 
\colhead{\tablenotemark{a}$L_{\nu_{e}}$} & 
\colhead{\tablenotemark{b}$t_{exp}$} & 
\colhead{\tablenotemark{c}$E_{exp}$} &
\colhead{\tablenotemark{d}$\dot{M}_{exp}$} & 
\colhead{\tablenotemark{e}$\overline{r}_{s}^{exp}$} 
\\ 
\colhead{} & 
\colhead{($10^{52}$ erg/s)} & 
\colhead{(ms)} & 
\colhead{($10^{51}$ erg)} &
\colhead{(\msun/s)} & 
\colhead{(km)} 
} 
\tablecolumns{6}
\startdata
\multicolumn{6}{c}{Slow contracting models - 2D}\\*
\hline\\*
M194A	&	1.943	&	161	&	0.973	&	0.282	&	657\\
M194B	&	1.943	&	184	&	1.038	&	0.265	&	765\\
M194C	&	1.943	&	202	&	0.933	&	0.254	&	820\\
M194D	&	1.943	&	186	&	1.025	&	0.260	&	847\\
M194E	&	1.943	&	167	&	1.145	&	0.274	&	744\\
M194F	&	1.943	&	200	&	0.925	&	0.258	&	755\\
M194G	&	1.943	&	166	&	1.080	&	0.276	&	759\\
M194H	&	1.943	&	207	&	0.937	&	0.248	&	867\\
M194I	&	1.943	&	194	&	1.105	&	0.260	&	732\\
M194J	&	1.943	&	148	&	1.054	&	0.288	&	721\\
\hline\\*
\multicolumn{6}{c}{Slow contracting models - 3D}\\*
\hline\\*
M187A	&	1.871	&	189	&	1.066	&	0.259	&	814\\
M187B	&	1.871	&	193	&	1.034	&	0.256	&	822\\
M187C	&	1.871	&	184	&	1.054	&	0.265	&	793\\
M187D	&	1.871	&	187	&	1.036	&	0.260	&	818\\
M187E	&	1.871	&	182	&	1.058	&	0.264	&	805\\
\hline\\*
\multicolumn{6}{c}{Fast contracting models - 2D}\\*
\hline\\*
M352A	&	3.528	&	109	&	1.074	&	0.315	&	697\\
M352B	&	3.528	&	114	&	1.037	&	0.311	&	711\\
M352C	&	3.528	&	115	&	1.089	&	0.310	&	724\\
M352D	&	3.528	&	112	&	0.995	&	0.312	&	707\\
M352E	&	3.528	&	110	&	0.990	&	0.311	&	702\\
M352F	&	3.528	&	114	&	1.139	&	0.313	&	707\\
M352G	&	3.528	&	110	&	0.968	&	0.316	&	667\\
M352H	&	3.528	&	107	&	1.221	&	0.317	&	666\\
M352I	&	3.528	&	117	&	0.962	&	0.305	&	772\\
M352J	&	3.528	&	113	&	1.019	&	0.309	&	729
\enddata
\tablenotetext{a}{Parameterized electron neutrino luminosity. The model anti-electron neutrino luminosity is 7.5\% larger.}
\tablenotetext{b}{Explosion time determined when the total positive binding energy is greater than $1\times10^{48}$ erg.}
\tablenotetext{c}{Total positive binding energy (i.e. explosion energy) evaluated at $t = 1.5$ s post-bounce.}
\tablenotetext{d}{Mass flow rate evaluated immediately upstream from the shock at $t = t_{exp}$.}
\tablenotetext{e}{Average shock radius evaluated at $t=t_{exp}$.}
\end{deluxetable*}
that most closely match the energetics of SN 1987A. The subset contains 10
models in two-dimensions for both slow and fast contracting families, and 5
slow contracting models in three-dimensions.

The model explosion times vary from about $200$ ms in the case of slow
contracting models to slightly above $100$ ms in the case of fast contracting
models. Every family of models shows intrinsic dispersion in both the
explosion times and the explosion energies. For example, the explosion times
for slow contracting models vary by about 30\% in 2D, and about 5\% in 3D. The
variations in explosion times are comparatively modest in the case of fast
contracting models in 2D, with observed variations not exceeding 10\%. The
corresponding variations in the energetics are about 20\% in 2D for slow and
fast contracting models. The observed variations of our 3D models does not
exceed 4\%.

We discuss the accretion rates and shock radii in Sections \ref{s:lvse} and
\ref{s:shock}, respectively.
\subsection{Effects of neutrino luminosity on explosion energy}\label{s:lvse}
\begin{figure}[t!]
\centering
\includegraphics[width=8cm]{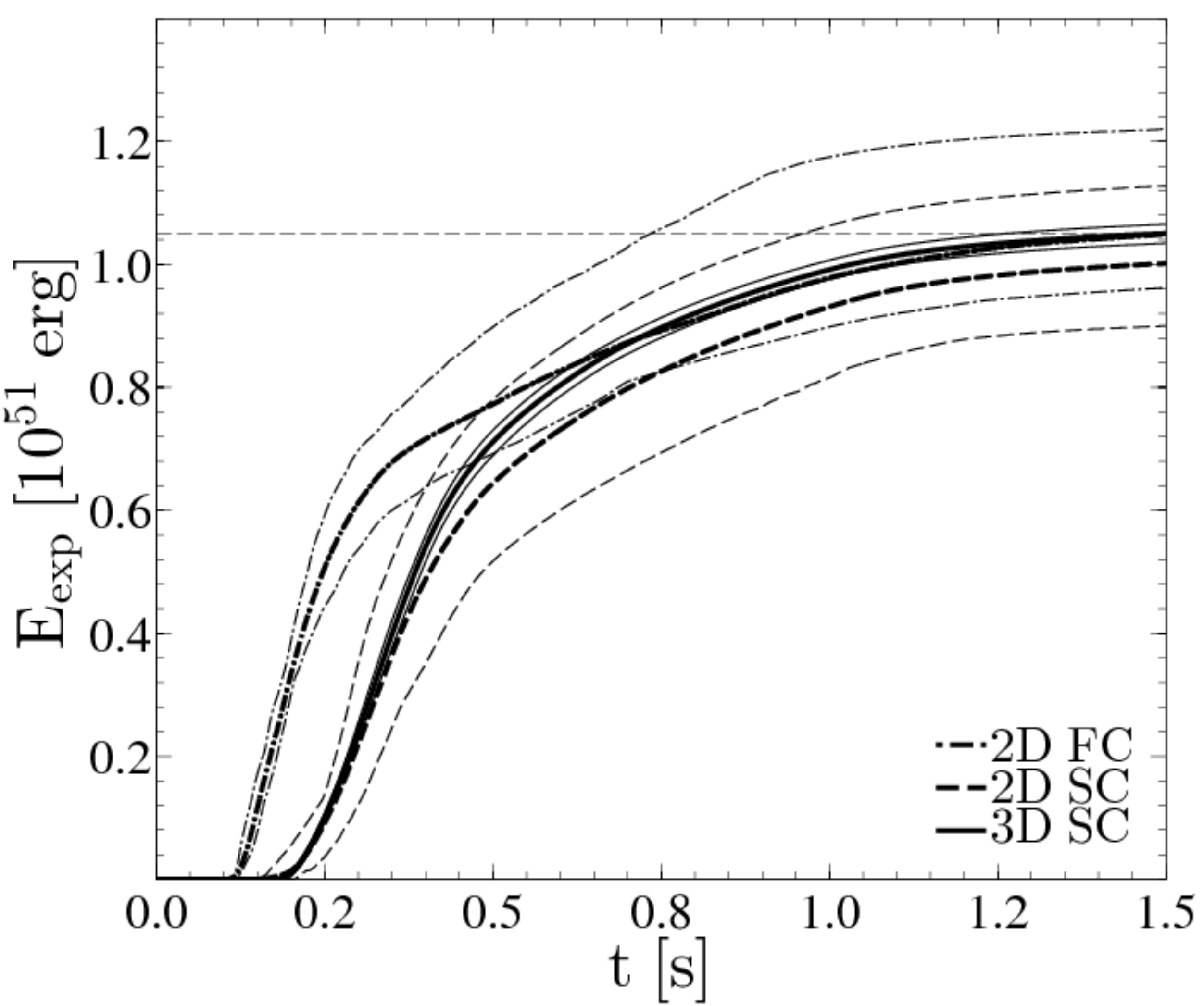}
\caption
{
Explosion energy for models tuned to the energetics of SN 1987A (saturation
at $\approx 1.05\times10^{51}$ erg). Thick curves denote the average over all
models in the group; thin curves denote the respective minimum and maximum
values. Note that the fast contracting proto-neutron star models begin
unbinding material in roughly half the explosion time of the standard
contracting models.
\label{f:exp_v_time}}
\end{figure}

Figure \ref{f:exp_v_time} 
shows the explosion energy in models with both slow
and fast contracting PNS cores. We show the mean explosion energy and the
explosion energy envelope for 2D FC models (dash-dot), 2D SC models (dash),
and 3D SC models (solid). In all cases, the explosion energy rises rapidly at
the onset of explosion and approximately saturates by the final time. In the
case of SC models, the explosion energy threshold is reached between 140 ms to
200 ms post-bounce depending on model realization. The two dimensional models
tend to explode faster and reach their final energies later than the three
dimensional models. The FC models explode significantly earlier, at
approximately 110 ms on average. The short explosion times found in FC
models, as compared to SC models, are not unexpected. This is because
significantly greater neutrino luminosities, approximately by a factor of two,
are required to energize the explosion in the presence of the deeper
gravitational potential well in the FC models. Regardless of the contraction
times of the PNS core, by the final time, $t_f = 1.5$ s, and given the
required explosion energetics and envelope mass, the average explosion energy
in our models reaches about $1.05\times10^{51}$ erg.

It is instructive to discuss our models in terms of the critical luminosity
curve \citep{burrows+93}. We shall note that finding the curve was not the goal
of our study, and therefore, we are not in a position to provide the exact
relation. Our results only provide an upper limit on the critical neutrino
luminosity for the (relatively small) range of accretion rates present in our
models (see Table \ref{t:models}).

\begin{figure*}[t!]
\centering
\includegraphics[width=8cm]{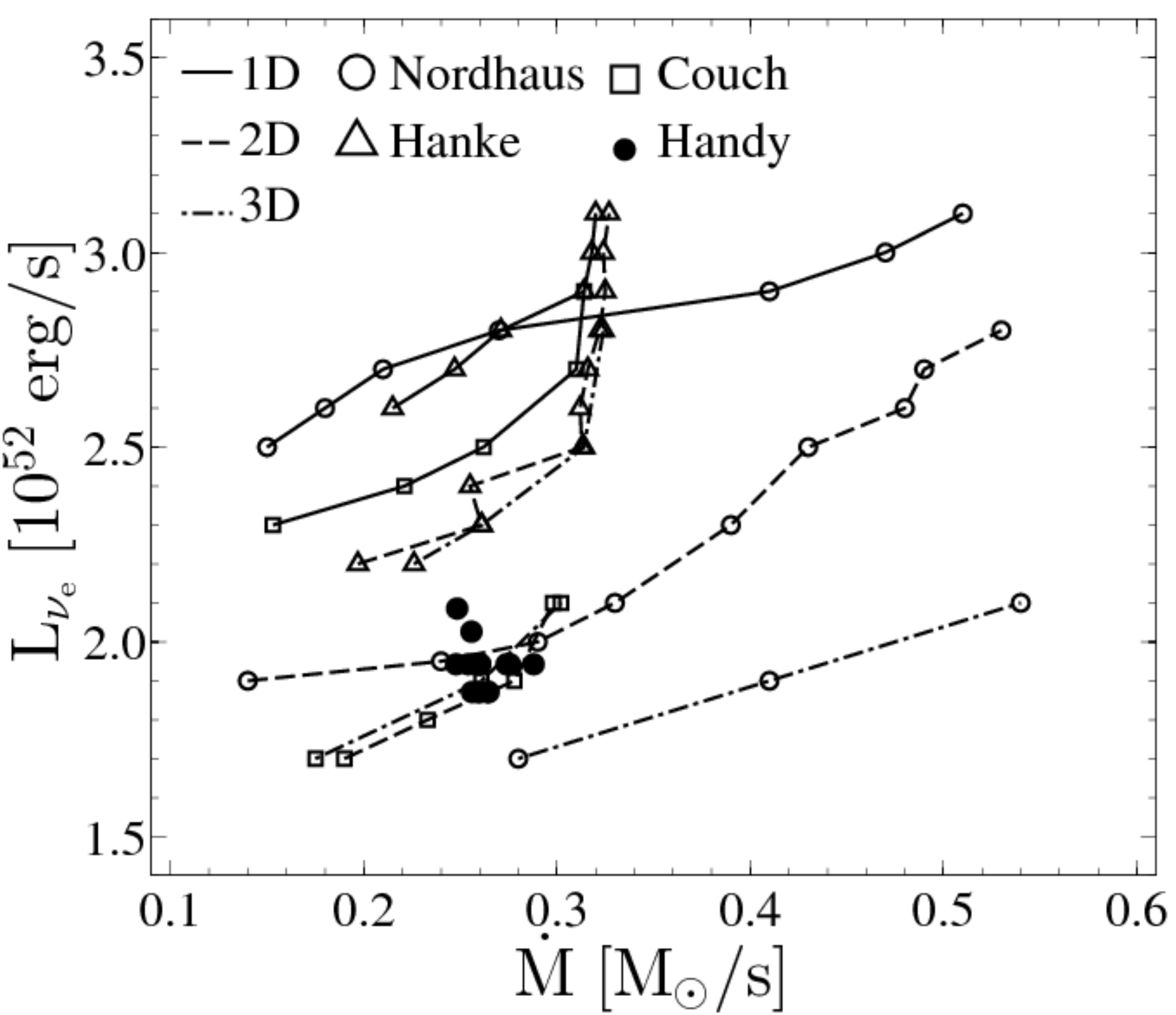}
\includegraphics[width=8cm]{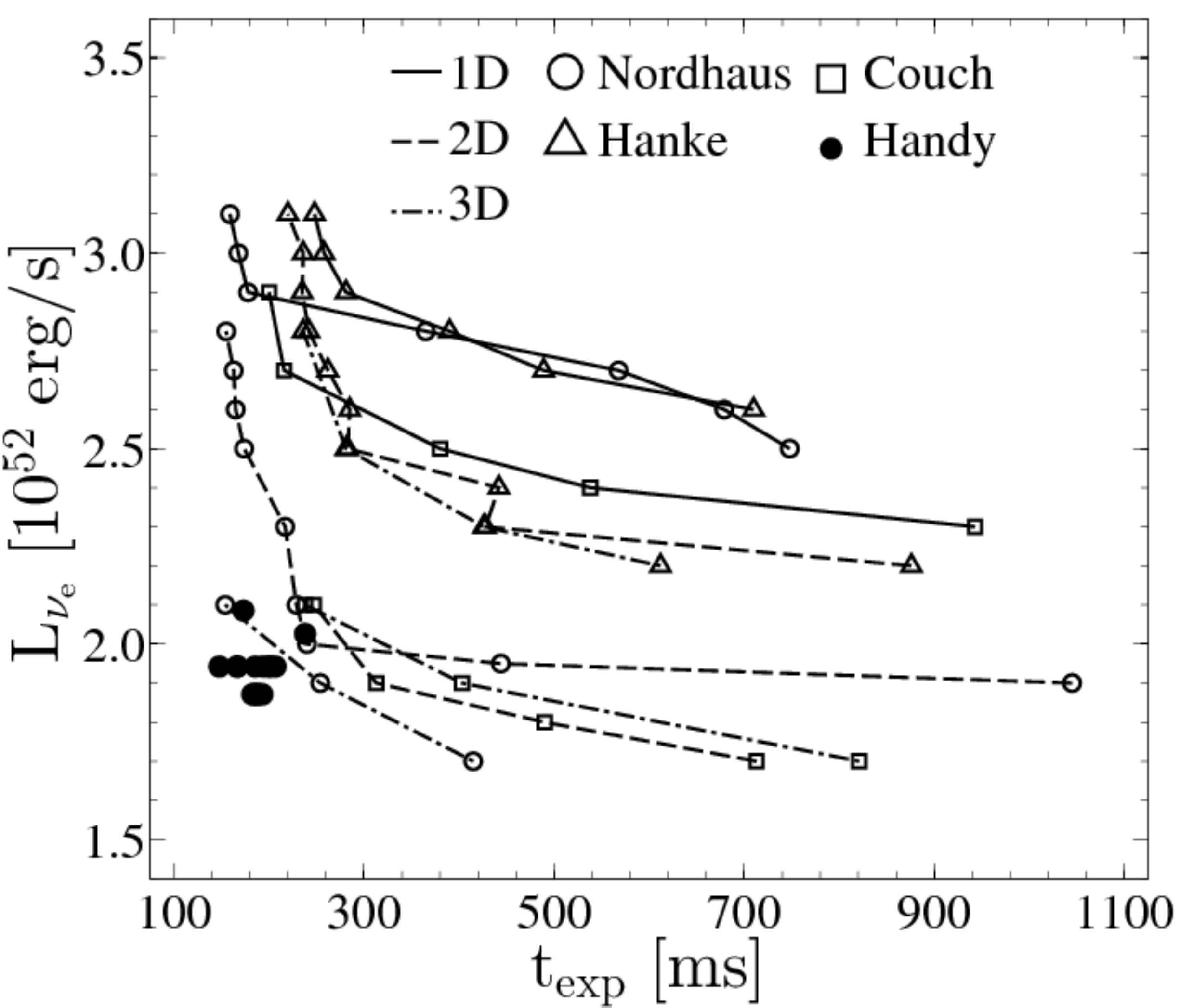}
\caption
{
Comparison of our explosion models with the results of critical neutrino
luminosity studies conducted by other groups. The results of individual
studies are shown with different symbols, while model dimensionality is
represented by line style (1D: solid; 2D: dashed; 3D: dash-dotted). (left
panel) The dependence of the critical neutrino luminosity on the accretion
rate at the time of explosion. We observe a systematic decrease in the
required neutrino luminosity as the model dimensionality increases. (right
panel) Parameterized electron neutrino luminosity as a function of explosion
time. For a given dimensionality, our models tend to explode sooner and
require lower neutrino luminosities.\label{f:lum}
}
\end{figure*}

Figure \ref{f:lum}
provides a comparison of our explosion models with the results of critical
neutrino luminosity studies conducted by other groups. In the figure, the
results of \cite{nordhaus+10}, \cite{hanke+12}, and \cite{couch13} are shown
with open circles, triangles, and squares, respectively, while our data are
shown with solid circles; the model dimensionality is indicated with line
style with solid, dashed, and dash-dotted lines connecting models obtained in
1D, 2D, and 3D, respectively. In the left panel, the neutrino luminosity
is shown as a function of the accretion rate, which is measured by integrating
the mass flux through the surface located just upstream of the shock at the
onset of explosion.

This figure indicates that our choice to constrain the final explosion energy
by observations required a relatively narrow range of neutrino luminosities,
compared to other studies. We also find that the required neutrino luminosity
systematically decreases as the dimensionality of the model increases.
Neutrino luminosities found in our models only provide an upper limit for the
critical luminosity for the mass accretion rates present in our simulations.
Furthermore, explosion times found in our models are shorter than those of
other groups for similar neutrino luminosities. The simplest explanation for
this finding is that the neutrino heating is more efficient in our models.
However, we cannot exclude the possibility that this observed, systematic
difference is due to differences in the progenitors. Detailed investigation of
possible differences between our findings and those of other groups is beyond
the scope of the current work.
\subsection{Neutron star recoil}

Since we excise the neutron star from the grid, a special method has to be
used to estimate its recoil velocity. To this end, we use the approach of
\cite{scheck+06}, and exploit momentum conservation. In this method, the
momentum of the neutron star balances the momentum of the surrounding envelope
(total momentum on the mesh in our simulations). An additional correction can
be applied to the neutron star recoil velocities by considering that the
neutron star will continue accreting material beyond the final time in our
simulations. We estimate the corrected velocity as
\begin{equation}
\label{e:v_inf}
v^{\infty}_{H} = v^{1.5}_{H} + v^{3-1.5}_{H},
\end{equation}
where $v^{1.5}_{H}$ is the recoil velocity obtained in our simulations at
$t_f=1.5$ s, and $v^{3-1.5}_{H}$ is the approximate correction due to late
time accretion onto the neutron star. To calculate the velocity correction, we
assumed the rate of the recoil velocity change obtained by \cite{scheck+06} in
his B18 model series between the times $t=1.5$ s and $t=3.0$ s 
\cite[see Figure 19 in][]{scheck+06},
\begin{equation}
\left\langle{\frac{v^{3-1.5}_{S}}{v^{1.5-1}_{S}}}\right\rangle\approx 0.862.
\end{equation}
Then the recoil velocity correction has been computed as
\begin{equation}
v^{3-1.5}_{H} = \left\langle{\frac{v^{3-1.5}_{S}}{v^{1.5-1}_{S}}}\right\rangle v^{1.5-1}_{H},
\end{equation}
In the above formulae superscripts denote post-bounce explosion times, and the
subscripts $S$ and $H$ denote the values from \cite{scheck+06} and the
current work, respectively. 

\begin{figure}[t!]
\centering
\includegraphics[width=8cm]{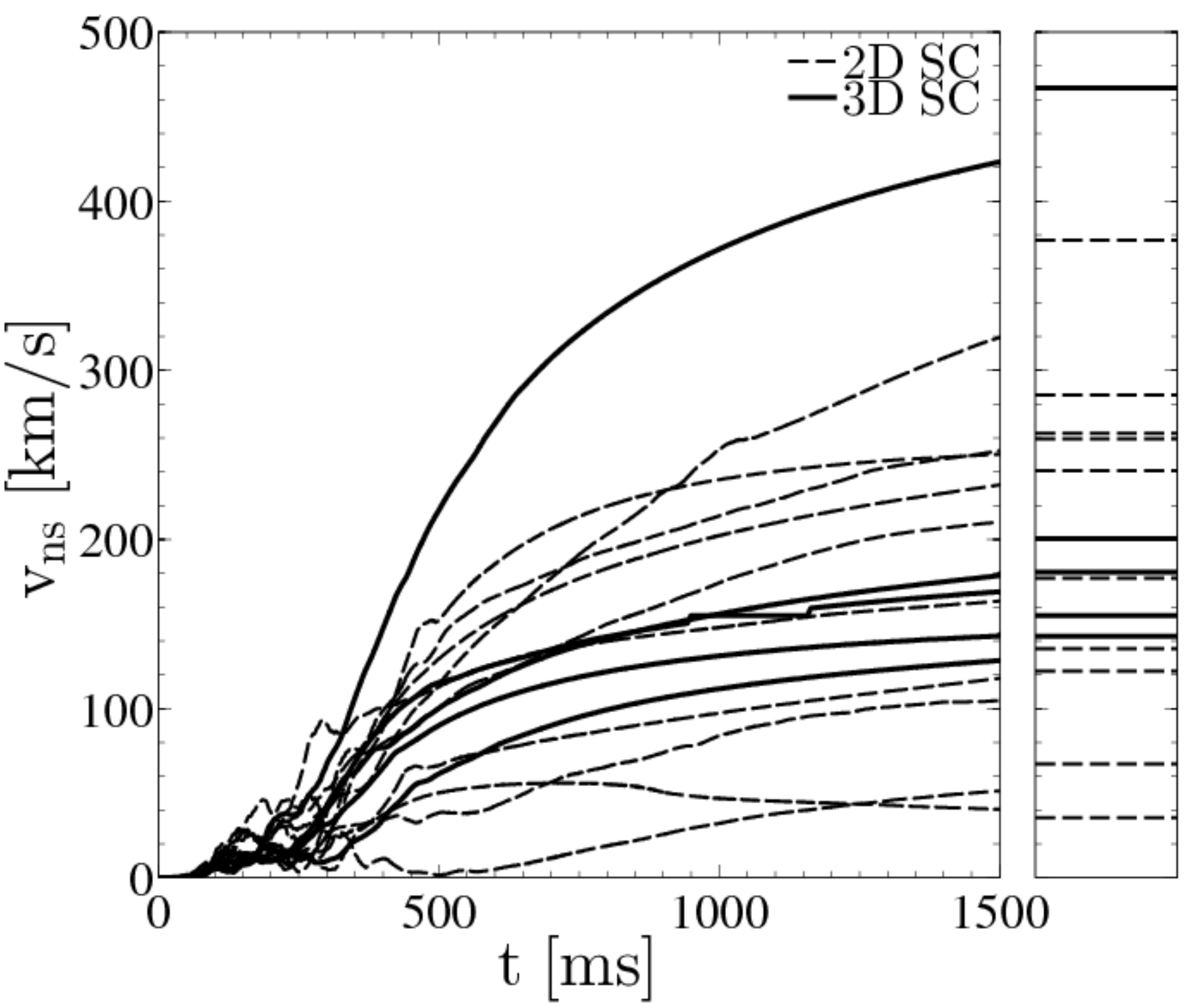}
\caption
{Neutron star kick velocity for our SC models, shown with dashed and solid
lines, respectively. The outset shows the estimated saturation velocities
based on estimates utilizing the results of \cite{scheck+06}.
\label{f:ns_vel}}
\end{figure}

Figure \ref{f:ns_vel} 
shows the evolution of recoil velocities in our models until the final time
(left panel), and the estimated saturation velocities based on Equation
\ref{e:v_inf} (right panel). Our results indicate relatively modest recoil
velocities compared to \cite{scheck+06}. The approximate final recoil
velocities are in the range between 100 km/s and 300 km/s, with the lowest and
highest recoil velocities of 30 km/s and 475 km/s. These results are in
qualitative agreement with the results reported by \citet[][Figure
20]{scheck+06} and \citet[][B-series models in Table 2]{wongwathanarat+13}.
\subsection{Gain region characteristics}
\label{s:grchar}

In this section we report the results following the analysis of a quasi-steady
state period in the pre-explosion epoch. If no such evolutionary stage is
reached then the entire pre-explosion phase could be considered a transient
phenomenon with little importance for the energetic core-collapse supernova
explosions considered here. If the opposite is true however, we can adopt
methods appropriate for analysis of quasi-steady state flows and apply them to
analyze the dynamics of the gain region.

\begin{figure*}[t]
\centering
\includegraphics[width=8cm]{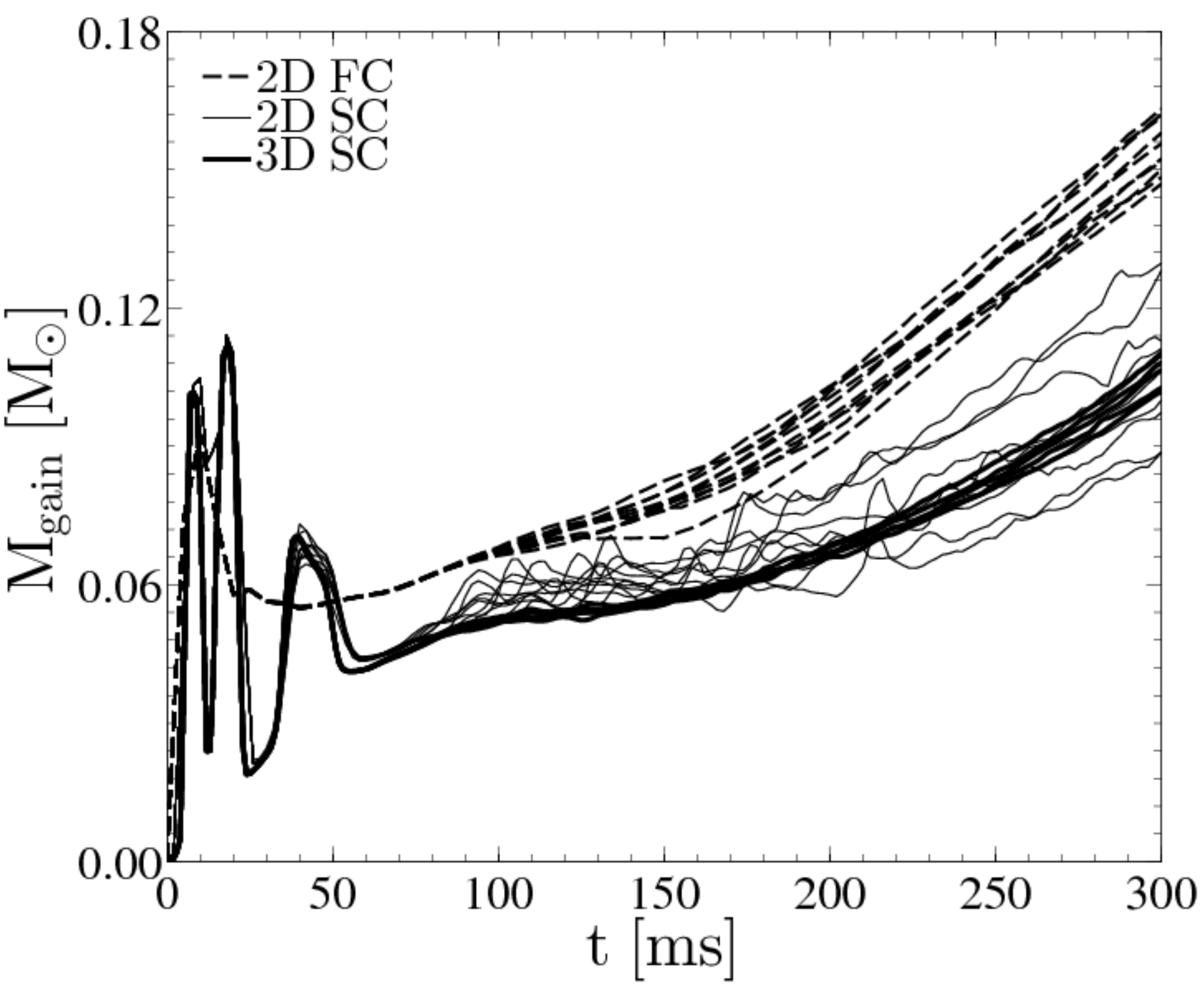}
\includegraphics[width=8cm]{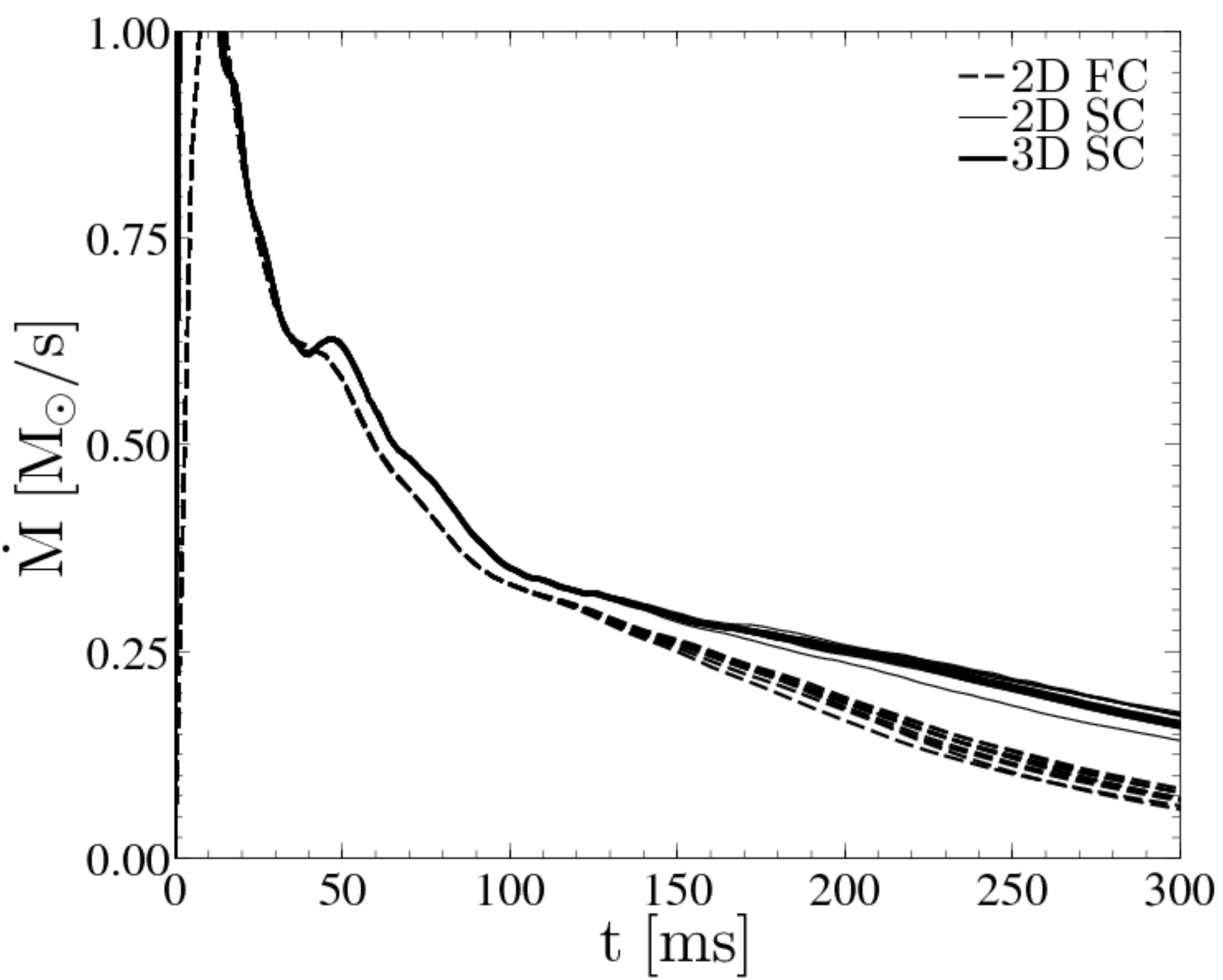}
\caption
{(left panel) Mass in the gain region. The evolution of mass in the gain
region is shown with solid lines for SC models (thin solid lines in 2D and
thick solid lines in 3D) and dashed lines for FC models. Note that the mass in
the gain region is, on average, greater in FC models than in SC models. Also,
after the initial transient oscillations, the mass in the gain region
stabilizes, indicating that evolution in the gain region has reached a quasi-steady 
state period. (right panel) Shock accretion rate. The line types are
associated with model family and model dimensionality as in the left
panel.Note that after the initial period of fast accretion the accretion rates
progressively decrease in both families of models. The accretion rates are, on
average, greater in SC models than in FC models after the initial transient.
See Section \ref{s:grchar} for details.}
\label{f:grmass}
\end{figure*}

We identify the quasi-steady state period in the post-shock flow evolution by
calculating the mass contained in the gain region, $M_{gain}$. Its time-dependence 
is shown in the left panel of Figure \ref{f:grmass}
for our set of multidimensional models. As one can see, the mass in the gain
region for the case of SC models (shown with solid lines) changes only
slightly between $t = 100$\ ms and $t = 150$\ ms. A similar period of modest
increase in the mass of the gain region can be found in the case of FC models
(shown with dashed lines) for times between $t=30$\ ms and $t=80$\ ms. We note
that in both cases, the moment when the mass in the gain region increases
precedes the explosion times, which is consistent with a scenario in which the
shock revival process occurs over an extended period of time in a quasi-steady
state fashion. Also note the presence of strong oscillations in the gain
region mass at early times in both families of models. Since, as we discuss
below (Section \ref{s:shock}), the shock radius is increasing steadily at
these early times, the observed oscillations in the gain region mass are
either due to density changes of the material entering the gain region or the
changes in the position of the gain radius during these times. Those
oscillations, however, have a transient character and do not play any role in
the subsequent evolution of the system, and cease as soon as the flow in the
gain region becomes multidimensional.

The mass accretion rate (shown in the right panel of Figure \ref{f:grmass}) as
measured at the position of the shock front shows a rapid decline during the
first 50 ms of the simulation time after which it steadily decreases at a
progressively slower rate. The evolution of accretion rate does not differ
significantly between SC and FC models. This is expected, as we consider only
a single progenitor model. At later times when the quasi-steady state is
reached, the accretion rates are between $0.29-0.35$ \msun\ s$^{-1}$ for SC
models and between $0.40-0.67$ \msun\ s$^{-1}$ for FC models. The FC models are
characterized by consistently lower accretion rates than the SC models for
times later than 50 ms.

To better understand the dynamics in the gain region in the context of
neutrino heating, we consider the advection timescale, $\tau_{adv}$,
\begin{equation}
\tau_{adv} = \frac{M_{gain}}{\dot{M}},
\end{equation}
where $M_{gain}$ is the mass in the gain region, and $\dot{M}$ is the
accretion rate. The advection time scale is the characteristic time a fluid
parcel spends in the gain region. This quantity neglects multidimensional
effects, which are known to be important in the process of shock revival. In
the same context, we consider the heating efficiency, $\eta$,
\begin{equation}
\eta = \frac{\tau_{adv}}{\int \rho\dot{Q}dV},
\label{e:eta}
\end{equation}
where $\dot{Q}$ is the net neutrino energy deposition rate, and the integral
in the above equation is taken over the gravitationally bound ($\frac{1}{2}u^2
+ \varepsilon + \Phi<0$) material inside the gain region. The quantity in the
denominator of the above equation is the heating timescale, i.e.\ the time it
takes to heat a gravitationally bound fluid parcel so that it becomes unbound.
The heating efficiency is a measure of the competition between the advection
and heating processes. In particular, $\eta>1$ implies that the material in
the gain region gains energy due to heating faster than it is removed by the
advection process.

\begin{figure}[t!]
\centering
\includegraphics[width=8cm]{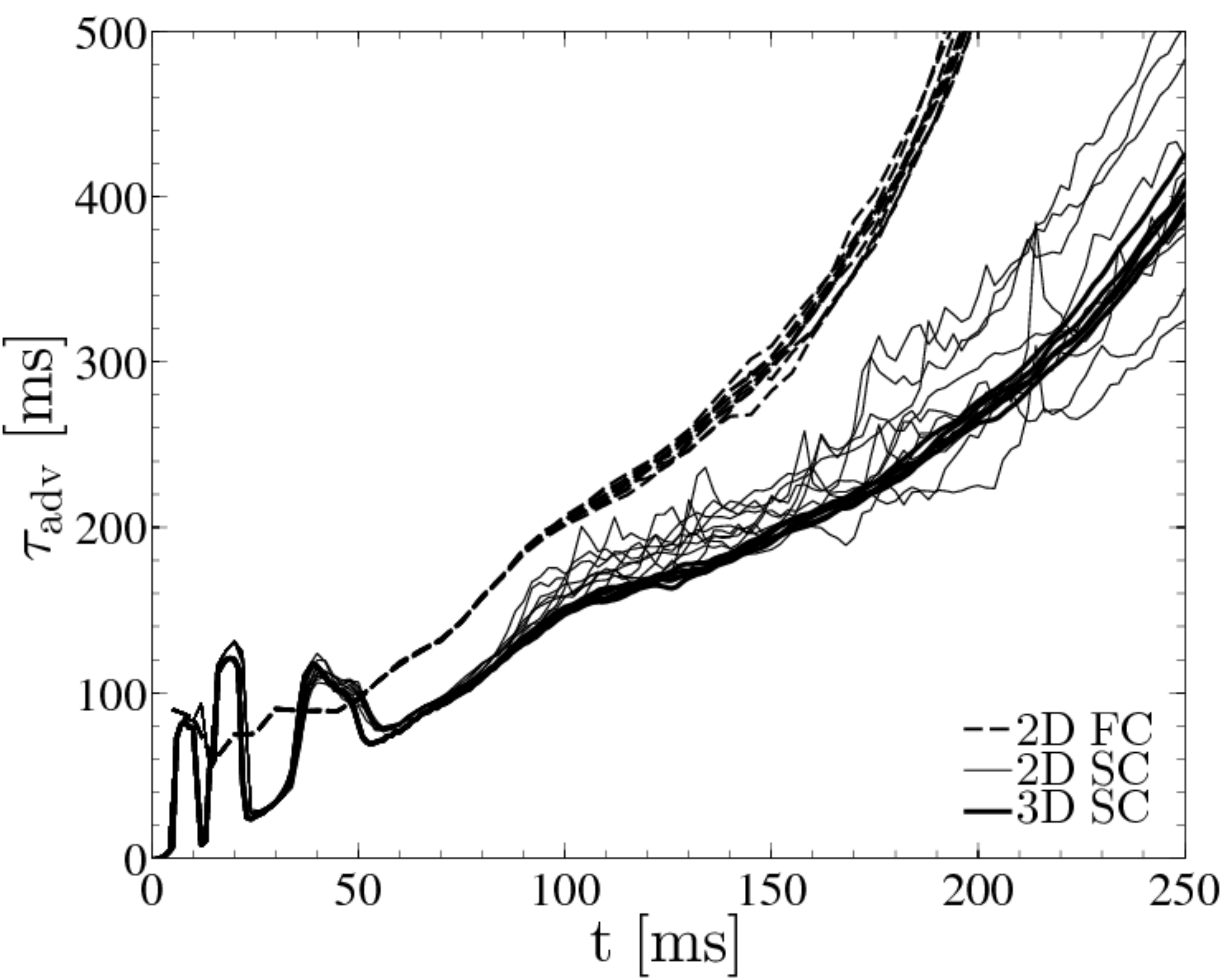}
\caption
{
Advective timescale for the gain region. The oscillatory behavior of the
advective timescale at early times for SC models likely has the same origin
as the transient oscillations observed in the gain region mass. Note that the
advective times are similar between model realizations in the same family. The
differences in advective times for different model realizations are
particularly large for 2D SC models that develop a few large-scale
nonuniformities in the gain region. On average, the advective times are longer
for FC models than for SC models. See Section \ref{s:grchar} for details.
\label{f:advtimes}}
\end{figure}

The time evolution of advective time for our sample of models is shown in
Figure \ref{f:advtimes}.
Since the advective time depends on the mass in the gain region, the
oscillations in the advective time visible in Figure \ref{f:advtimes} prior to
50 ms are simply a reflection of the oscillations in the gain region mass
(c.f. left panel in Figure \ref{f:grmass}). At later times, both SC and FC
families (shown with solid and dashed lines in Figure \ref{f:advtimes}) show
qualitatively similar behavior. Again, this is expected due to a common
progenitor model used in these calculations. Also, and as we discussed above,
the variations between model realizations are due to differences that develop
in multidimensional flow structure during the shock revival process.

We can compare the advective time to the length of the quasi-steady state
phase of the evolution. For the SC models, we estimate the steady state lasts
about 50 ms, while the advective time in this case is about 160 ms. In the
case of FC models, we estimate the steady state lasts about 50 ms, but the
advective time is somewhat shorter (about 100 ms). We conclude that in our
models, on average, fluid parcels enter the gain region and reside in it
during the entire process of shock revival. This conclusion is consistent with
the history of individual fluid parcels (see Section \ref{s:particle} below).

The advective times found in our simulations are 3 to 10 times larger than
those reported by \cite{mueller+12}. We initially suspected that this
significant difference in advective times might be due to general relativistic
effects in the potential carefully accounted for by M\"{u}ller et al.\ in
their model. However, we believe the most probable explanation is the
difference in the model energetics obtained in the two studies. This is
because in the models of M\"{u}ller et al.\ the shock resides during the shock
revival phase for long times at much lower radii ($\approx100$ km) than in our
models ($\approx400$--$500$ km). Therefore, provided the accretion rates are
similar, the gain region has much lower mass in their models, resulting in
correspondingly shorter advective times.

\begin{figure}[t!]
\centering
\includegraphics[width=8cm]{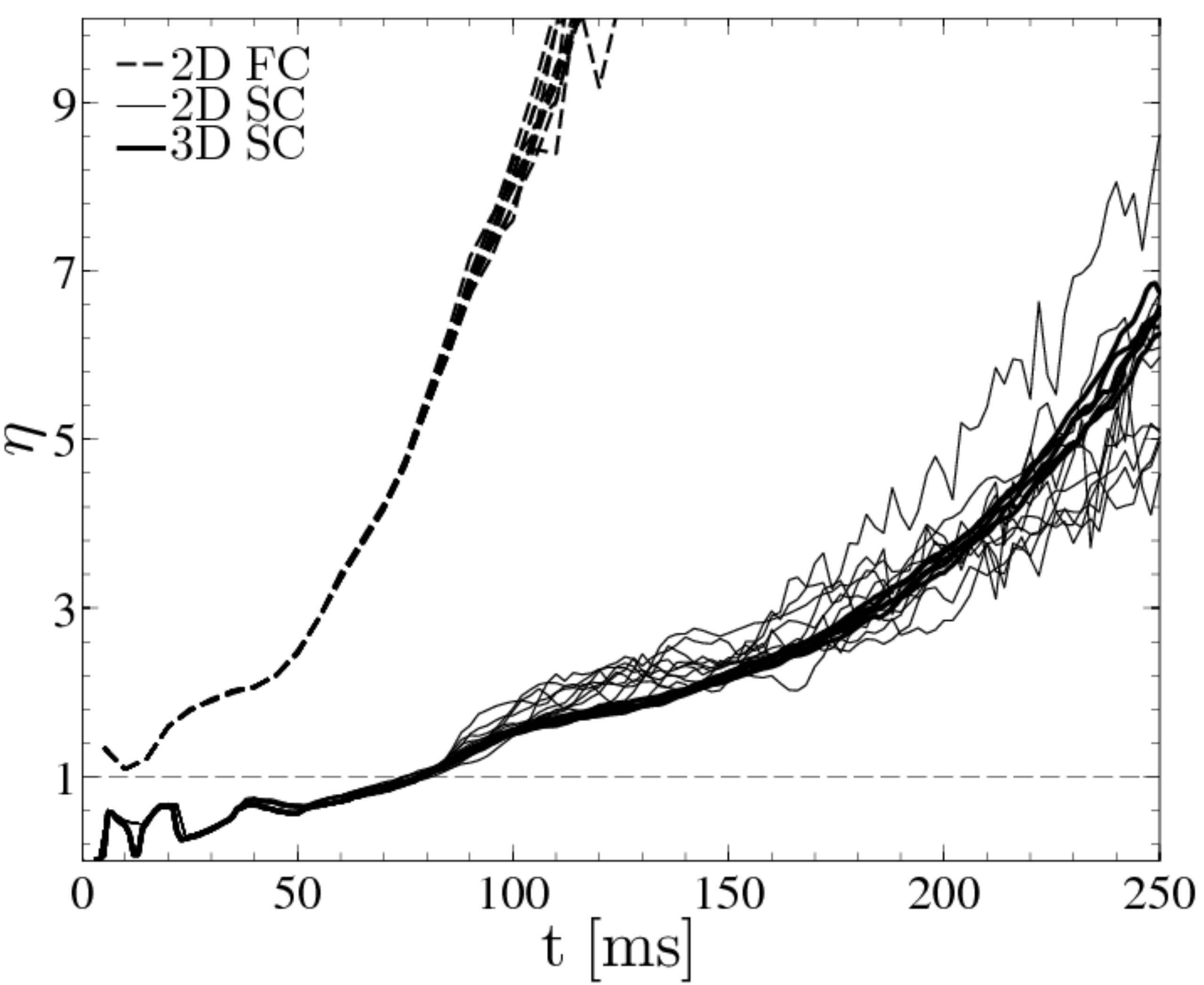}
\caption
{Heating efficiency for the gain region. The evolution of heating efficiency is
shown for SC and FC models with solid and dashed lines, respectively. The
results from the 2D SC models are shown with thin solid lines. Note that the
heating efficiency in the FC models is always greater than one. This indicates
that a strong, immediate neutrino heating is required in order to produce
energetic explosions in this case. On the other hand, the heating becomes
efficient in the SC models only once the quasi-steady state is established.
Again, as in the case of the mass in the gain region and advective times, the
evolution of heating efficiency is qualitatively similar between SC and FC
families.
\label{f:efficiency}}
\end{figure}

The evolution of heating efficiency is shown in Figure \ref{f:efficiency}.
It is interesting to note a qualitative similarity between the SC and FC model
families (shown with solid and dashed lines in the figure). For example, we
note a period of modest increase in heating efficiency between $t=20$ ms and
$t=50$ ms in FC models, and $t=50$ ms and $t=90$ ms in SC models. This
increase in the heating efficiency is associated with the onset of convection
prior to the quasi-steady state phase (see Section \ref{s:convection}). At
later times, after the quasi-steady state is established, the heating
efficiency continues to steadily increase in the SC models while the increase
in the heating efficiency is significantly greater in the FC models.
Nevertheless, the heating efficiency continues to evolve in a qualitatively
similar manner in both families of models. As we noted before, the
similarities between SC and FC models can be easily explained by the fact that
we consider only a single progenitor model in this work.

On the quantitative level, the heating efficiency in the FC models is greater
than one from the beginning of the evolution. On the other hand, the heating
becomes efficient in the SC models only after the quasi-steady state is
established. We conclude that a strong, immediate neutrino heating is required
in order to produce energetic explosions.

\subsection{Shock evolution}
\label{s:shock}

The shock revival process can also be studied by analyzing the time evolution
of the shock radius. In particular, prolonged periods of shock stagnation may
indicate the onset of the standing accretion shock instability (SASI).
Conversely, a steady increase in the shock radius may indicate that other
processes (e.g. neutrino-driven convection) operate efficiently. Additionally,
the unbinding of shocked matter is directly dependent on radius (through the
gravitational potential), and large shock radii may aid in reaching the
explosion threshold. Finally, hydrodynamic perturbations present at early
times can be imprinted on the shock and affect the morphology of the supernova
ejecta. To this end, in the following discussion we use the shock aspect ratio
to quantify the explosion asymmetry.

\begin{figure}[t!]
\centering
\includegraphics[width=8cm]{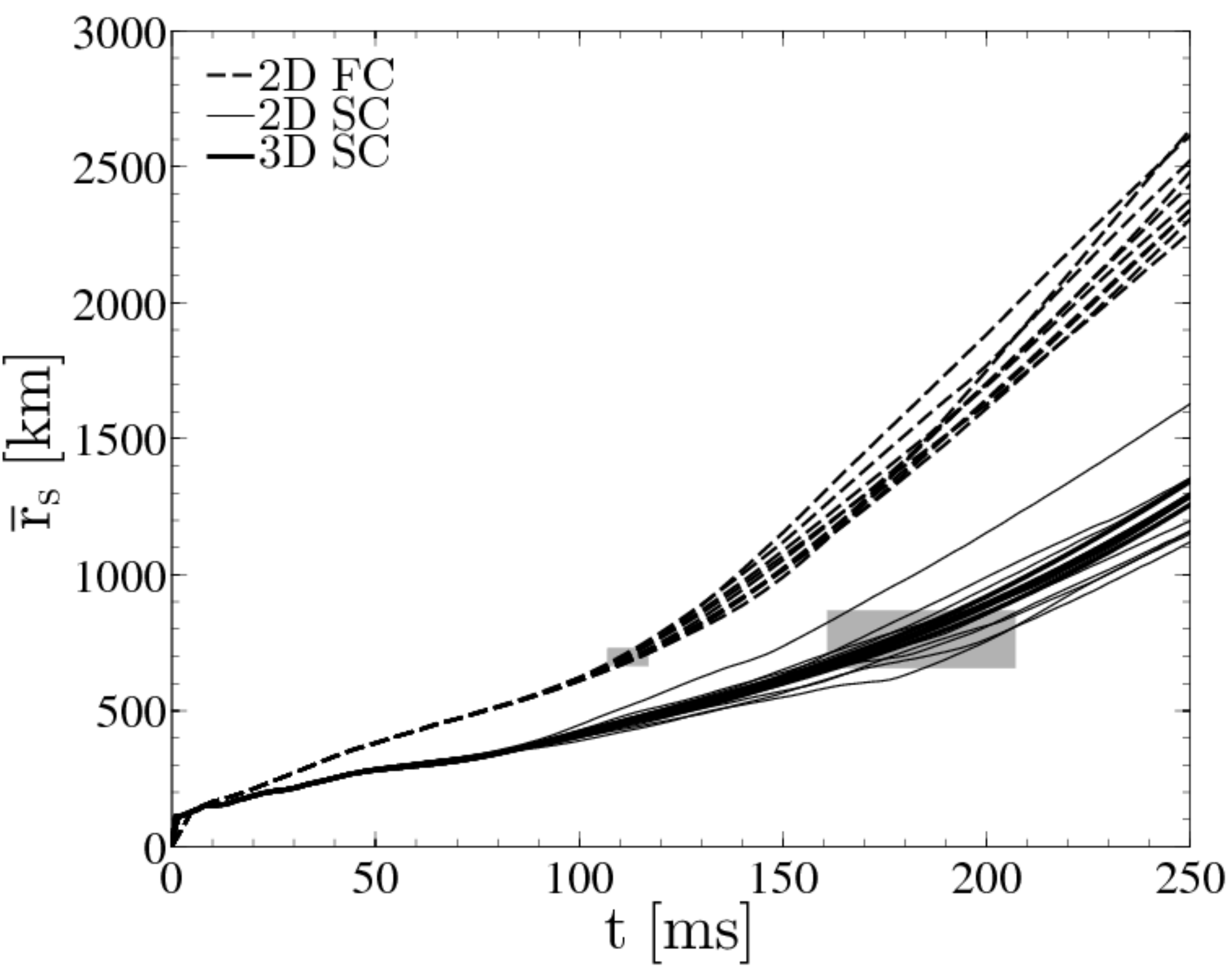}
\caption
{Evolution of the average shock radius. The average shock radius is shown with
solid and dashed lines for SC and FC models, respectively, with 2D SC models
shown with thin solid lines. Shaded areas denote regions where
($t_{exp},\overline{r}_{s}$) pairs for SC and FC model families are located.
Note that the FC and SC models explode at similar shock radii but at
significantly different times. The shock expansion rate for FC models is, on
average, twice that of the SC models.
\label{f:shockradius}}
\end{figure}
The evolution of the average shock radius, $\overline{R}_{s}$, in our models
is shown in Figure \ref{f:shockradius} with solid and dashed lines for the SC
and FC models, respectively. The shock radius increases steadily in both
families of models, although the shock expansion is about 25\% faster in the
FC models. At the time when the early convection sets in, the shock expands to
about 270 km in the FC models and about 200 km in the SC models. By the time
the shock is revived, its average radius for different model realizations
varies between 645 km and 855 km in the 2D SC models, and between 790 km and
810 km in the 3D SC models. We believe the greater variation in the average
shock radius in the 2D models compared to 3D may hint at differences in the
dynamics of the gain region between 2D and 3D. Moreover, 2D FC models display
a similar amount of variation in the average shock position as 2D SC models.
Also, in that case, the average shock position at the time of explosion varies
between 640 km and 730 km. Overall, the average shock radius at the time of
explosion is similar in SC and FC families. One should keep in mind, however,
that the FC models explode, on average, twice faster than the SC models (c.f.
Table\ \ref{t:models}).

\begin{figure}[t!]
\centering
\includegraphics[width=8cm]{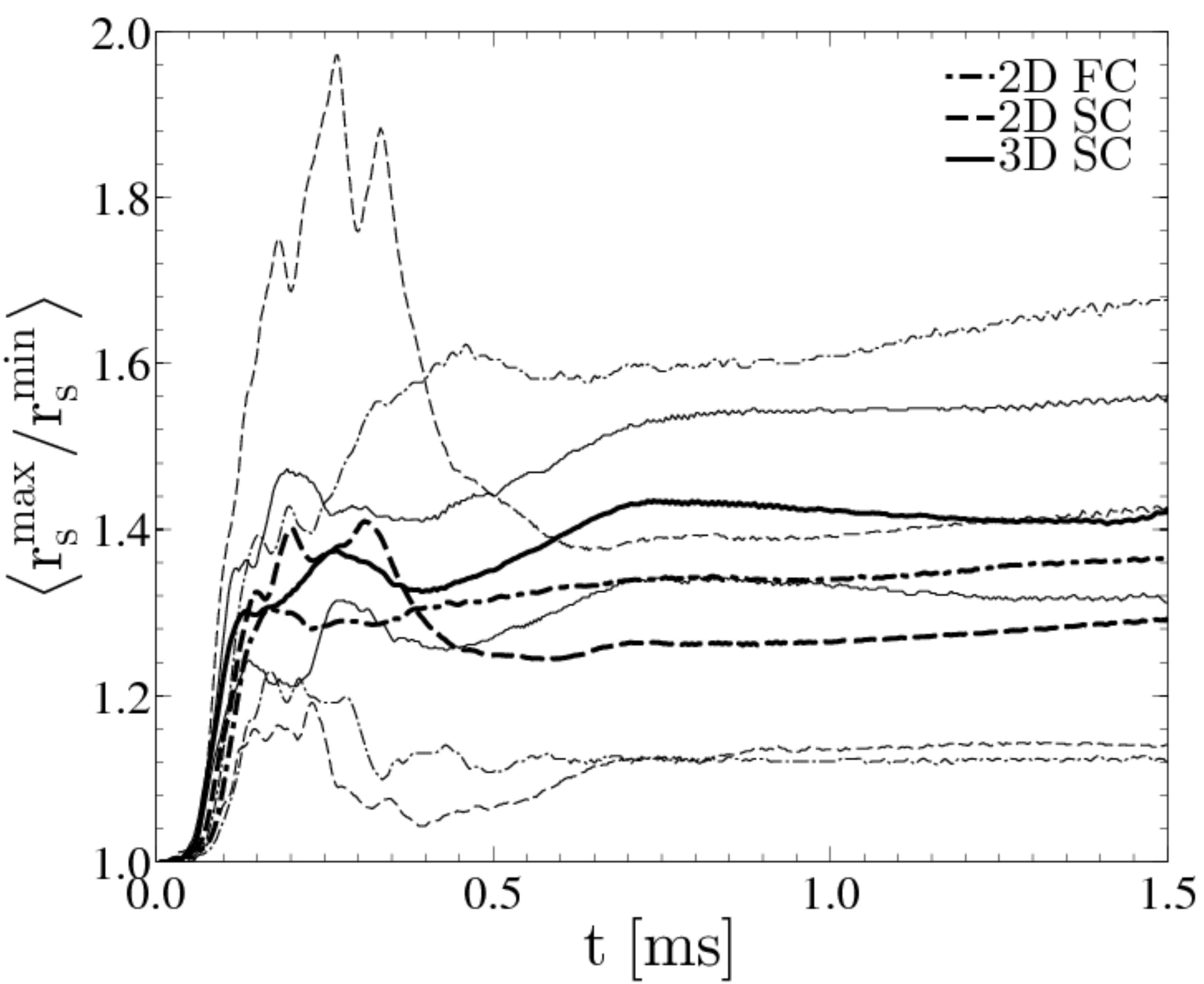}
\caption
{
Shock aspect ratio for our models. Thick lines denote the average over all
models in the group. Thin lines denote the minimum and maximum values for the
particular families. Our results show relatively mild asymmetries
($r_{s}^{max}/r_{s}^{min}\approx 1.3$). Visual inspection of SASI-producing
models from \cite{mueller+12} show aspect ratios upwards of 3 to 4.
\label{f:aspectratio}}
\end{figure}

The shock aspect ratio, $r_{s}^{max}/r_{s}^{min}$, found in our models (see
Figure \ref{f:aspectratio})
appears relatively modest compared to the shock asymmetry reported in other
studies \citep{hanke+12,mueller+12}. Although an extreme shock aspect ratio of
$2$ was found in 2D SC simulations, the average aspect ratios are only 1.25 in
2D and 1.4 in 3D. This should be compared to the shock aspect ratios between
$3$ and $4$ we estimated based on the data presented by \cite{mueller+12}. 

The most striking feature of the evolution of the shock aspect ratios in our
models are strong asymmetries observed in the 2D SC models at early times.
These asymmetries seem to rapidly decrease for times $>400$ ms. In contrast,
the shock aspect ratios in the 2D FC models appear to evolve more smoothly and
the most deformed FC models seem to retain their asymmetric character as time
progresses. Qualitatively, however, the degree to which the shock is deformed
in 2D models is modest and on average is $\approx 1.3$. In 3D models, the
average degree of shock asymmetry at late times is somewhat greater ($\approx
1.4$), and the variations between the individual 3D model realizations are
less than $20$\%.

We speculate that the smooth evolution and persistent character of
deformations found in the 2D FC models might be due to weaker convection which
operates over a shorter period. Conversely, in 2D SC models convection has
more time to organize the flow inside the gain region. Occasionally, the
dominant $l=1$ mode develops in those models (cf.\ Figure
\ref{f:entropy_texp}(a)) and results in a strong asymmetry. In other 2D SC
models, the higher order modes make the shock more spherical (cf.\ Figure
\ref{f:entropy_texp}(b)). The situation is similar in the case of 3D SC
models, in which the post-shock region shows richer convective structure and
evolves on a similar timescale as in 2D.

The evolution of shock radius also provides information that is helpful in the
context of the standing accretion shock instability (SASI). Specifically, SASI
manifests itself as low-order ($l=1,2$) oscillations of the shock radius
\citep{foglizzo+06,laming07,scheck+08,foglizzo09}. Therefore, the first step
in the SASI analysis of the shock revival is to decompose the shock radius in
terms of spherical harmonics. The expansion coefficients are given by
\begin{equation}
\label{e:sphericalharmonic}
a_{lm} = \int_{\Omega} Y^{*}_{lm}\left(\theta,\phi\right)f\left(\theta,\phi\right)d\Omega,
\end{equation}
where the spherical harmonic $Y^{*}_{lm}$ is normalized by a factor
$\sqrt{\frac{2l+1}{4\pi}\frac{(l-m)!}{(l+m)!}}$, and
$f\left(\theta,\phi\right)=r_{s}\left(\theta,\phi\right)$.

\begin{figure*}[t!]
\centering
\includegraphics[width=8cm]{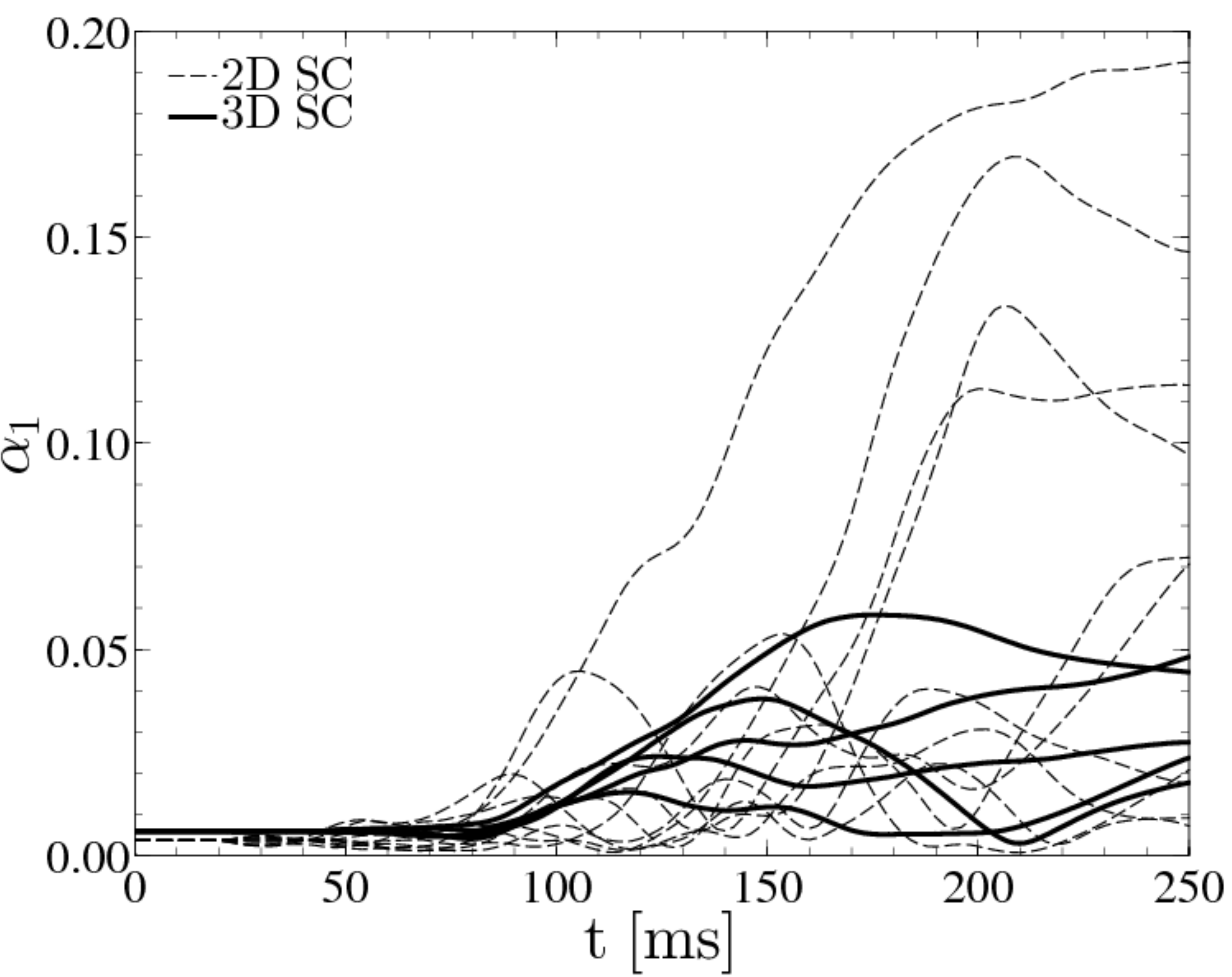}
\includegraphics[width=8cm]{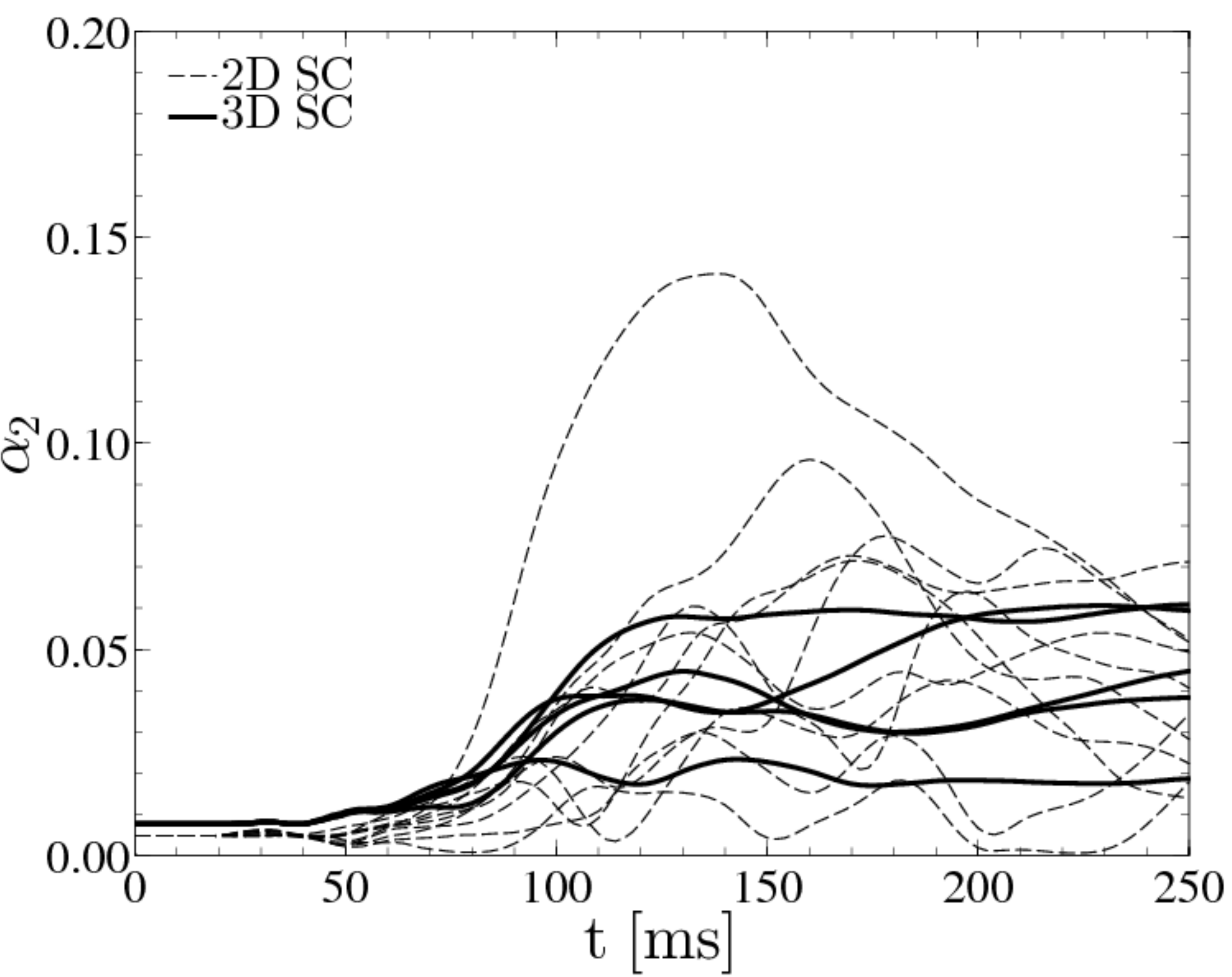}
\caption
{Evolution of the leading coefficients in the spherical harmonic decomposition
of the shock radius. Only the results of the decomposition for SC models are
shown. The runs of the $\alpha_1$ (left panel) and $\alpha_2$ (right panel)
coefficients are presented with solid and dashed lines for 3D and 2D models,
respectively. Prior to the onset of convection ($t\approx 50$ ms), the shock
front remains essentially spherically symmetric. The emergence of low order
perturbations around $t\approx 80$ ms is due to buoyant convective plumes
deforming select sections of the shock front rather than SASI. Note that the
degree of shock perturbation is relatively modest compared to some recent
results \citep[see, e.g., ][]{mueller+12}.
\label{f:modaldecomp}}
\end{figure*}

In order to enable direct comparison between the 2D ($m=0$) and 3D ($m=-l\ldots l$) cases, we
consider a suitably normalized contribution of the $m$-mode coefficients \citep{ott+13},
\begin{equation}
\alpha_{l} = \frac{1}{a_{00}}\sqrt{\sum_{m=-l}^{l}a_{lm}^2}.
\end{equation}
The evolution of $\alpha_{1}$ and $\alpha_{2}$ coefficients is shown in Figure \ref{f:modaldecomp}.
The larger values of the coefficients indicate that the 2D SC models, on
average, exhibit stronger variability than their three-dimensional
counterparts. However, we also observe a certain dichotomy among 2D SC models
with one subgroup showing variability distinctly larger than the remaining set
of models. 2D SC models also show relatively weak $l=1$ mode contributions.
perturbations, while others rise to a moderate fraction of the average shock
position. The 2D FC models (data not shown in Figure \ref{f:modaldecomp}) show
a similar degree of variability as the 2D SC models at all times.

Prior to explosion, the 2D models exhibit at most one or two weak
oscillations. This is a quantitatively different behavior than found in the
simulations of \cite{mueller+12}, who reported strong, multiple cycles leading
up to shock revival. In addition, the shock perturbations in our models are
weak compared to those considered as evidence for SASI. Furthermore, we do not
find qualitative differences between the 2D and 3D models, as we mentioned
above. Note that most evidence for SASI presented in the literature is
essentially restricted to 2D, non-exploding models.

In our SC models, the low order modes emerge at about $t=75$ ms, shortly
before the quasi-steady state is established in the gain region. Visual
inspection of the flow morphology during the quasi-steady state provides no
evidence for large-scale ``sloshing'' motions in the gain region, considered a
defining signature of SASI. We conclude it is unlikely the SASI plays any
important role in the evolution of our models.

\subsection{Morphology of the gain region}
\label{s:morphology}

As soon as the neutrino-driven convection sets in and the related
perturbations reach the shock, one faces a difficult problem of disentangling
various physics processes participating in the shock revival, including fluid
instabilities and neutrino-matter interactions. Analysis of the overall flow
morphology is the first step in the process of understanding the explosion
mechanism. It provides initial evidence for the physics of fluid flow
participating in re-energizing the stalled shock \citep{herant+94}. It also may
provide evidence for the possible role of model parameters, such as assumed
symmetries and discretization errors, e.g. near the symmetry axis
\citep{scheck+06,gawryszczak+10}, on the flow dynamics. In this section we
present the morphological evolution of our SC models in two and three
dimensions. In particular, we are interested in identifying when the fluid
flow instabilities imprint their structure on the just-formed inner core of
the supernova ejecta.

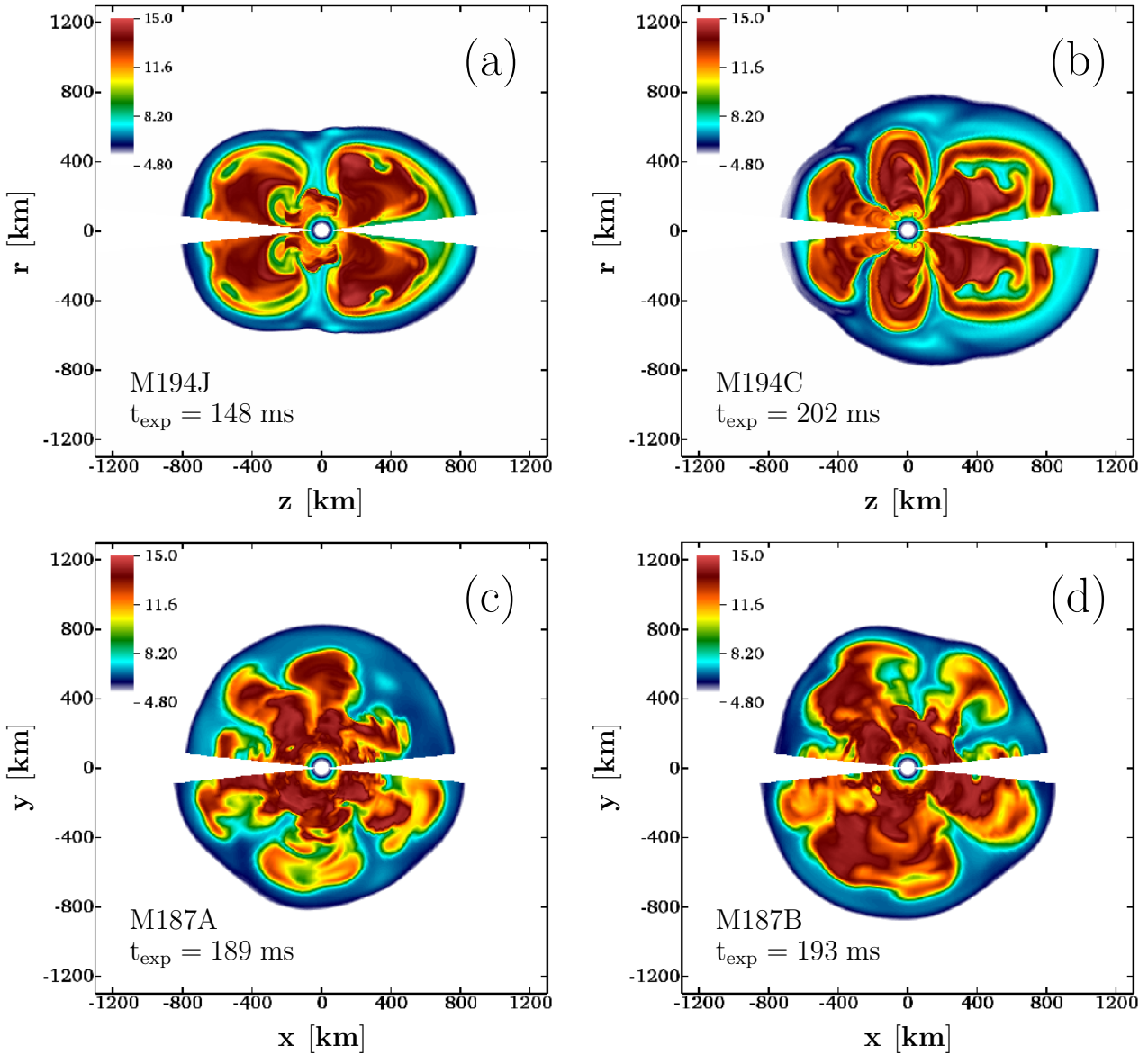
\begin{figure*}[ht!]
	\begin{center}
    \begin{tabular}{cc}
		%
		%
		%
		\begin{overpic}[height=7cm,viewport={115 115 985 921},clip]{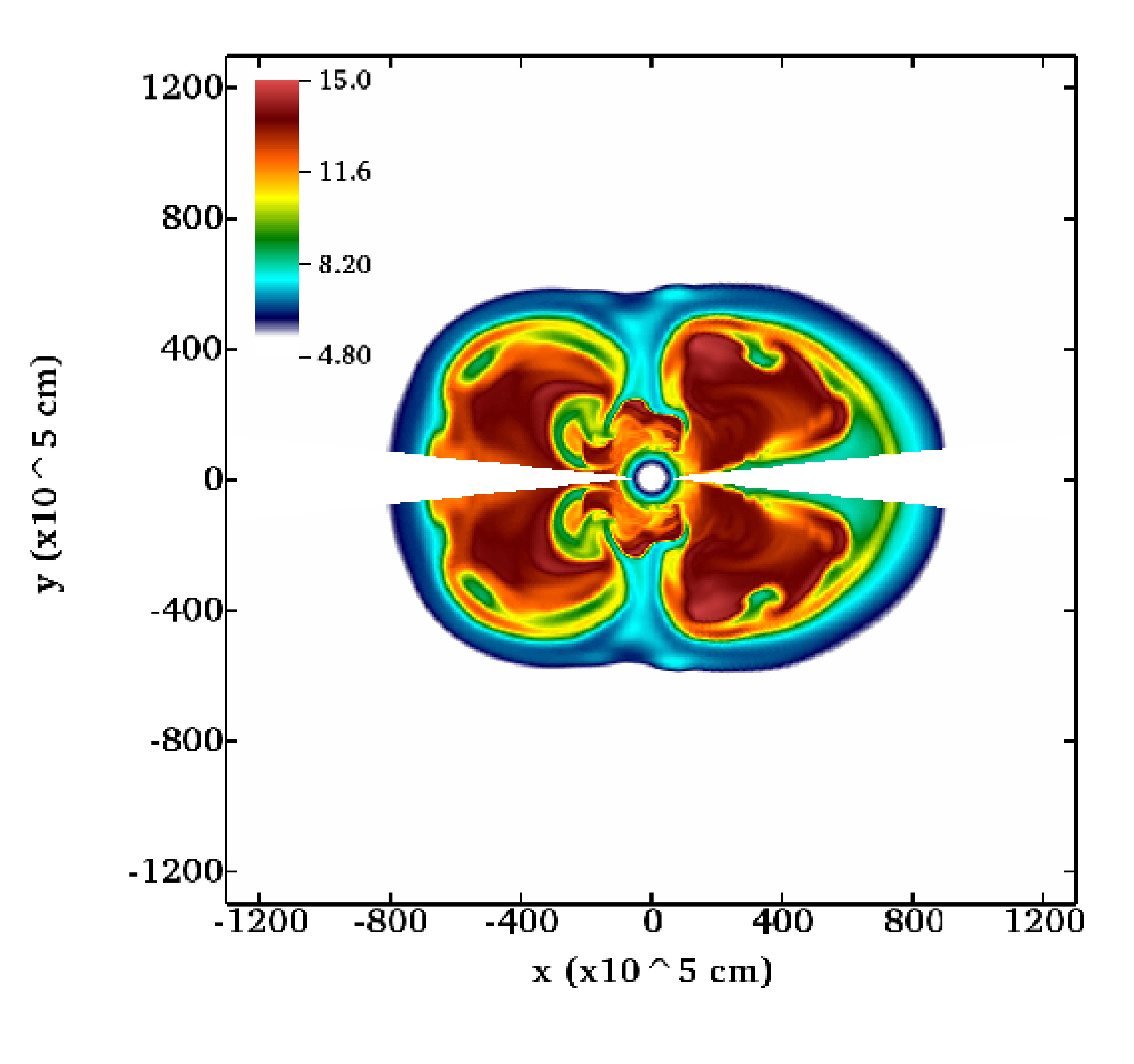}
		\put(82,78){\huge $\mathrm{\left(a\right)}$}
		\put(17,16){\large $\mathrm{M194J}$}
		\put(17,10){\large $\mathrm{t_{exp}=148\;ms}$}
		\put(-7,39){\large \rotatebox{90}{$\mathbf{r\;\left[km\right]}$}}
		\put(46,-7){\large {$\mathbf{z\;\left[km\right]}$}}
		\end{overpic} 
		\vspace{1.5em}
		\hspace{1.5em}
		&
		%
		%
		%
		\begin{overpic}[height=7cm,viewport={115 115 985 921},clip]{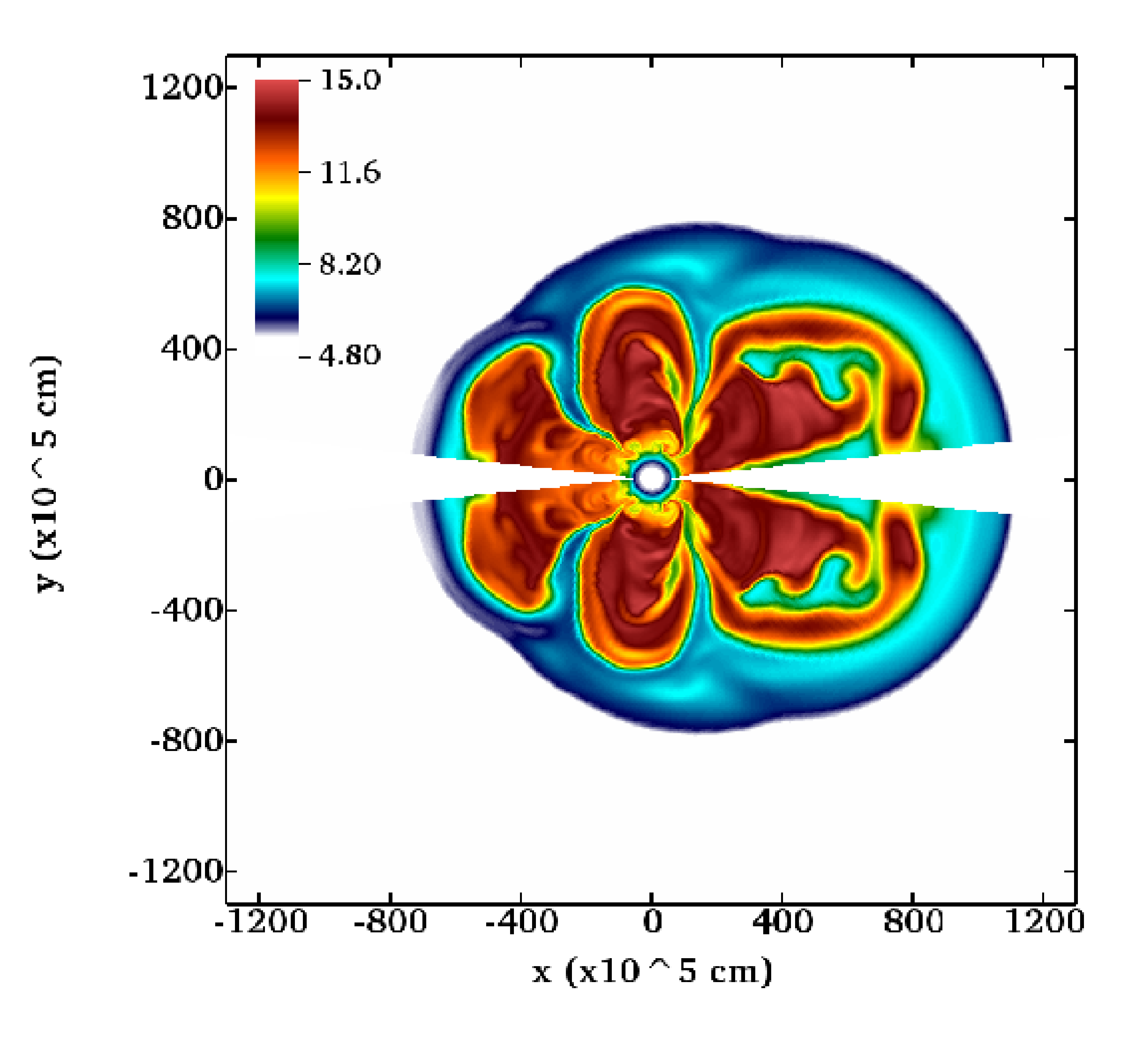}
		\put(82,78){\huge $\mathrm{\left(b\right)}$}
		\put(17,16){\large $\mathrm{M194C}$}
		\put(17,10){\large $\mathrm{t_{exp}=202\;ms}$}
		\put(-7,39){\large \rotatebox{90}{$\mathbf{r\;\left[km\right]}$}}
		\put(46,-7){\large {$\mathbf{z\;\left[km\right]}$}}
		\end{overpic} 
		\vspace{1.5em}
		\\
		%
		%
		%
		\begin{overpic}[height=7cm,viewport={115 115 985 921},clip]{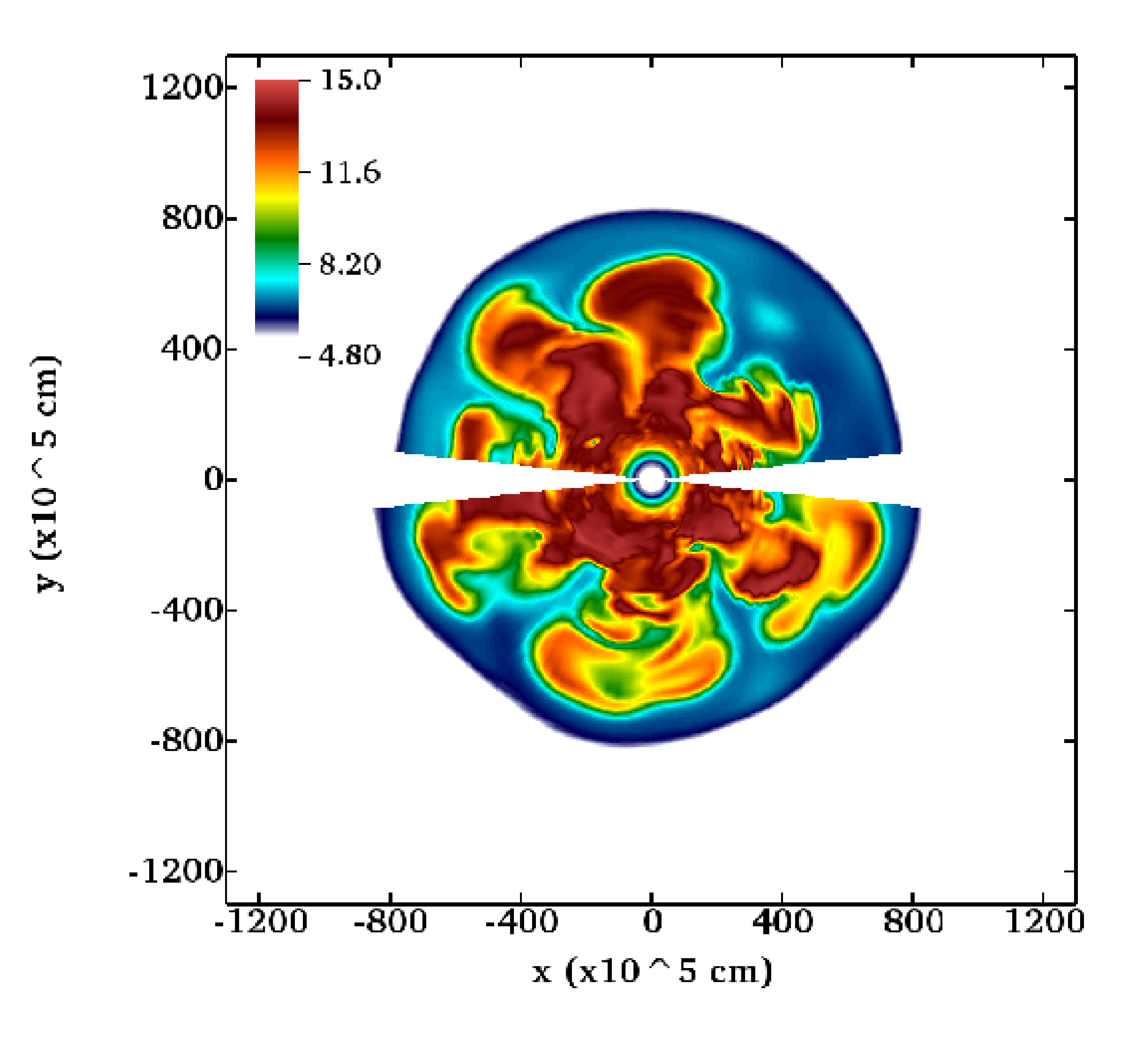}
		\put(82,78){\huge $\mathrm{\left(c\right)}$}
		\put(17,16){\large $\mathrm{M187A}$}
		\put(17,10){\large $\mathrm{t_{exp}=189\;ms}$}
		\put(-7,39){\large \rotatebox{90}{$\mathbf{y\;\left[km\right]}$}}
		\put(46,-7){\large {$\mathbf{x\;\left[km\right]}$}}
		\end{overpic} 
		\vspace{1.5em}
		\hspace{1.5em}
		&
		%
		%
		%
		\begin{overpic}[height=7cm,viewport={115 115 985 921},clip]{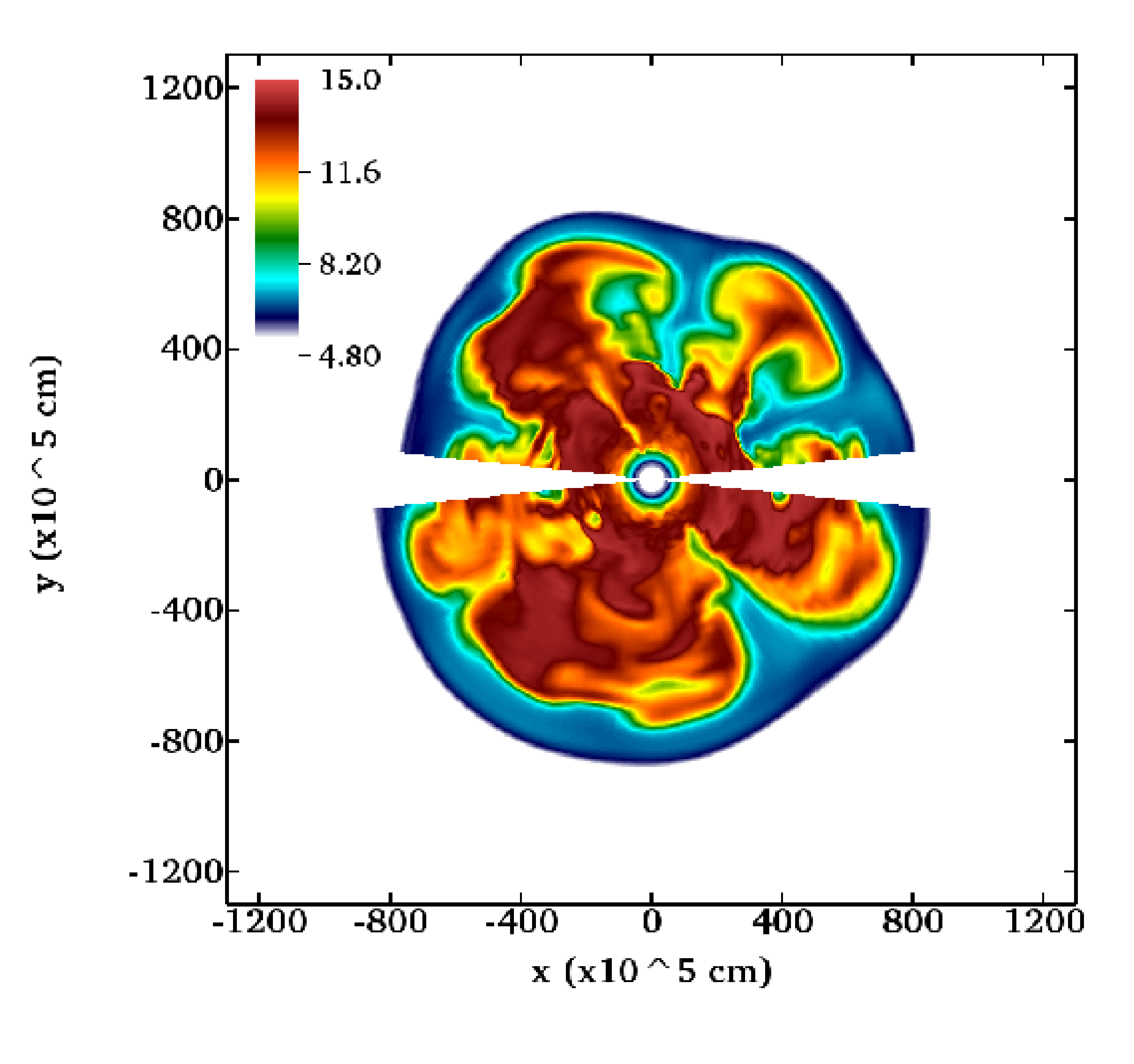} 
		\put(82,78){\huge $\mathrm{\left(d\right)}$}
		\put(17,16){\large $\mathrm{M187B}$}
		\put(17,10){\large $\mathrm{t_{exp}=193\;ms}$}
		\put(-7,39){\large \rotatebox{90}{$\mathbf{y\;\left[km\right]}$}}
		\put(46,-7){\large {$\mathbf{x\;\left[km\right]}$}}
		\end{overpic} 
		\vspace{1.5em}
    \end{tabular}
	\caption
	{
	Entropy distribution in the post-shock region at the explosion time for
	select two-dimensional and three-dimensional models. Entropy distribution
	is shown with pseudocolor maps for models M194J (panel a), M194C (panel
	b), M187A (panel c), and M187B (panel d). The entropy distribution maps
	for two-dimensional models (panels (a), (b)) are reflected across the
	symmetry axis. For three-dimensional models, the entropy is shown for a
	slice through the computational domain at the equatorial plane. In two
	dimensions models tend to have either dipolar (panel a) or quadrupolar
	(panel b) ejecta structures which result in an oblate shape of the shock.
	In three dimensions the ejecta appear to have more regular structure, and
	the shock is less deformed. See Section \ref{s:morphology} for discussion.
	\label{f:entropy_texp}}
	\end{center}
\end{figure*}

Entropy pseudocolor maps of the gain region for two two-dimensional and
two three-dimensional SC models at their respective explosion times are shown in
Figure \ref{f:entropy_texp}.
Prior to reaching the explosion threshold, the 2D SC models evolve to have
either dipolar ($l=1$, Figure \ref{f:entropy_texp}(a)) or quadrupolar ($l=2$,
Figure \ref{f:entropy_texp}(b)) $m=0$ ejecta morphology. These flow features
emerge when strong downflows in the gain region are formed shortly after the
onset of convection. Furthermore, our models indicate that these flow
structures do not evolve once they set in and continue to persist after the
explosion is launched, evolving in a self-similar fashion at later times (see
below). Additionally, these downflows carry accreted material deep into the
gain region close to the gain radius. This behavior does not seem to be as
extreme in 3D SC models, because a comparable amount of accreted material is
transported through many more downflows (see Figure \ref{f:entropy_texp}(c) and
(d)). The presence of many downflows in three-dimensions reflects the fact
that the flow structure is inherently different in 2D and 3D. We quantify this
behavior later in Section \ref{s:convection}.
 
\begin{figure*}[ht!]
	\begin{center}
    \begin{tabular}{ccc}
		%
		%
		%
		\begin{overpic}[width=4.5cm,viewport={139 105 924 855},clip]{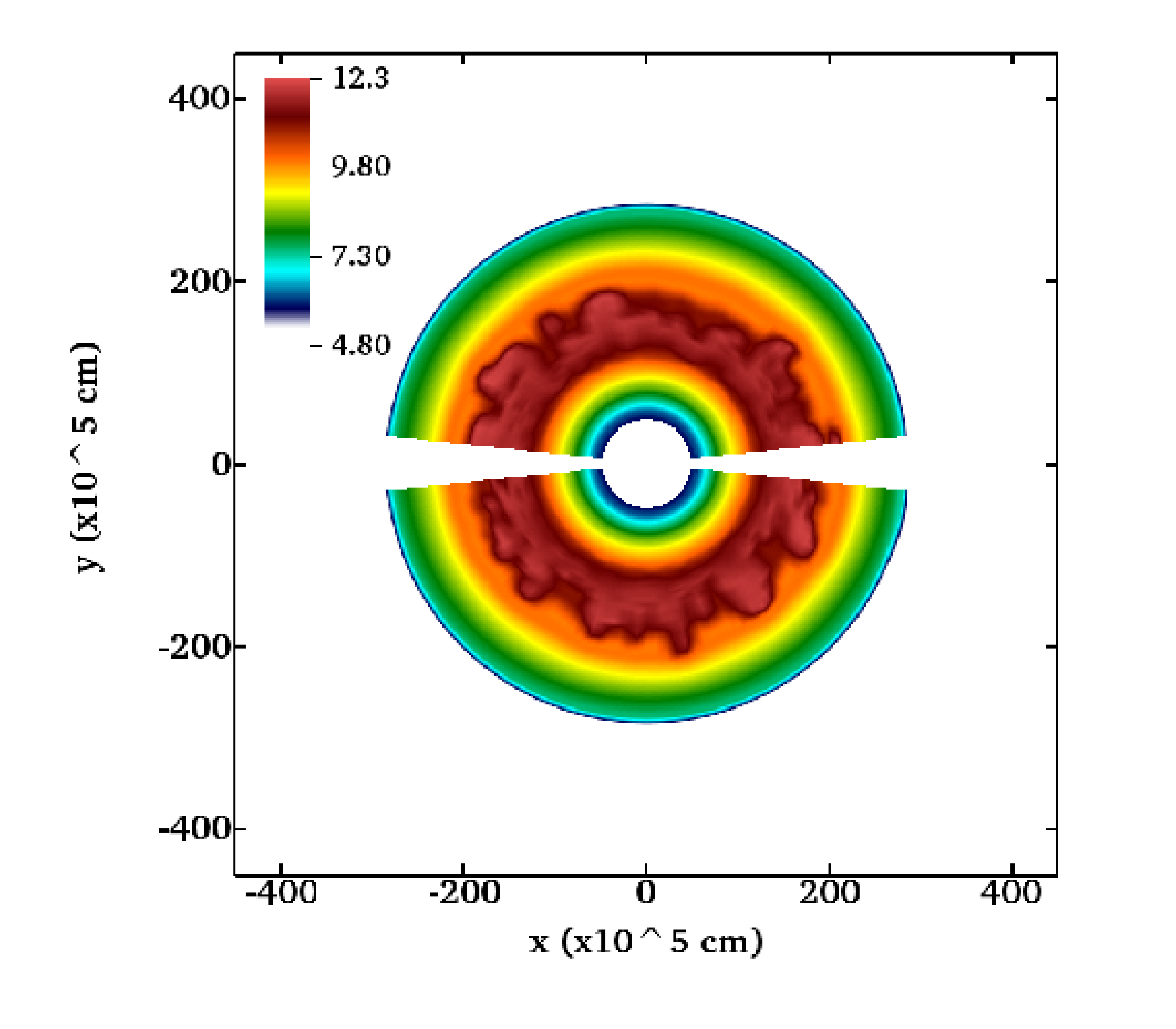}
		\put(80,80){\large $\mathrm{\left(a\right)}$}
		\put(15,10){\large $\mathrm{t=50\;ms}$}
		\put(-10,32){\rotatebox{90}{$\mathbf{y\;\left[10^5\;cm\right]}$}}
		\put(37, -7){{$\mathbf{x\;\left[10^5\;cm\right]}$}}
		\end{overpic}
		\hspace{.5em}
		\vspace{.75em}
		&
		%
		%
		%
		\begin{overpic}[width=4.5cm,viewport={139 105 924 855},clip]{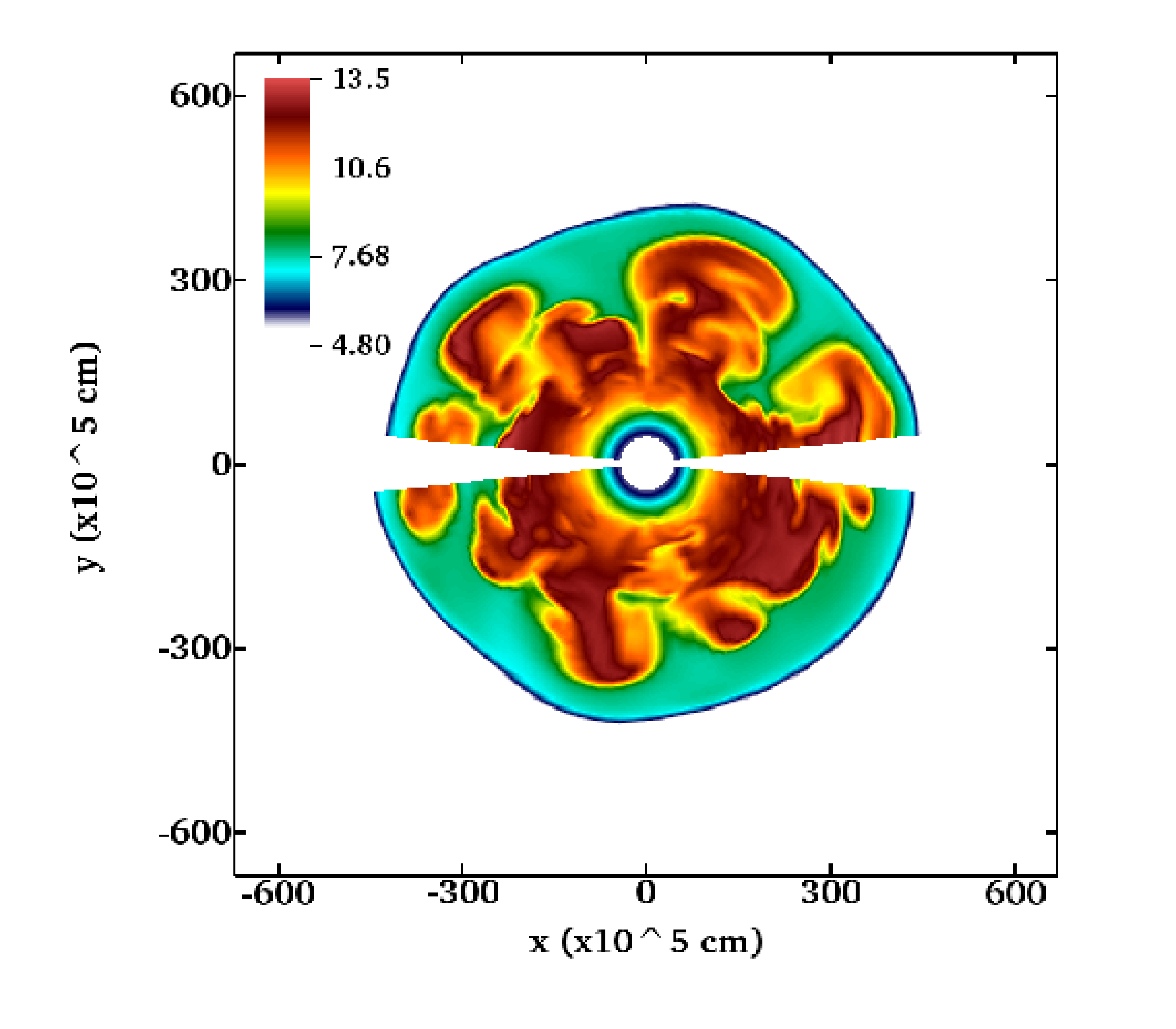}
		\put(80,80){\large $\mathrm{\left(b\right)}$}
		\put(15,10){\large $\mathrm{t=100\;ms}$}
		\put(-10,32){\rotatebox{90}{$\mathbf{y\;\left[10^5\;cm\right]}$}}
		\put(37, -7){{$\mathbf{x\;\left[10^5\;cm\right]}$}}
		\end{overpic}
		\hspace{.5em}
		\vspace{.75em}
		&
		%
		%
		%
		\begin{overpic}[width=4.5cm,viewport={139 105 924 855},clip]{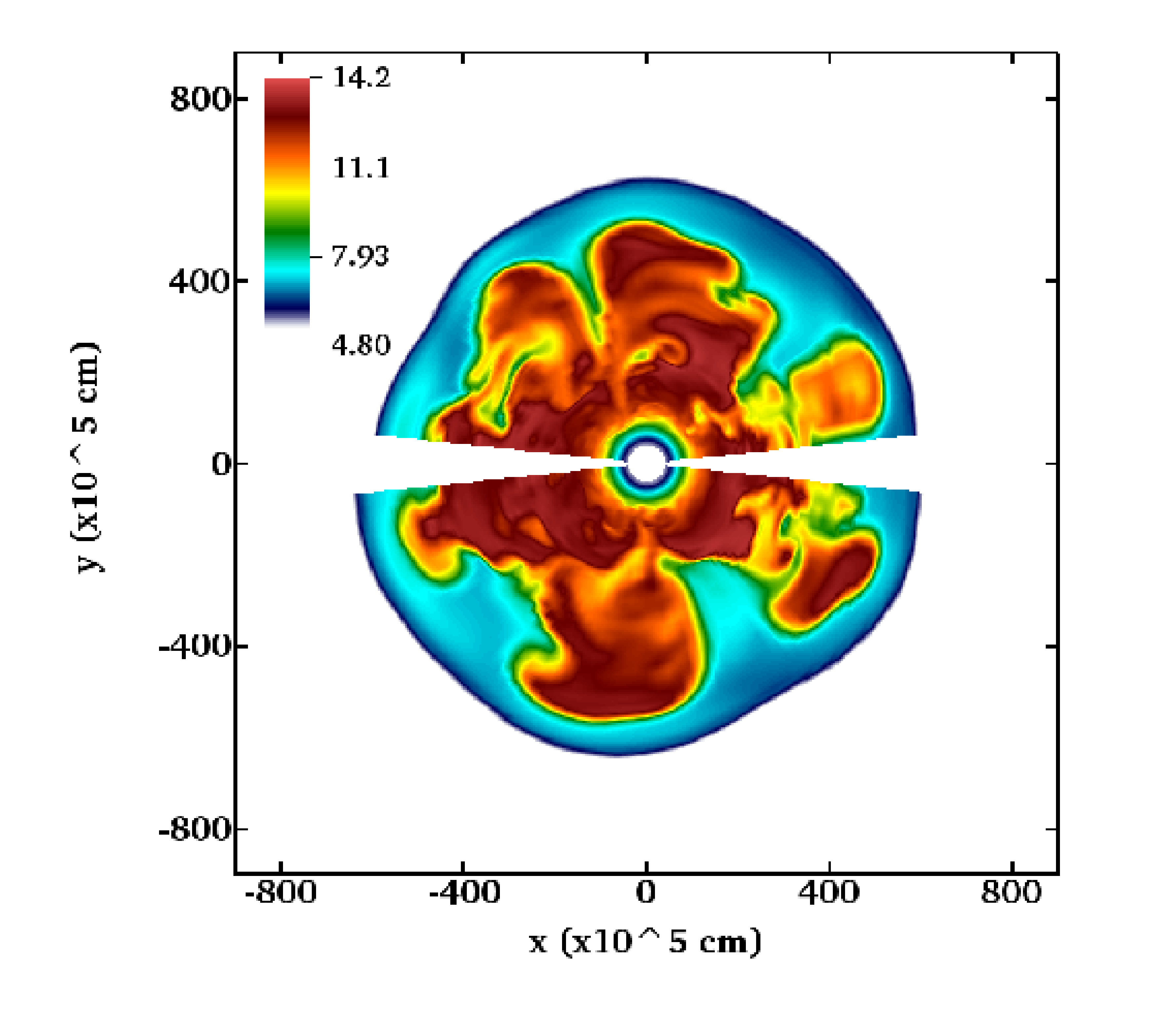}
		\put(80,80){\large $\mathrm{\left(c\right)}$}
		\put(15,10){\large $\mathrm{t=150\;ms}$}
		\put(-10,32){\rotatebox{90}{$\mathbf{y\;\left[10^5\;cm\right]}$}}
		\put(37, -7){{$\mathbf{x\;\left[10^5\;cm\right]}$}}
		\end{overpic} 
		\vspace{.75em}
		\\
		%
		%
		%
		\begin{overpic}[width=4.5cm,viewport={139 105 924 855},clip]{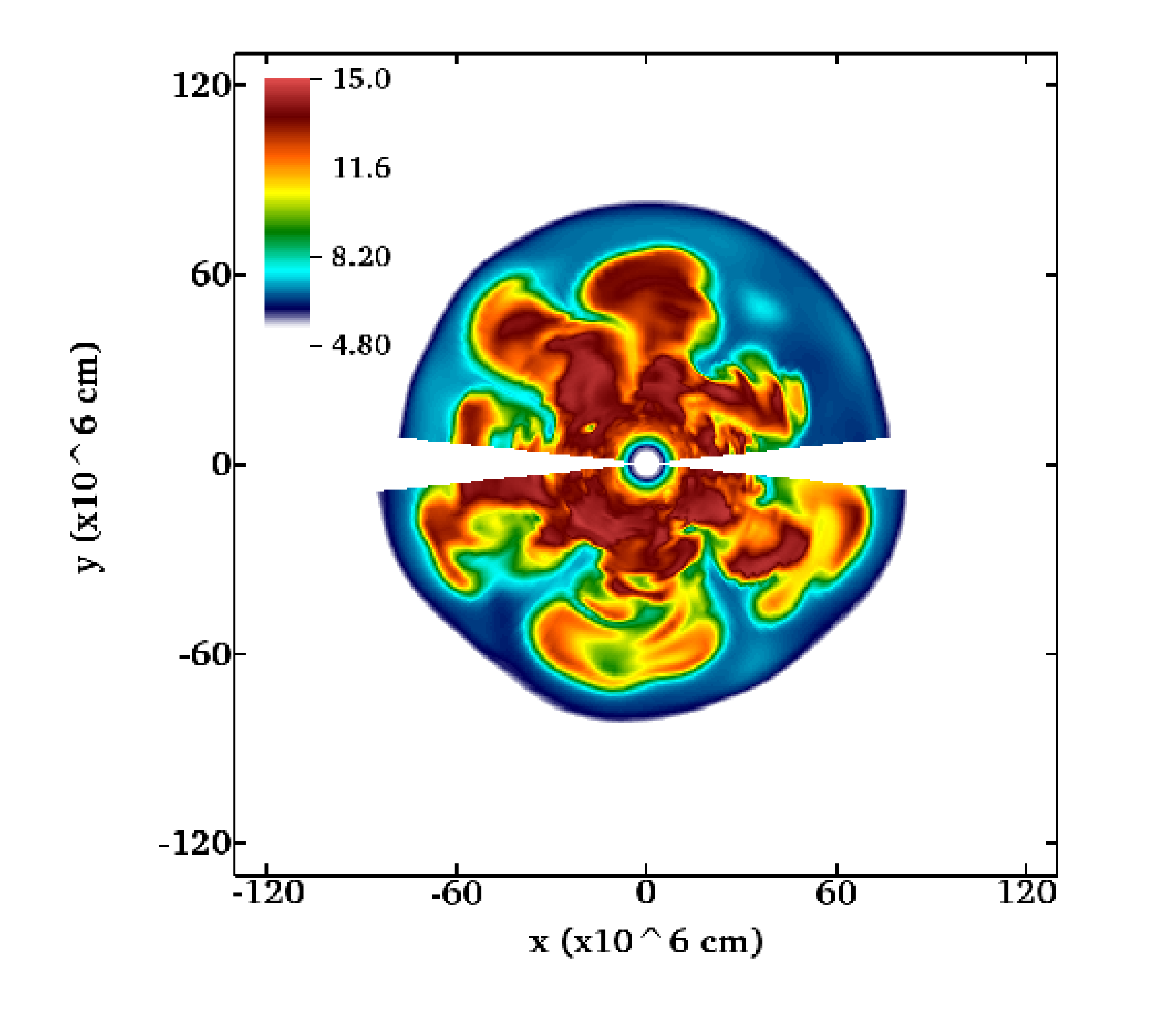}
		\put(80,80){\large $\mathrm{\left(d\right)}$}
		\put(15,10){\large $\mathrm{t=189\;ms}$}
		\put(-10,32){\rotatebox{90}{$\mathbf{y\;\left[10^6\;cm\right]}$}}
		\put(37, -7){{$\mathbf{x\;\left[10^6\;cm\right]}$}}
		\end{overpic}
		\hspace{.5em}
		\vspace{.75em}
		&
		%
		%
		%
		\begin{overpic}[width=4.5cm,viewport={139 105 924 855},clip]{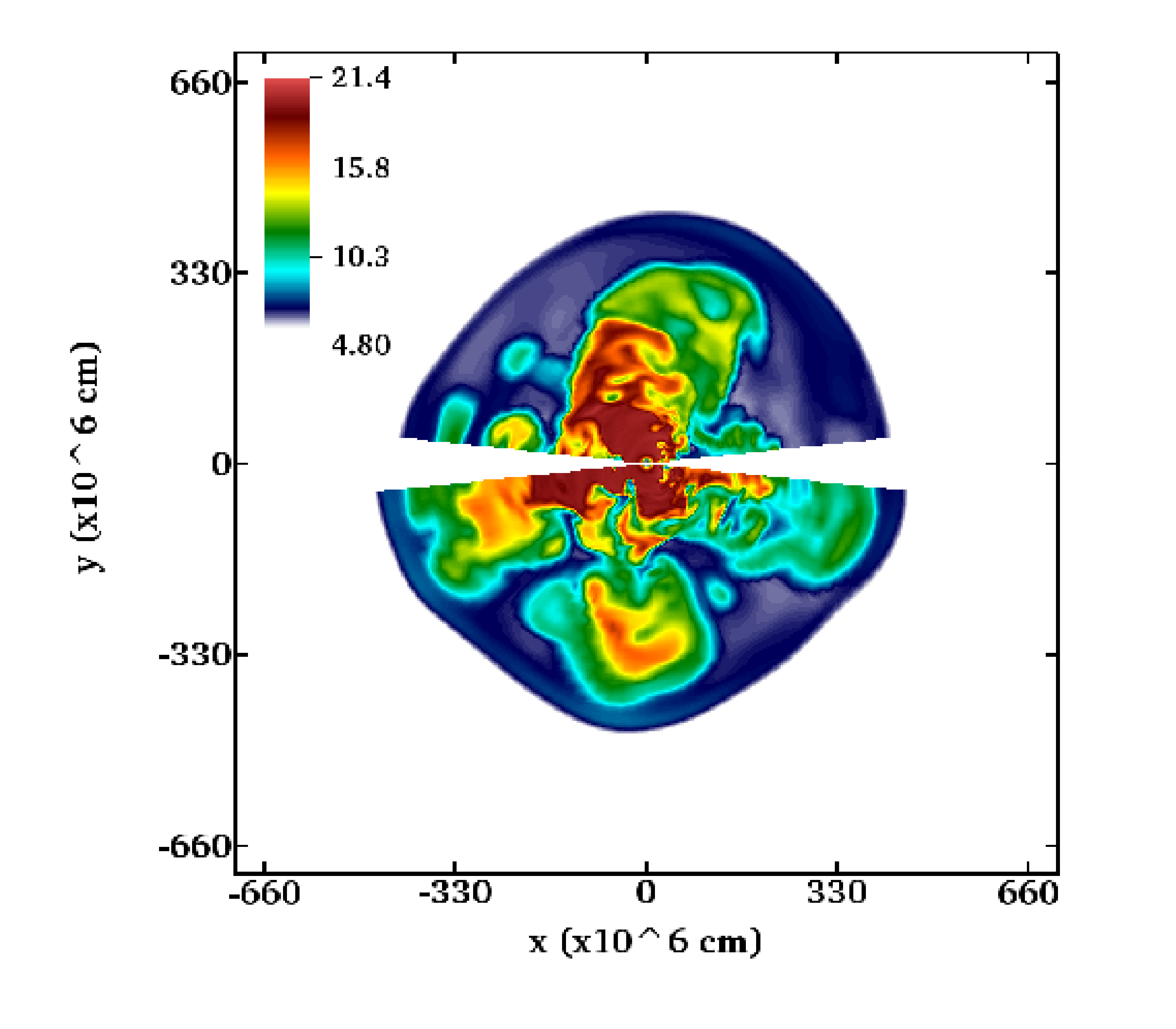}
		\put(80,80){\large $\mathrm{\left(e\right)}$}
		\put(15,10){\large $\mathrm{t=500\;ms}$}
		\put(-10,32){\rotatebox{90}{$\mathbf{y\;\left[10^6\;cm\right]}$}}
		\put(37, -7){{$\mathbf{x\;\left[10^6\;cm\right]}$}}
		\end{overpic}
		\hspace{.5em}
		\vspace{.75em}
		&
		%
		%
		%
		\begin{overpic}[width=4.5cm,viewport={139 105 924 855},clip]{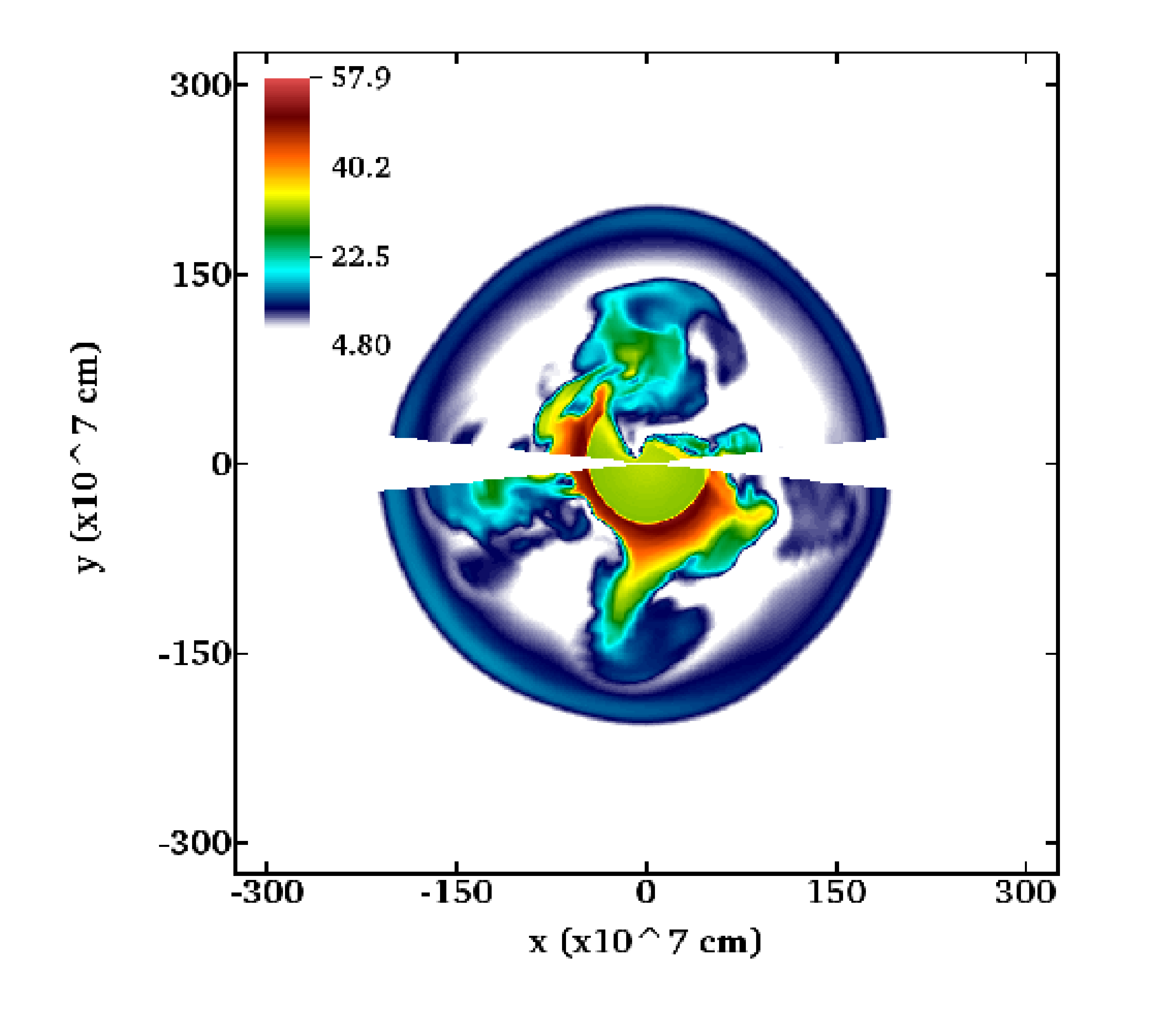}
		\put(80,80){\large $\mathrm{\left(f\right)}$}
		\put(15,10){\large $\mathrm{t=1500\;ms}$}
		\put(-10,32){\rotatebox{90}{$\mathbf{y\;\left[10^7\;cm\right]}$}}
		\put(37, -7){{$\mathbf{x\;\left[10^7\;cm\right]}$}}
		\end{overpic} 
		\vspace{.75em}
    \end{tabular}
	\caption
	{Distribution of entropy in the 3D SC model M187A. The entropy is shown
	with pseudocolor maps for a slice through the computational domain at the
	equatorial plane. The entropy distribution is shown at $t=50, 100, 150,
	189, 500, \mathrm{and}\, 1500$ ms in panels (a)-(f). The development of
	convection can be seen as early as $t=50$ ms (panel a). The flow structure
	becomes progressively more complex at intermediate times (panels (b) and
	(c)) and the explosion is launched at $t=189$ ms (panel (d)). Relatively
	little change in the structure of the gain region takes place at later
	times (panels (e) and (f)).
	\label{f:entropy_evol}}
	\end{center}
\end{figure*}

The entropy distribution in the 3D SC model M187A is shown in Figure
\ref{f:entropy_evol} for select times.
At early times, the flow evolves away from initially spherically symmetric
accretion, with the first signs of convective instability emerging around
$t=50$ ms (Figure \ref{f:entropy_evol}(a)). At this time, the shock is not
affected by convection. At intermediate times ($t=100$ ms, Figure
\ref{f:entropy_evol}(b)), the initial convective plumes gradually merge into
larger structures and begin deforming large segments of the shock front. As
time progresses, the process of bubble merging continues (Figure
\ref{f:entropy_evol}(c)) and the shock eventually is launched around $t=189$
ms (Figure \ref{f:entropy_evol}(d)).

The large-scale morphology of the post-shock region appears essentially frozen
after the explosion commences. For example, three ejecta plumes seen in the
southern hemisphere (Figure \ref{f:entropy_evol}(d)), can still be easily
identified more than 300 ms later (Figure \ref{f:entropy_evol}(e)). However,
during the same period the convective bubbles in the northern hemisphere show
significant evolution and appear to merge into a single structure. This
morphology persists until the end of the simulation and the central region
becomes filled with the neutrino-driven wind (Figure \ref{f:entropy_evol}(f)).
\section{Convection}
\label{s:convection}
Our analysis of the explosion process presented above revealed that the
explosion is preceded by a relatively short (compared to other studies)
quasi-steady state inside the gain region (Section \ref{s:grchar}) during
which the shock only slowly expands and SASI does not seem to a major role
(Section \ref{s:shock}). We also found that the heating becomes efficient
shortly before the quasi-steady state is established (Section \ref{s:grchar}).

There exists a large body of evidence from numerical simulations of core-collapse 
supernova explosions indicating that multidimensionality, and
specifically the presence of convection, decreases the requirements for the
neutrino luminosity produced by the contracting core of the proto-neutron star
\citep{janka+96,nordhaus+10,hanke+12,couch13}. In this section we focus on
characterization and quantification of the effects of neutrino-driven
convection during the quasi-steady state of the gain region. We introduce and
apply novel methods dedicated to investigating this aspect of the supernova
explosion process.
\subsection{Minimum resolution requirements}
\label{s:resolution}
We perform a series of three-dimensional simulations in order to assess the
dependence of model physics operating inside the gain region on numerical
resolution. In particular, we are interested in finding conditions required to
suppress neutrino-driven convection on large scales. If indeed convection is a
critical component driving the explosion, one could expect that by suppressing
convection (by whatever means) the explosion would not occur, even in models
with neutrino luminosities that in other situations are sufficiently large
enough to revive the shock. This expectation is strongly supported by numerous
calculations which show that, for example, much higher luminosities are
required to produce an explosion in one-dimension than in multidimensions
(see, e.g., \cite{janka+96}, Section \ref{s:grchar}, and Figure \ref{f:lum}
for discussion and compilation of relevant recent results).

In our series of 3D simulations, we keep the radial resolution fixed while the
angular resolution is gradually decreasing by a factor of 2 between the
models, from 3\degree\ (our base resolution) down to 24\degree. In addition, we
compute a 2\degree\ resolution model. For each model, we recorded the explosion
times and explosion energies at $t=250$ ms. The basic characteristics of
models obtained in this series is given in Table \ref{t:angulartable}.
\begin{deluxetable}{ccc}
\tablecaption{Characteristics of the M187A model series with varying angular resolution. \label{t:angulartable}}
\tablehead{
\colhead{Angular} &
\colhead{$t_{exp}$} &
\colhead{$E_{exp}\left(t=250 \mathrm{\,ms}\right)$} 
\\
\colhead{Resolution} &
\colhead{(ms)} &
\colhead{($10^{51}$ erg)} 
} 
\startdata
2\degree	&	183	&	0.114	\\*
3\degree	&	183	&	0.107	\\*
6\degree	&	175	&	0.106	\\*
12\degree	&	198	&	0.077	\\*
24\degree	&	216	&	0.028
\enddata
\end{deluxetable}
The inspection of data shown in the table reveals that the explosion times and
the explosion energies change little as long as the angular resolution is not
worse than 6\degree. For coarser angular meshes, we observe a significant
increase in the explosion times and decrease in the explosion energies. No
explosions were found at resolutions coarser than 24\degree.

\begin{figure*}[ht!]
	\begin{center}
    \begin{tabular}{cc}
		%
		%
		%
		\begin{overpic}[width=7cm,viewport={133 113 975 914},clip]{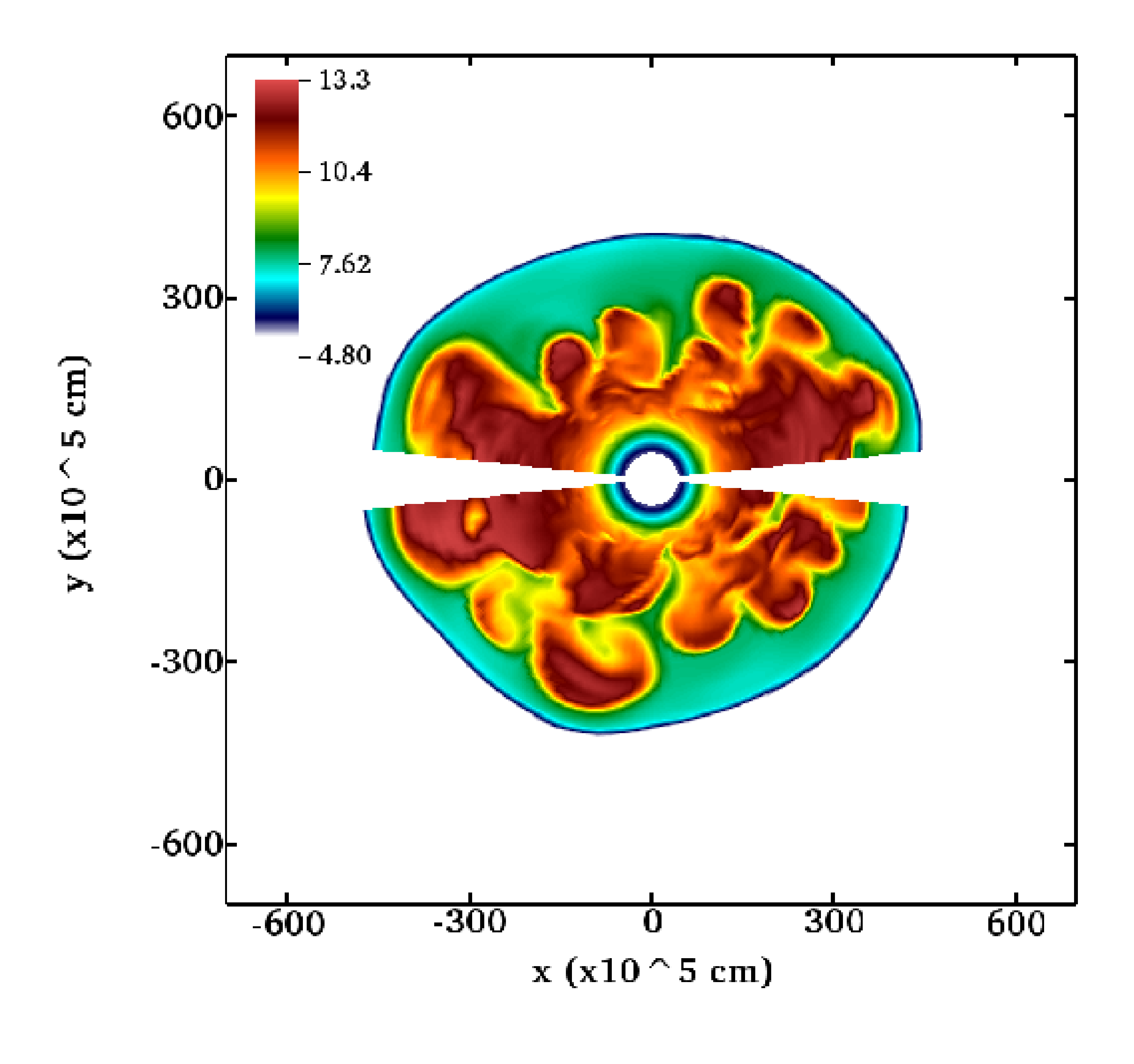}
		\put(82,82){\huge $\mathrm{\left(a\right)}$}
		\put(17,12){\huge $\mathrm{3\degree}$}
		\put(-9,40){\large \rotatebox{90}{$\mathbf{y\;\left[km\right]}$}}
		\put(45,-7){\large {$\mathbf{x\;\left[km\right]}$}}
		\end{overpic} 
		\vspace{1.5em}
		\hspace{1.5em}
		&
		%
		%
		%
		\begin{overpic}[width=7cm,viewport={133 113 975 914},clip]{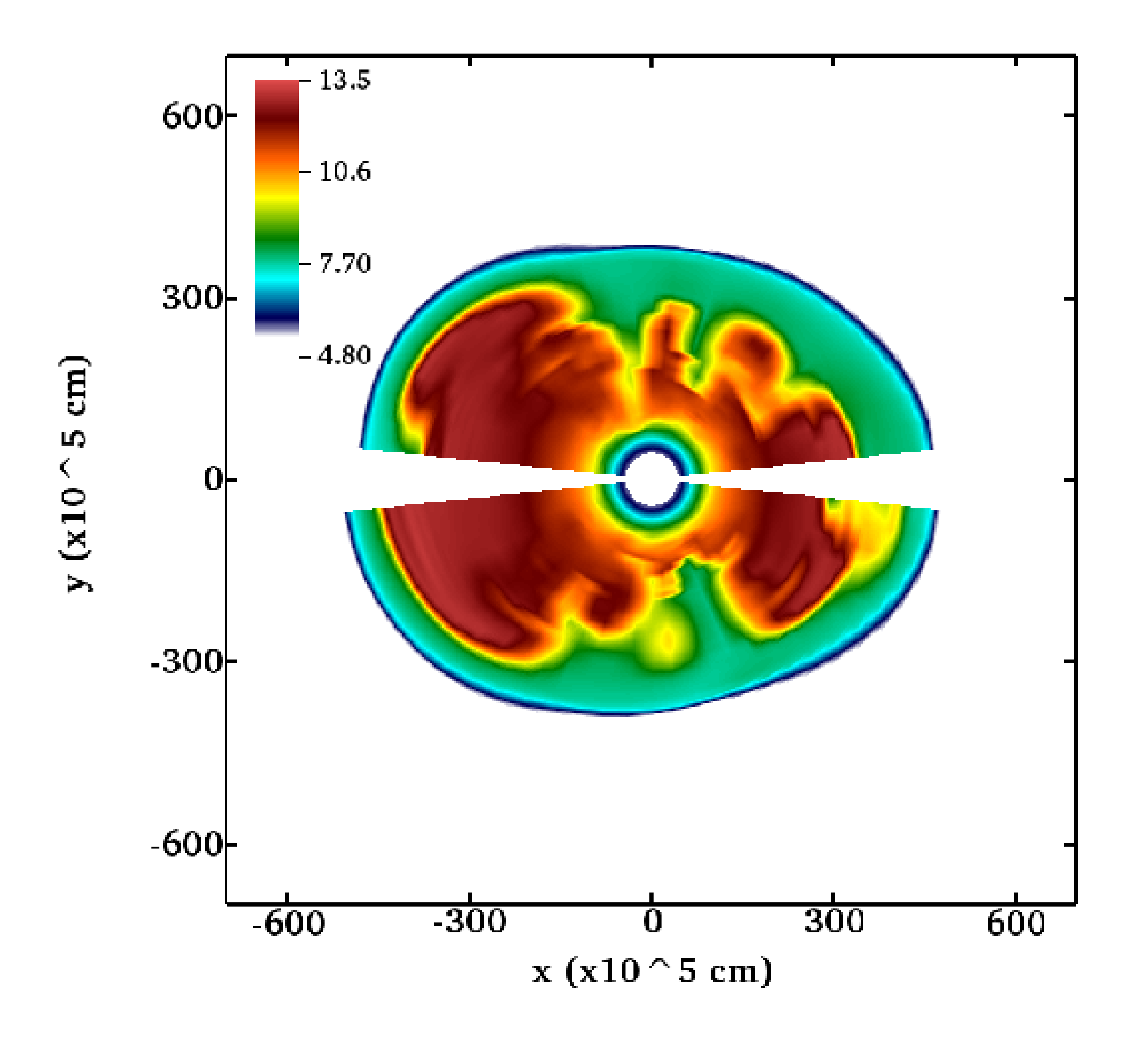}
		\put(82,82){\huge $\mathrm{\left(b\right)}$}
		\put(17,12){\huge $\mathrm{6\degree}$}
		\put(-9,40){\large \rotatebox{90}{$\mathbf{y\;\left[km\right]}$}}
		\put(45,-7){\large {$\mathbf{x\;\left[km\right]}$}}
		\end{overpic} 
		\vspace{1.5em}
		\\
		%
		%
		%
		\begin{overpic}[width=7cm,viewport={133 113 975 914},clip]{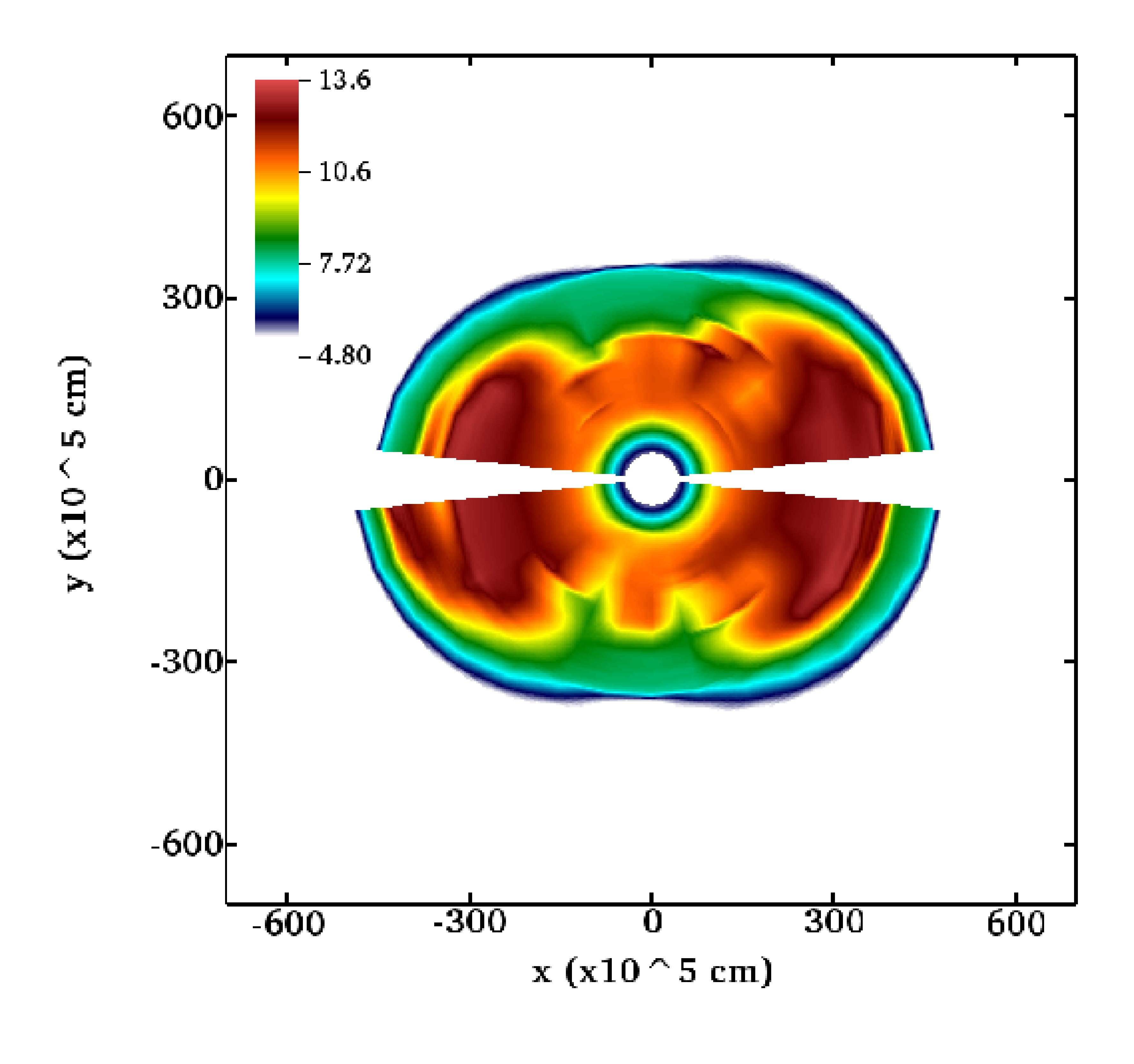}
		\put(82,82){\huge $\mathrm{\left(c\right)}$}
		\put(17,12){\huge $\mathrm{12\degree}$}
		\put(-9,40){\large \rotatebox{90}{$\mathbf{y\;\left[km\right]}$}}
		\put(45,-7){\large {$\mathbf{x\;\left[km\right]}$}}
		\end{overpic} 
		\vspace{1.5em}
		\hspace{1.5em}
		&
		%
		%
		%
		\begin{overpic}[width=7cm,viewport={133 113 975 914},clip]{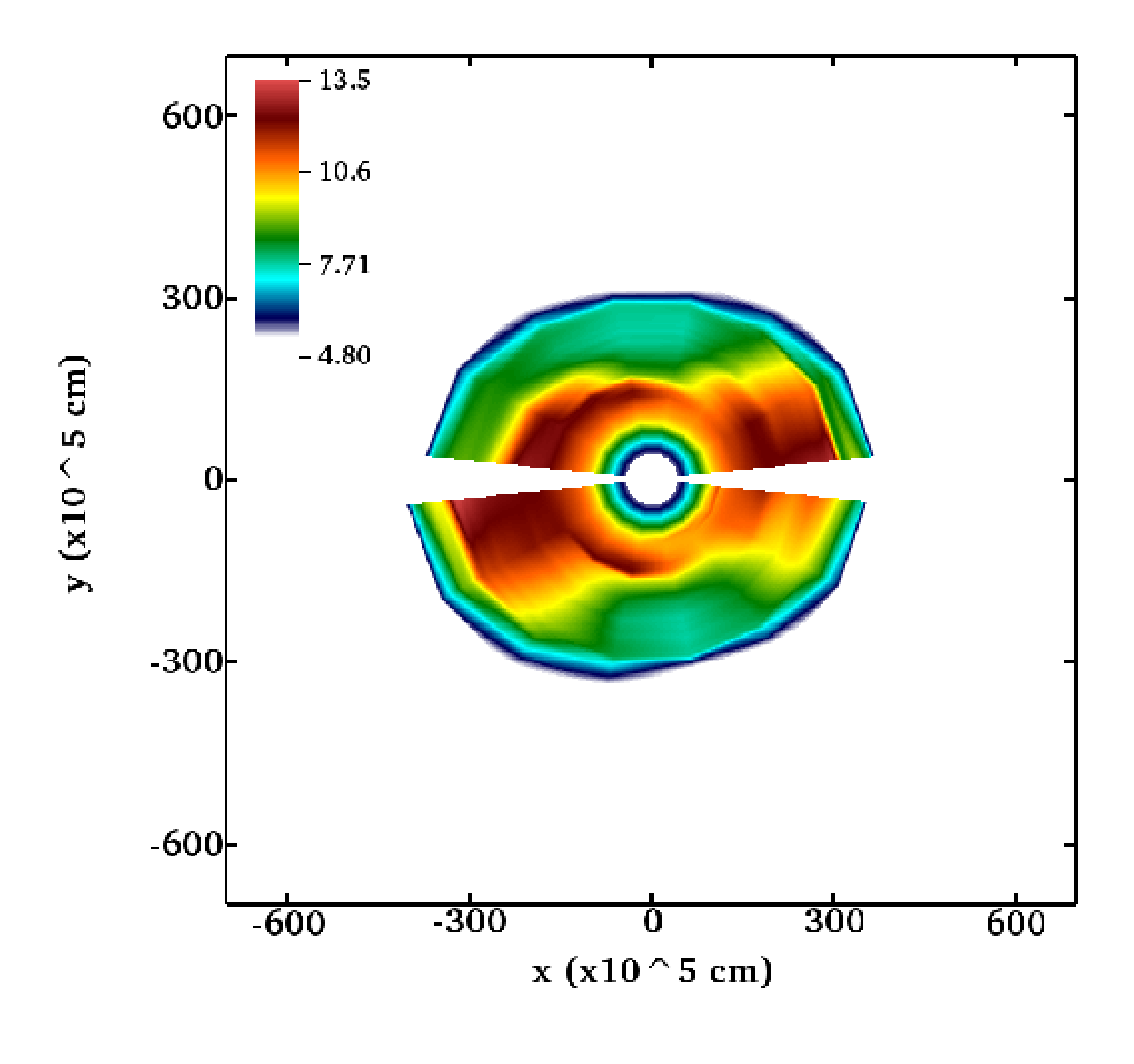}
		\put(82,82){\huge $\mathrm{\left(d\right)}$}
		\put(17,12){\huge $\mathrm{24\degree}$}
		\put(-9,40){\large \rotatebox{90}{$\mathbf{y\;\left[km\right]}$}}
		\put(45,-7){\large {$\mathbf{x\;\left[km\right]}$}}
		\end{overpic} 
		\vspace{1.5em}
    \end{tabular}
	\caption
	{The distribution of entropy in the M187A series of models with varying
	 angular resolution. The entropy is shown for a slice taken at the
	 equatorial plane at $t=100$ ms in models with angular resolutions of (a)
	 3\degree, (b) 6\degree, (c) 12\degree, and (d) 24\degree. Note the
	 overall appearance of convection is degraded as the angular resolution
	 decreases, indicating a drop in the efficiency of the neutrino-driven
	 convective engine. The shock radius is significantly smaller in the
	 coarsest model. See Section \ref{s:resolution} for discussion.
	\label{f:angular_res}}
	\end{center}
\end{figure*}
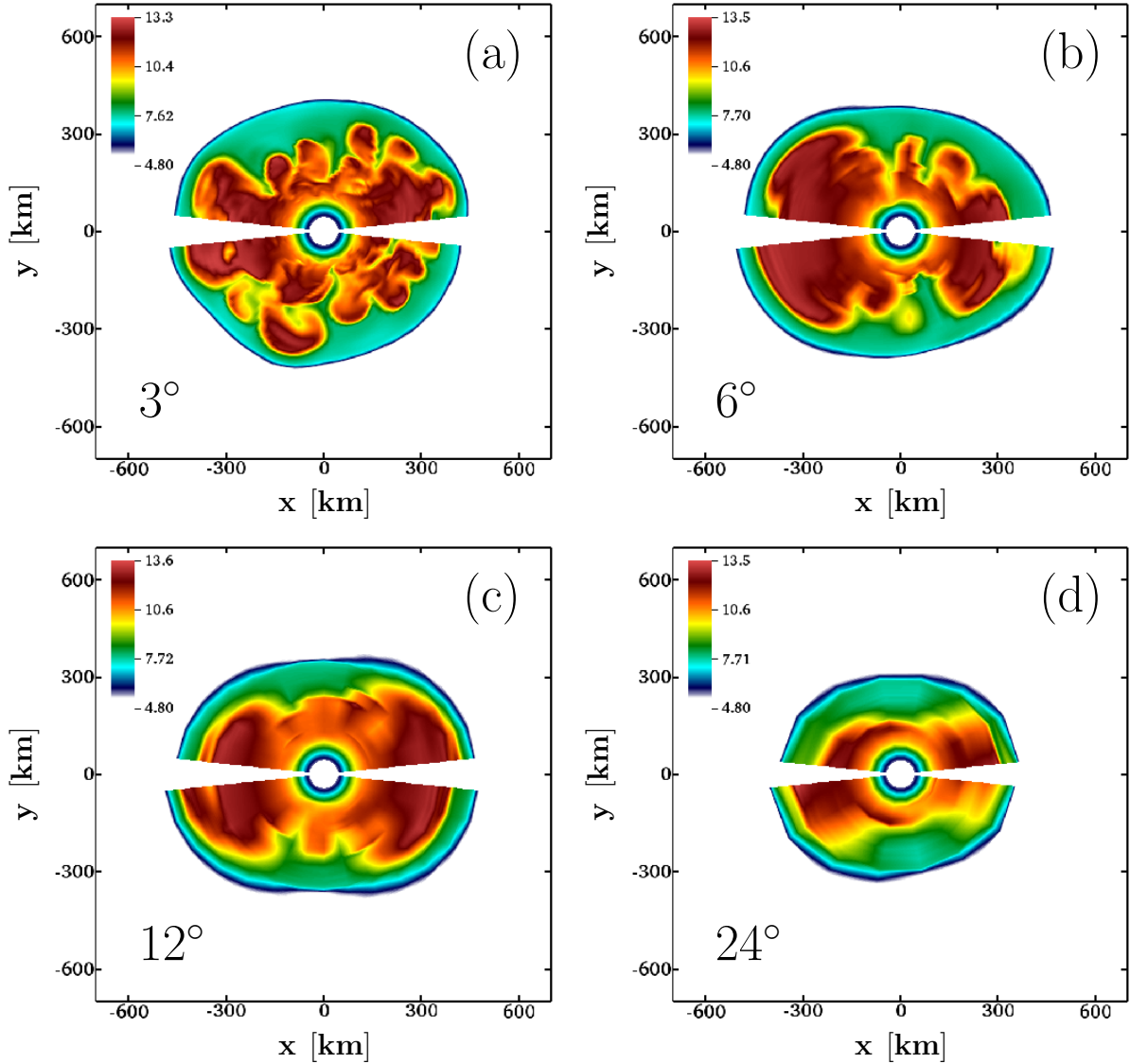

It is conceivable that the observed changes in global model characteristics
should be correlated with changes in the flow structure inside the gain
region. Figure \ref{f:angular_res}
shows the entropy maps in a subset of the models used in this resolution
study. One can easily note profound structural changes in the morphology of
the gain region as the angular model resolution decreases. Large- and small-scale 
convective bubbles can clearly be seen in the 3\degree\ resolution model
(panel (a)). There are significantly fewer small, convective bubbles present
in the 6\degree\ model (panel (b)). In those two models the overall shock
radius appears comparable. In contrast, no well-defined, small-scale bubbles
can be identified in the 12\degree\ or 24\degree\ models (shown in panels (c)
and (d), respectively). In addition, the shock radius is visibly smaller in
the 24\degree\ model.

We interpret the observed dependence of energetics and explosion timing on
angular resolution described above as follows. Consider that the hydrodynamic
solver used in our simulations is the PPM scheme \citep{colella+84}. PPM is
nominally third order (and practically second order) accurate in space, and
uses piece-wise parabolic interpolation to describe profiles of the
hydrodynamic state. Therefore, the minimal number of mesh resolution elements
required to resolve a convective bubble is about 3 mesh cells. In our
12\degree\ and 24\degree\ resolution models there are only 13 and 7 cells in
angle, respectively. This implies our hydrodynamic solver can only represent
between 2 to 4 large-scale structures in those models. The number of those
structures does not represent the actual number of bubbles, as some mesh cells
are also required to describe the flow structure between the bubbles. This
explains the lack of bubbles on smaller scales and the overall degraded
appearance of the neutrino-driven convection in these cases.

We demonstrated through this resolution study that the neutrino-driven
convection is a critical component of the explosion mechanism in our models.
Specifically, we have shown that one can turn an energetic, multidimensional
model into a failed multidimensional model simply by degrading its angular
resolution. Therefore, to correctly capture the efficiency of the convective
engine one needs to resolve its basic components. This picture is also
consistent with the analysis of \cite{herant+94}, who argued that, independent
of dimensionality, large scales will play the dominant role in convection.
Furthermore, it is important to resolve not only structures on the largest
scales, but also the structures on $\sim 1$\degree\ are important. We conclude
with the somewhat obvious statement that capturing the relevant physics of the
problem requires adequate numerical resolution. It is conceivable that the
next generation of ccSNe models with much higher resolution will begin
uncovering new physics effects that cannot be observed in the current
generation of models due to their insufficient quality.
\subsection{Dynamics}
\label{s:convdynamics}
\begin{figure}[t!]
\centering
\includegraphics[width=8cm]{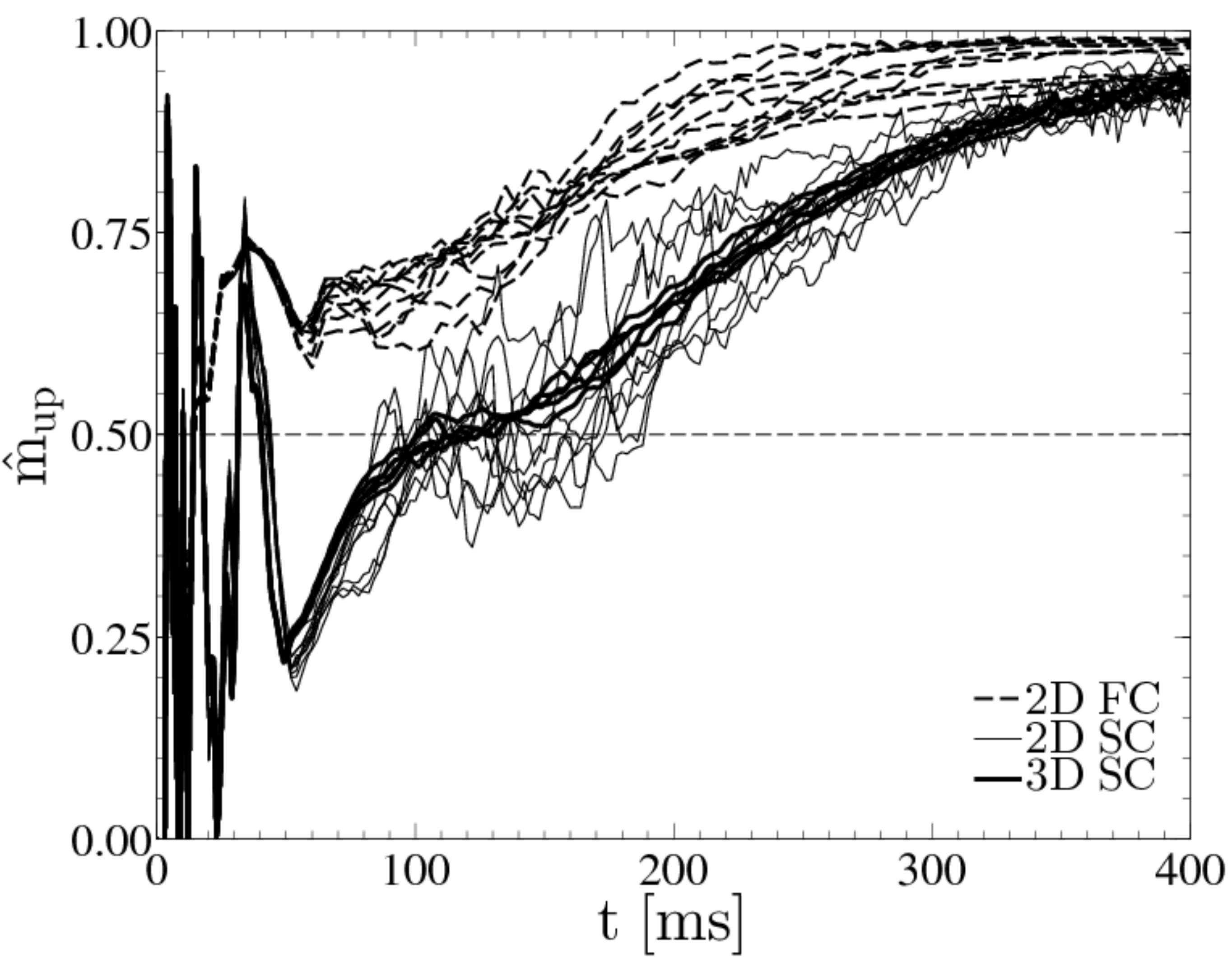}
\caption
{	
	The temporal evolution of the upflowing mass in the gain region. The
	fraction of the total mass inside the gain region with
	$\left(u_r>0\right)$ is shown with dash-dotted, dashed, and solid lines
	for 2D FC, 2D SC, and 3D SC models, respectively. See Section
	\ref{s:convdynamics} for discussion.
\label{f:massfrac}
}
\end{figure}

One of the quantities of interest in the context of dynamics of convection is
the amount of mass contained inside rising, convective bubbles. To
characterize that quantity, we introduce the upflow mass fraction,
\begin{equation}
\hat{m}_{up}\left(t\right) = \frac{\iiint\limits_{gain}\rho|_{u_{r}>0}dV}{M_{gain}},
\label{e:mhat}
\end{equation}
where the integration is performed over the gain region, and takes into
account only the fluid elements (grid zones) with positive radial velocity.
Thus, the upflow mass fraction characterizes the mass inside the gain that is
moved away from the proto-neutron star toward the shock. The evolution of the
upflow mass fraction in our models is shown in Figure \ref{f:massfrac}.
As one can see, the initial behavior of the upflowing mass is characterized by
a transient period until $t\approx 50$ ms. Soon after that, the amount of
upflowing mass begins to steadily increase. It levels off during the quasi-steady 
state phase lasting approximately between $85$--$140$ ms for SC models
(shown with solid lines in Figure \ref{f:massfrac}) and $65$--$105$ ms for FC
models (shown with dashed lines in Figure \ref{f:massfrac}), and rises again
after the shock is launched. Interestingly, the net mass flux inside the gain
region is close to zero in the SC models (the horizontal dashed line at
$\hat{m}_{up}=0.5$), while upflows dominate in the FC models. This can be
qualitatively understood considering that energetic explosions in the case of
fast contracting proto-neutron star cores require much higher neutrino
luminosities in order to unbind a sufficiently large amount of material inside
the gain region. This indicates that perhaps convection plays a smaller role
in FC models compared to SC models. Also note that the amount of upflowing
mass begins to differ between various model realizations shortly before the
quasi-steady state is established. Those variations are greater in 2D than in
3D (for SC models). Ultimately, after the explosion is in progress, the amount
of upflowing mass steadily increases and eventually levels off, signifying the
fact of the overall expansion of the material inside the gain region. 

It is interesting to note that the observed behavior in the upflowing mass
fraction, in particular its steady increase after the transient period, is
closely correlated with the evolution of the heating efficiency (cf.\ Figure
\ref{f:efficiency}). For example, the heating efficiency becomes greater than
1 in the SC models before the amount of upflowing material becomes comparable
to the amount of accreted material. In addition, the heating efficiency is
always greater than 1 in the case of FC models, where upflows dominate at all
times.

\begin{figure}[t!]
\centering
\includegraphics[width=8cm]{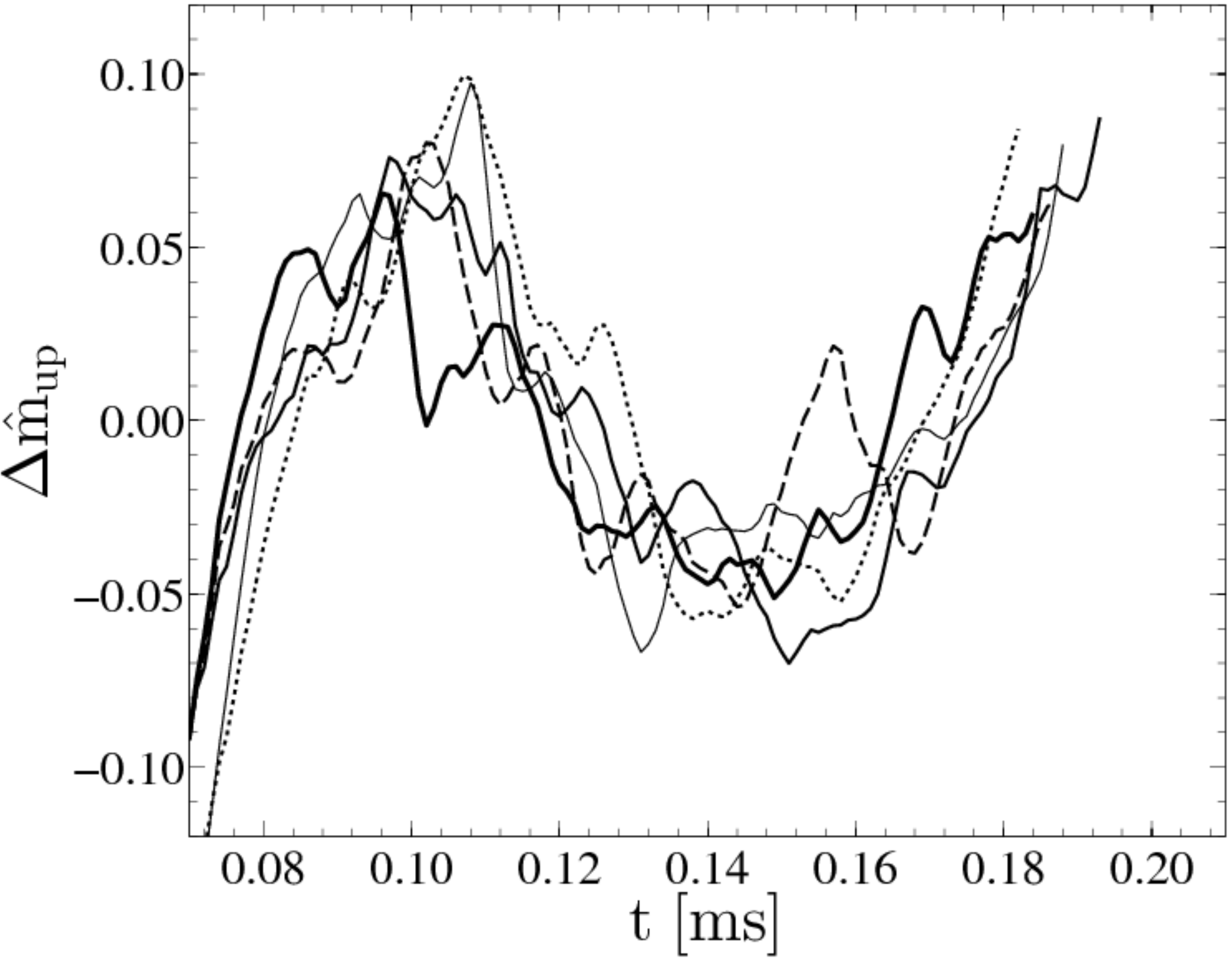}
\caption
{	
	The evolution of relative mass fluctuations inside the gain region shortly
	before and during the quasi-steady state phase in the 3D SC explosion
	models. Fluctuations are calculated relative to a linear fit of the
	upflowing mass in the gain region from the beginning of the quasi-steady
	state phase until the explosion time, where individual curves terminate.
	(thin solid) M157A; (medium solid) M157B; (thick solid) M157C; (dashed)
	M157D; (dotted) M157E. See Section \ref{s:convdynamics} for discussion.
}
\label{f:mup}
\end{figure}

Figure \ref{f:mup} shows the time evolution of the fluctuations in the
upflowing mass in the gain
region in the 3D SC models as a function of time around the quasi-steady
phase. It is interesting to note that the increase of the outflowing mass in
these models is non-monotonic. One can identify several short ($\Delta t
\approx 10$ ms) episodes during which the amount of mass in upflows
fluctuates. These fluctuations have an amplitude of several percent. Given
that the typical number of convective bubbles, and therefore also the number
of downflows, in those models during the quasi-steady state phase is about
$10$ to $25$ (see Section \ref{s:bubbles}), it is conceivable that perhaps
individual downflows are responsible for the observed fluctuations.
\subsection{Characterization of convective structures}
\label{s:bubbles}
Having in mind the continuing debate regarding the role of dimensionality in
numerical models of ccSNe explosions, we seek information about the dependence
of large-scale structure of convection inside the gain region on model
dimension. As the first step in our analysis,
we wish to measure the net radial momentum over a fraction of the gain region, 
\begin{equation}
\label{e:imgmap}
I_{mom}\left(\theta,\phi\right) = 
\begin{cases} 
\mbox{1,} &\int\limits_{r_{mid}\left(\theta,\phi\right)-\alpha w}^{r_{mid}\left(\theta,\phi\right)+\alpha w} \rho u_{r}dr > \mbox{0} 
\\ 
\mbox{0,} &\mbox{otherwise} 
\end{cases} 
\end{equation}
where $r_{mid}=\left(r_{s}+r_{g}\right)/2$ is the angle-dependent midpoint
between the shock and gain radii ($r_{s}$ and $r_{g}$, respectively), $w =
r_{s}-r_{g}$ is the angle-dependent gain region width, and $\alpha$ is a
fractional offset between $0$ and $0.5$ which controls the extent of the gain
region included in the integration. We use the above equation to project
average properties of the radial flow and distinguish between regions
dominated by either upflows ($I_{mom} = 1$) or downflows ($I_{mom} = 0$).
Recall that our models are essentially free of possible SASI contributions,
and therefore this method should be applicable to any convective flows.

Apart from the situation when the shock and gain radius are constant, the
radial extent of the integration bounds in the above equation will vary with
angle. In this way, we avoid possible ambiguities that may arise near those
two regions, and make the results of this method insensitive to flow
asymmetries naturally developing especially along the shock front. In the
above equation we heuristically set $\alpha=0.1$ so that our results are based
on the data over the middle 20\% of the gain region (at a given angle).

We use the maps produced with the help of the above equation to create
connected clusters of upflow-like and downflow-like sections. We identify the
upflow-dominated clusters with the rising convective bubbles. The solid angle
spanned by each bubble is simply computed by integrating the area of the
upflow-dominated cluster.

\begin{figure}[t!]
\centering
\includegraphics[width=8cm]{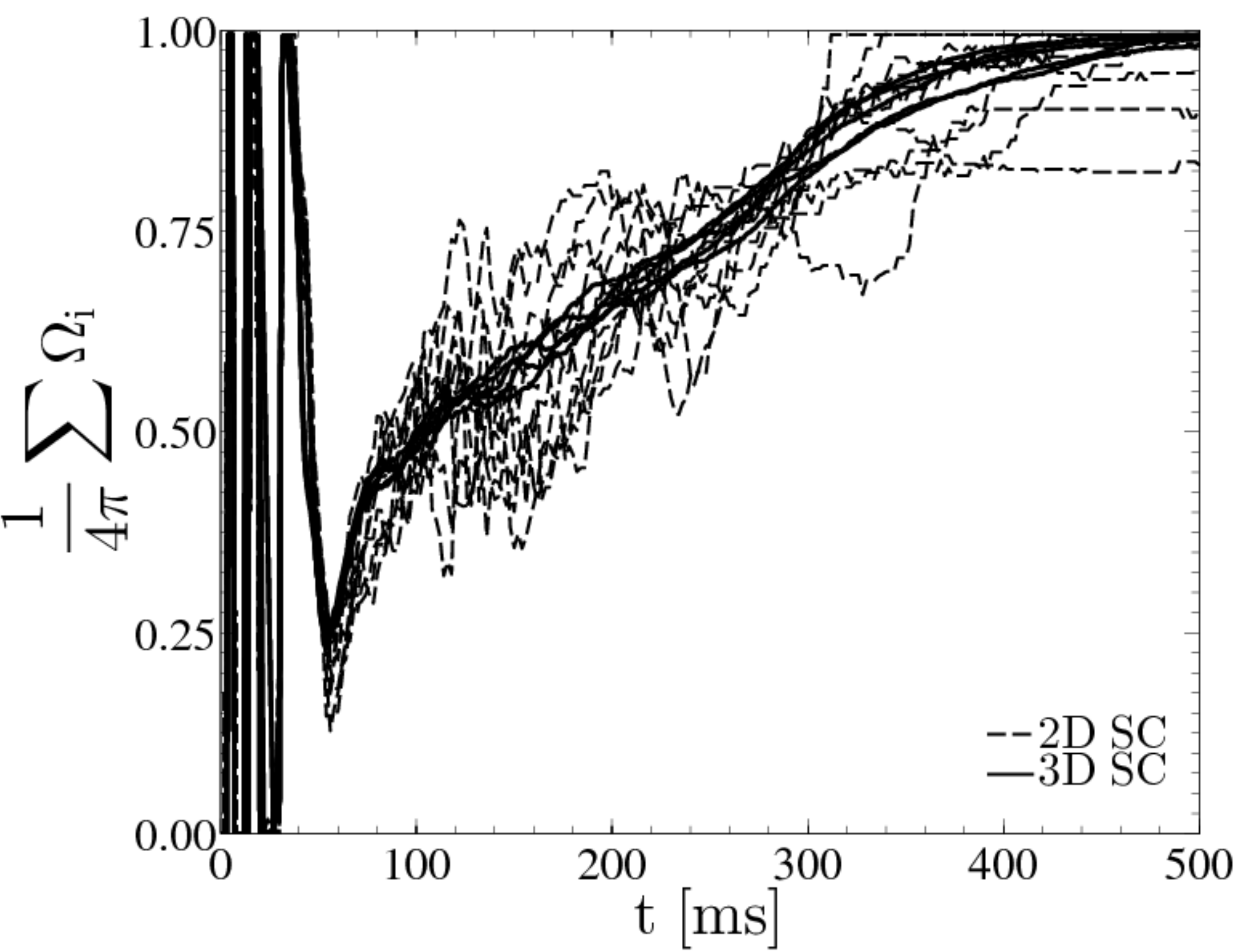}
\caption
{
	Estimate of the total solid angle spanned by buoyant, rising material. The
	onset of convection at $t\approx 50$ ms results in a rapid increase in the
	spanned solid angle as the first convective plumes begin to form. During
	the quasi-steady convective phase the plumes continue to grow. By the time
	the explosion is launched they cover roughly 2/3 of the total solid angle.
	See Section \ref{s:bubbles} for discussion.
\label{f:clustersize}
}
\end{figure}

The time evolution of the total solid angle occupied by bubbles for the SC
models is shown in Figure \ref{f:clustersize} (with data for 2D and 3D models
shown by dashed and solid lines, respectively).
Prior to the onset of convection ($t<50$ ms), the whole gain region oscillates
radially. As convective structures begin to emerge, the solid angle occupied
by bubbles starts to increase. This process starts to slow down at the
beginning of the quasi-steady state phase ($t\approx 80$ ms). However, and
unlike in the case of other diagnostics we have discussed above, the evolution
of the solid angle occupied by bubbles does not provide any clear signature of
the explosion time. Rather, the growth continues at a similar pace until about
150 ms after the explosion commences. It starts to level off soon after that
time, and by 400 ms after bounce the whole gain region is filled-in with
upflowing material.

Apart from large-amplitude oscillations, the evolution of the solid angle
occupied by bubbles in 2D SC models is qualitatively similar to that seen in
3D, with the average value similar between the models. However, in a few 2D
cases persistent downflows develop (see also Section \ref{s:morphology}). For
these models the solid angle occupied by bubbles does not reach $1$ by the end
of the period we analyzed here. Note that our diagnostics can be use to
identify situations when strong accretion onto the proto-neutron star
continues at late times. 

\begin{figure}[tH]
\centering
\includegraphics[width=8cm]{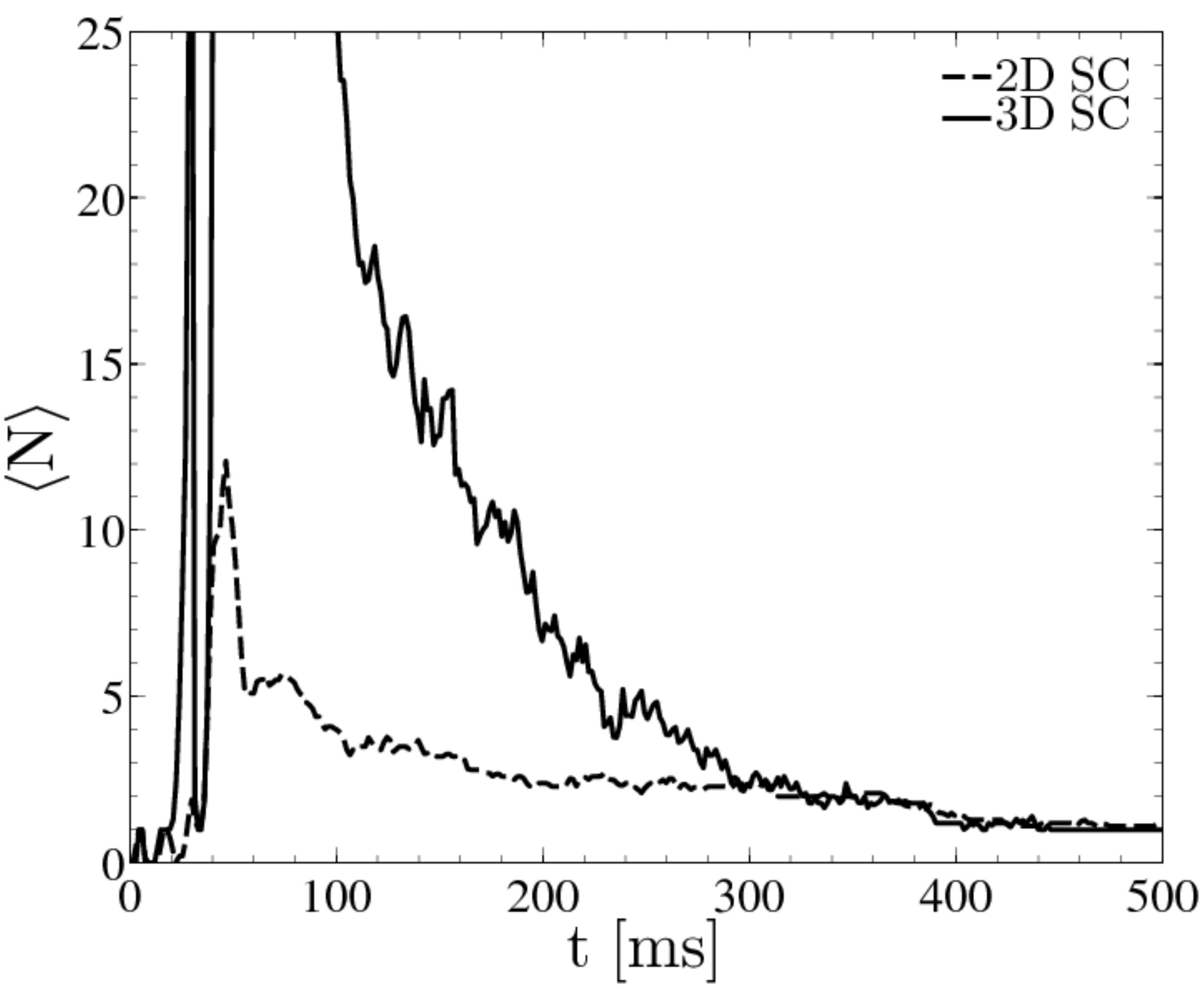}
\caption
{
	Evolution of the number of rising bubbles inside the gain region. The
	number of bubbles is averaged separately for 2D and 3D SC model
	realizations (shown with dashed and solid lines, respectively). For the
	clarity of presentation, the vertical scale is limited to
	$\left\langle{N}\right\rangle=25$. The maximum average number of bubbles
	found in 3D is $\approx 200$. See Section \ref{s:bubbles} for discussion.	
}
\label{f:numclusters}
\end{figure}

Figure \ref{f:numclusters} 
shows the time evolution of the number of bubbles inside the gain region
averaged separately for 2D and 3D SC models (shown with dashed and solid
lines, respectively). Initially, our method identifies only a single bubble or
finds no bubble during the initial transient ($t<50$ ms), in agreement with
the radial oscillations of the gain region during that period as we discussed
earlier. Soon after convection sets in, the number of distinct upflows sharply
increases, reaching $\approx 12$ in 2D and $\approx 200$ in 3D. (Note, for the
purpose of presentation we have limited the scale in Figure \ref{f:numclusters}
to 25.) Given that at those early times the solid angle occupied by bubbles is
small, we conclude that the bubbles are initially small and grow in angular
extent over time. By the time of the explosion, the bubbles are roughly 4
times smaller in 3D than in 2D.

Recall that in 2D the bubbles are truly tori rather than quasi-spherical
plumes. We recognize this may have certain implications for the bubble
dynamics \citep[see, e.g.,][]{kifonidis+03,hammer+10,couch13}. Furthermore,
the observed evolution of bubble sizes from small to large \citep[see
also][]{herant+94} offers an interesting parallel between the evolution of the
structure of neutrino-driven convection (and perhaps convection in general),
and that of bubble merger observed in a multi-mode Rayleigh-Taylor instability
\citep[see, e.g., ][]{alon+94,miles04}. The study of the process of merging
convective bubbles is beyond the scope of the present work.

The growth of convective bubbles after $t=50$ ms coincides with time at which
the heating efficiency starts to increase (cf.\ Figure \ref{f:efficiency}) and
accumulation of mass in the gain region (cf.\ Figure \ref{f:grmass}). We
believe this indicates that early, small-scale convection begins to interact
with the incoming accretion flow, resulting in rather rapid increase in the
gain region mass. Subsequently, the advection time scale also increases. We
believe this provides evidence for the direct dependence of the heating
efficiency on the intensity of convection.
\subsection{Convective energy transport}
\label{s:convenergytrans}
In our discussion of convective energy transport, we adopt the approach
originally introduced for the analysis of stellar convection by
\cite{hurlburt+86}. In this approach, the individual components of the energy
transport equation, Equation \ref{e:energy}, are averaged over lateral directions.
Then, one computes radial distributions of deviations away from the lateral
averages, $f^\prime \equiv f - \bar{f}$. In the discussion here, we use the
notation of \cite{mocak+09}. We define the convective flux, $F_C$, kinetic
flux, $F_K$, buoyant work, $P_A$, and expansion work, $P_P$, as follows:
\begin{equation}
\label{e:flux_c}
F_C = \oint{u_{r}\rho \left(\varepsilon+\frac{P}{\rho}\right)^\prime r^{2}d\Omega},
\end{equation}
\begin{equation}
\label{e:flux_k}
F_K = \oint{u_{r}\rho \left(\frac{1}{2}\vect{u}\vect{\cdot}\vect{u}\right)^\prime r^{2}d\Omega},
\end{equation}
\begin{equation}
\label{e:flux_pa}
P_A = -\oint{u_{r}\rho^\prime\frac{\partial\overline{\Phi}}{\partial r}r^{2}d\Omega},
\end{equation}
\begin{equation}
\label{e:flux_pp}
P_P = \oint{\left(\vect{\nabla}\vect{\cdot}\vect{u}\right)P^\prime r^{2} d\Omega},
\end{equation}
where the integrals are taken over the suitable solid angle (in our models
with the excised $12$\degree\ cone around the symmetry axis this amounts to
$\approx 99.5$\% of the full solid angle). These terms describe the total
energy transported per unit time through a surface of a sphere of radius $r$.

\begin{figure*}[t!]
\centering
\includegraphics[width=8cm]{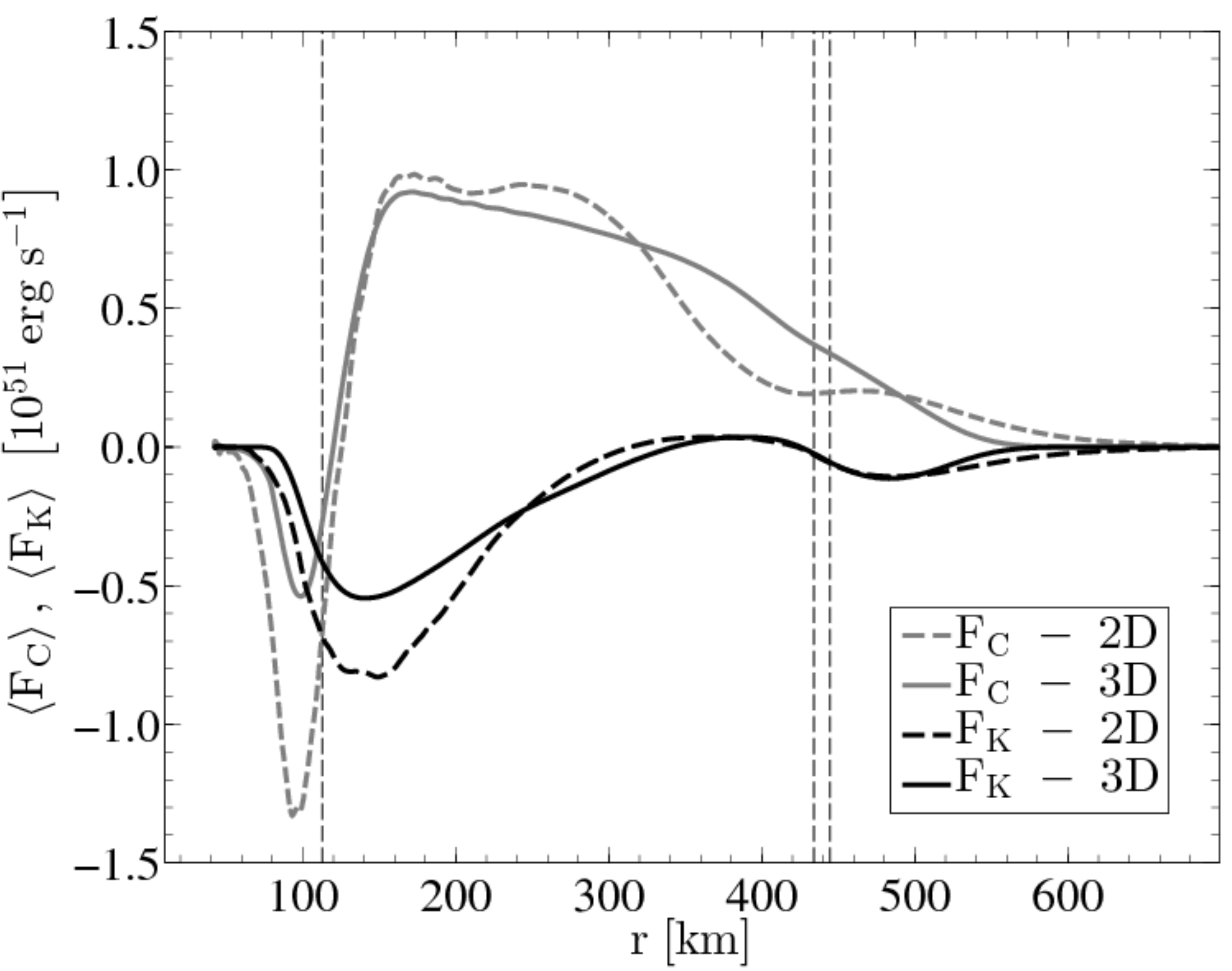}
\includegraphics[width=8cm]{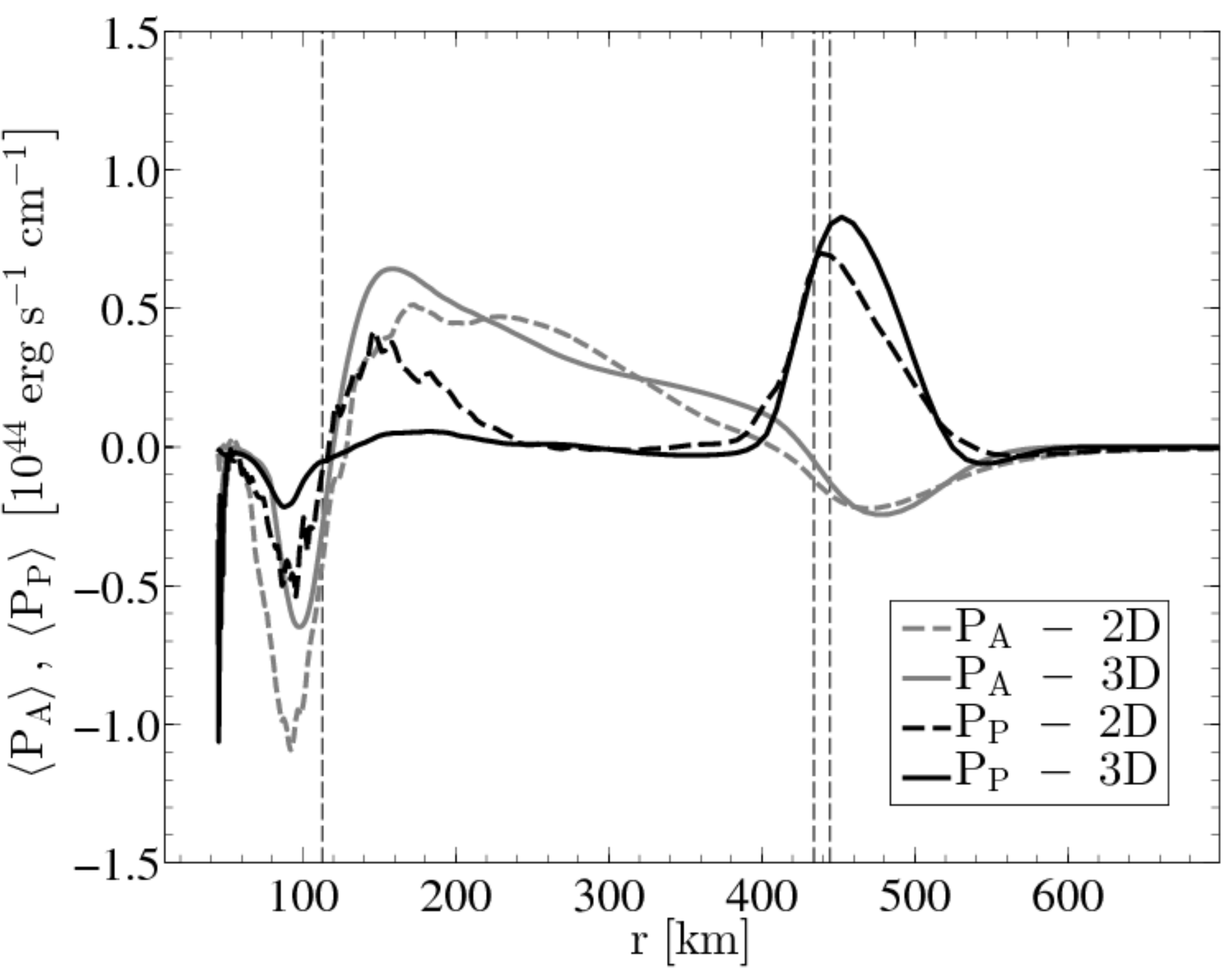}
\caption
{
	Decomposition of the total energy flux and the distribution of the related
	work source terms in the SC models. Quantities are averaged over time over
	the period between $t=110$ ms and $t=120$ ms. The individual curves
	represent averages taken over all model realizations for a particular
	model dimension (shown with dashed and solid lines for 2D and 3D models,
	respectively). The vertical lines denote the approximate locations of the
	gain radius and the minimum shock radii for the two families. (left panel)
	Convective flux, $F_C$, and kinetic flux, $F_K$. (right panel) Buoyant
	($P_A$) and expansion ($P_P$) work. See Section \ref{s:convenergytrans}
	for discussion.
\label{f:fluxes}
}
\end{figure*}

Figure \ref{f:fluxes} 
shows the time averages of convective and kinetic energy fluxes (left panel)
and the buoyant and expansion work terms (right panel) for the SC models. The
main characteristics of the energy transport follow from the direction of the
energy transport and the relative contributions of individual terms. We begin
our analysis at the gain radius located at $\approx 115$ km in both 2D and 3D
models (marked with the solid, vertical line). There, the convective flux is
initially negative but rapidly increases and reaches the maximum at $\approx
15$\% of the radial extent of the gain region. Further out, the convective
flux gradually decreases and ultimately ceases well upstream of the shock. On
the other hand, the kinetic energy flux is negative below the gain radius and
remains negative through most of the gain region. The observed run of the
convective flux provides clear evidence for convective transport in which
material heated near the gain radius buoyantly rises toward the shock.

The relation between buoyancy and convection also manifests itself through the
buoyant work, $P_A$ (shown with gray lines in the right panel of Figure
\ref{f:fluxes}). The work done by buoyancy becomes positive starting at the
gain radius, indicative of buoyancy driving the convection upwards. It first
rapidly increases through the bottom layers of the gain region, and reaches
its maximum at $\approx 10$\% of the gain region extent. Further out, the work
done by buoyancy gradually decreases and eventually vanishes as expected at
the shock, where the flow becomes, on average, spherically symmetric. 

As we noted, the buoyant work is positive throughout the gain region. This can
be caused by either overdense material sinking or underdense material rising.
Since the work due to buoyancy is negative below the gain region, therefore
the opposite scenario with either overdense, cold material is rising or
underdense, hot material is sinking. Since we do not observe sinking of the
underdense, hot material in that region, we conclude that the negative work
due to buoyancy indicates the presence of rising overdense and cold matter. In
our models, this rising, overdense material originates from the downflows that
are turned around by the pressure gradient. This process is known as buoyant
braking \citep{brummell+02}.

The most conspicuous difference between 2D and 3D lies in the distribution and
value of the expansion work source term, $P_{P}$ (shown with solid lines in
the right panel of Figure \ref{f:fluxes}). When integrated over time and
space, this term represents the $PdV$ work of fluid elements. Below the gain
radius this term is negative, indicative of the work done in the process of
compressing the fluid. It changes sign across the gain radius, and remains
positive through the lower one-third of the gain region. This indicates that
the fluid is expanding in that region. We quantify the amount of work due to
expansion by integrating Equation \ref{e:flux_pp} over the portion of domain
between $r\approx 60$ km and $r\approx 250$ km, where the expansion work
source term is significant and differs most between 2D and 3D. Assuming a
characteristic timescale of $100$ ms, we estimate the (kinetic) energy change
due to $PdV$ work to be $\approx 1\times10^{50}$ erg and $\approx
-3\times10^{49}$ erg in 2D and 3D, respectively, or about a few percent of the
explosion energy.

We believe the difference in the $PdV$ work, as we described above, reflects
the fact that the structural integrity of upflows and downflows in 2D is much
greater than in 3D. The structural integrity of flow features depends on the
surface-to-volume ratio. For a given volume, 2D structures have less surface
area than their 3D counterparts. Thus, flow structures in 2D are less
susceptible to (surface) perturbations. We alluded to this property in our
discussion of morphological differences between 2D and 3D simulations in
Section \ref{s:morphology}. Consequently, in 3D, the downflows do not
penetrate as deep, which can be clearly seen in the left panel of Figure
\ref{f:fluxes}, which shows that both the kinetic fluxes are negative and of
comparable magnitude in both 2D and 3D around $r\approx 250$ km. However, as
one moves into deeper layers the kinetic flux much more rapidly decreases in
2D than in 3D, and vanishes in the layers closer to the proto-neutron star
surface. The convective fluxes in 2D and 3D behave in a qualitatively similar
way, although they are positive and equal closer to the gain radius, at
$r\approx 150$ km. From there inward, the convective flux decreases at a
significantly greater rate across the gain radius and ceases closer to the
proto-neutron star surface in 2D than in 3D. Furthermore, the greater
structural integrity of upflows and their coherent nature in 2D makes the
expansion process dramatically more efficient than in 3D. This is evidenced by
much greater work done by expansion above the gain radius, as we discussed
earlier.

As one can see in the right panel of Figure \ref{f:fluxes}, the buoyant and
expansion terms are both negative below the gain region. This is because below
the gain region the system is no longer energy conserving due to intense
neutrino heating. This situation is different, however, across the shock
front. There, both terms are of different sign, which is consistent with the
fact that the neutrino heating is weak and the energy is approximately
conserved in that region.

Finally, we would like to note an interesting possibility that the difference
in efficiency of the convective engine between 2D and 3D might be another
consequence of the difference in surface-to-volume ratio between those models.
This is because in 3D the surface area of the interface between hot and cold
material where the mixing, and thus also heat exchange, takes place is
relatively larger. Quantifying this possibly important effect is, however,
beyond the scope of this paper.
\section{Turbulence}
\label{s:turb}
Motivated by the recent discussion of the possible role of turbulence in the
process of shock revival \citep{murphy+11,hanke+12}, we present spectra of the
kinetic energy in a lateral direction and turbulent Reynolds stresses
characteristic of the gain region for our SC models. In particular, we will
determine if the kinetic energy spectra are consistent with model
dimensionality, and discuss the qualitative differences in the Reynolds
stresses found between 2D and 3D.
\subsection{Spectra}
\label{s:spectra}
We investigate the spectral properties of turbulence by considering the
spherical harmonic decomposition of Equation \ref{e:sphericalharmonic} with
$f\left(\theta,\phi\right)=\sqrt{\rho}u_{\theta}$, where $u_{\theta}$ is the
polar component of the velocity vector. Our choice of this particular velocity
component hinges on the assumption that the flow is isotropic in the lateral
directions, which is supported by our analysis of the Reynolds stresses
presented below in Section \ref{s:reynoldsstresses}.

To ensure that the energy contained in the spherical harmonic modes adequately
represents the lateral kinetic energy, we include the factor $\sqrt{\rho}$
\citep{endeve+12}. Following \cite{hanke+12}, we define the lateral kinetic
energy spectrum, $E\left(l\right)$, as
\begin{equation}
\label{e:spectra}
E\left(l\right) = \sum_{m=-l}^{l} \left[\int_{\Omega}Y_{lm}^{*}\left(\theta,\phi\right)\sqrt{\rho}u_{\theta}d\Omega\right]^{2},
\end{equation}
where the integral is taken over the solid angle included in our simulations.
In order to account for spatial and temporal variations of the lateral kinetic
energy spectrum, we average the spectrum over a shell of thickness $25$ km
centered at the midpoint of the gain region ($r_{mid}\approx 290$ km). In
addition, in our analysis the spectra are averaged over snapshots from the
time interval between $t=120$ ms and $t=130$ ms (soon after the quasi-steady
state phase). In passing we note that there exists other choices of physical
quantities for the analysis of turbulence, for example density fluctuations
\citep{borriello+13}, which are more suited for the analysis of
neutrino-matter interactions rather than fluid flow dynamics.

\begin{figure}[t!]
\centering
\includegraphics[width=8cm]{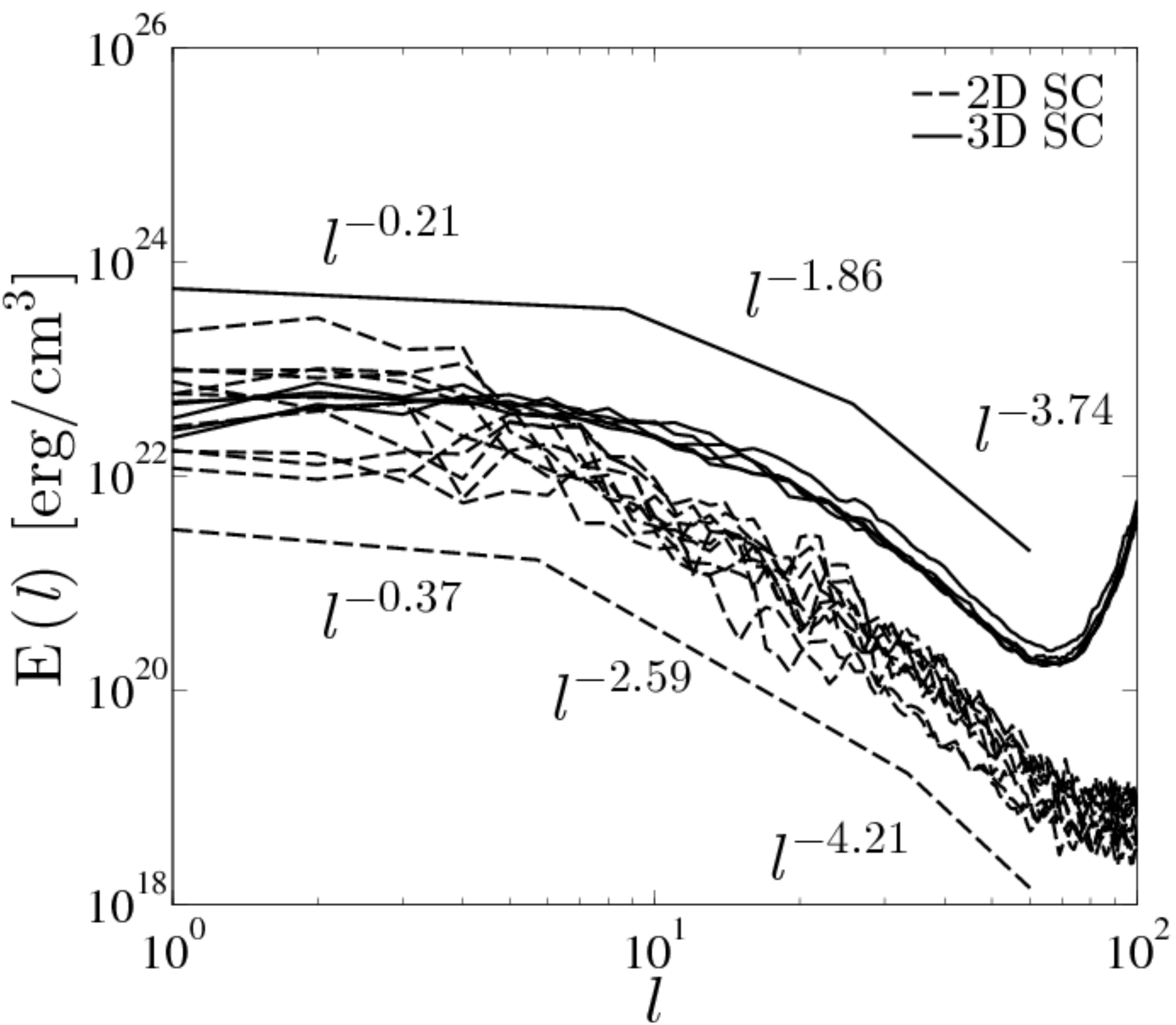}
\caption
{
	Lateral kinetic energy spectra for SC models. The spectra are shown with
	dashed and solid lines for 2D and 3D models, respectively. The data are
	averaged spatially over a spherical shell 25 km thick centered at 290 km,
	and over the time period between $t=120$ ms and $t=130$ ms. Piecewise
	powerlaw functions are fit separately to spectra for 2D and 3D model
	families, and are shown with dashed and solid lines for 2D and 3D,
	respectively. The fit process allows for optimal locations of the break
	points in the piecewise powerlaw functions. See Section \ref{s:spectra}
	for discussion.
\label{f:spectra}
}
\end{figure}

Figure \ref{f:spectra} 
shows the lateral kinetic energy spectra for our SC models with 2D and 3D
models shown with dashed and solid lines, respectively, averaged over time and
space as described above. We find that one can identify three distinct
spectral regions in the lateral kinetic energy spectrum. The first region
occupies large scales ($l \lesssim$ a few), the intermediate region extends up
to $l \lesssim$ several 10s, and the third region extends toward still smaller
scales ($l \gtrsim 60$). This resembles spectra routinely found in turbulence
studies, with the highest end of our spectrum closely resembling the region
found in numerical simulations. It is customary to characterize those regions
by a piecewise-linear combination of powerlaws, $E\left(l\right) \propto
l^\alpha$. In particular, it is expected that in the intermediate regime, in
which the energy is transported from large to small scale (and possibly also
in the opposite direction in 2D situations), a cascade develops with the power
law exponent, $\alpha$, chiefly dependent on the particular characteristics of
the system.

To quantify the shape of the power spectra found in our simulations, we fit
power law functions inside each region. We found that the power law exponent
that best characterizes the shape of the spectrum at large scales is
$\alpha\approx -0.2$ (in 3D) to $\alpha\approx -0.4$ (in 2D). On the other
extreme, the spectrum steepens around $l \approx 35$ with $\alpha\approx -4$.
At still smaller scales ($l \gtrsim 60$), the spectrum abruptly changes its
shape presumably due to numerical dissipation.

Of the greatest interest from the point of view of turbulence are energy-transporting 
intermediate scales (a few $ \lesssim l \lesssim 35$). We find
that in this region the power spectra in 2D simulations are significantly
steeper ($\alpha = -2.59$) than in 3D ($\alpha = -1.86$). This is
qualitatively consistent with the results of turbulence studies in 2D. We also
note that the power law exponent found in 3D is not too dissimilar from that
of the classic Kolmogorov spectrum ($\alpha = -5/3$; \cite{monin+71}).

On a final note, the Reynolds number estimated for our models is quite small
($Re \simeq 10$s $\ldots 100$s, see Appendix \ref{s:reynoldsnumber}),
characteristic of perturbed laminar flow rather than turbulence. We caution
one should be careful with drawing any conclusions regarding the turbulent
character of such underresolved numerical models
\citep{murphy+08,murphy+11,hanke+12}. Furthermore, our models do not account
for additional physics effects that may play an important role at the base of
the convective region, such as neutrino viscosity (see discussion in Section
3.1 of \cite{fryer+07}). These effects may actually reduce the physical
Reynolds number down to $\approx 100$, which is comparable to the numerical
Reynolds number found in our simulations. Studying such effects, however, is
beyond the scope of the present work.
\subsection{Turbulent Reynolds stresses}
\label{s:reynoldsstresses}

\begin{figure}[t!]
\centering
\includegraphics[width=8cm]{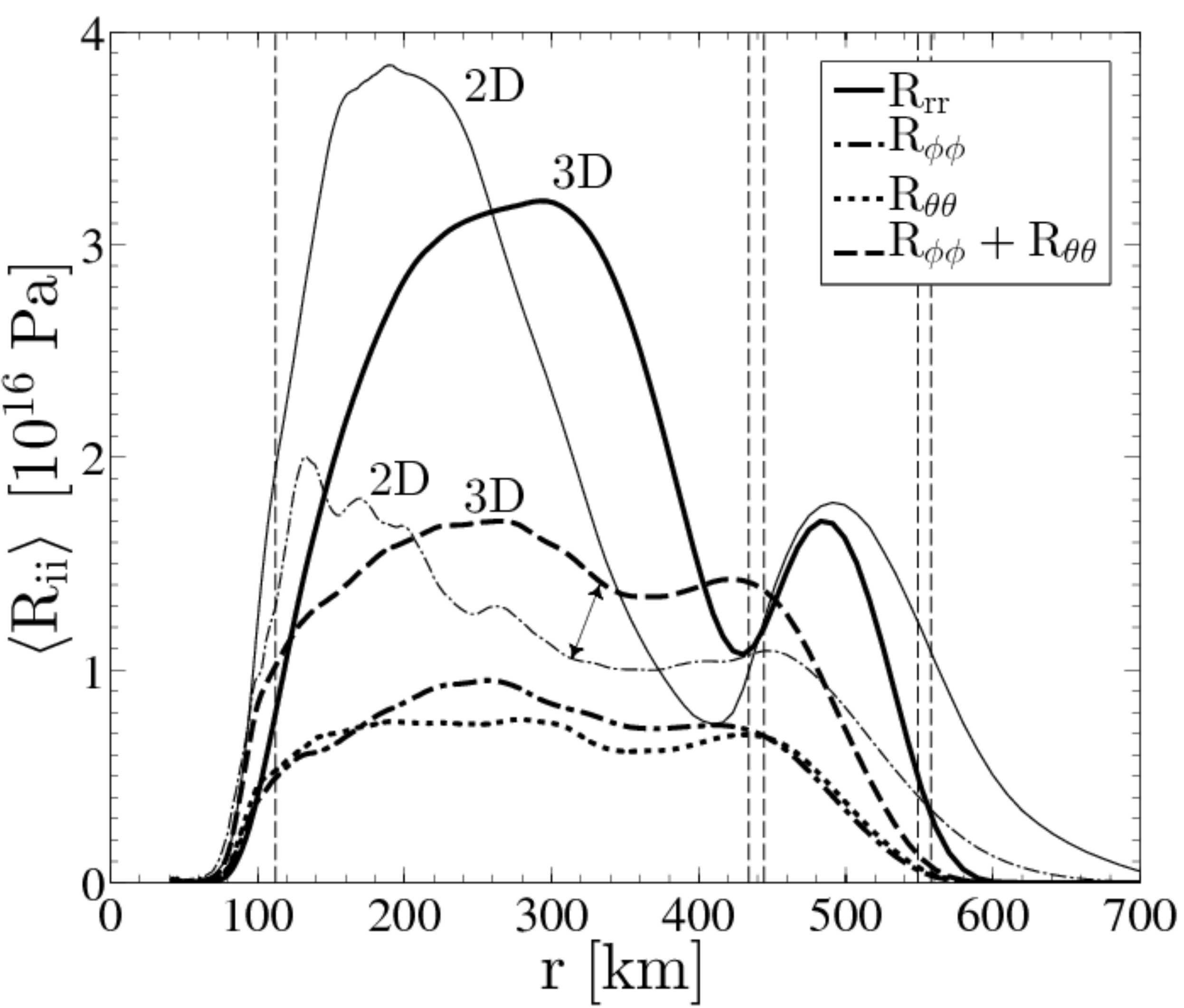}
\caption
{Distributions of Reynolds stresses in the gain region for the SC models.
The data shown in the figure is obtained by first averaging stress
distributions for individual model realizations over the period between
$t= 110$ ms and $t=130$ ms, and then by calculating mean values based on
the time-averages separately for 2D and 3D. The results for 2D and 3D
families are shown with thin and thick lines, respectively. The dashed
vertical lines at $r\approx 110$ km, $400$ km, and $570$ km mark the
approximate positions of the gain radius, minimum shock radius, and
maximum shock radius, respectively. Note that the stresses for 2D models
rise much more rapidly across the gain radius region than in 3D. See
Section \ref{s:reynoldsstresses} for discussion.
\label{f:reynoldsstresses}}
\end{figure}

Figure \ref{f:reynoldsstresses}
shows the time- and model realization-averaged turbulent Reynolds stress
components for the SC models. The data shown in the figure were generated 
as follows. First, the tensor components were computed for individual model
realizations using 
\begin{equation}
\label{e:reynoldsstresses}
R_{ij} = \frac{\langle\rho u_{i}u^\prime_{j}\rangle}{\rho_0},
\end{equation}
where $\langle\cdot\rangle$ denotes the averaging operation over solid angle,
$u^\prime_{j} = u_{j}-\langle u_{j} \rangle$ is the perturbation away from the
background of the $j$-th component of velocity, and $\rho_0 =
\langle\rho\rangle$ is the background density. After averaging over the solid
angle, the stress components were then averaged over the period between $t=110$
ms and $t=130$ ms. In the final step, we used these averages to compute the
model-averaged values of turbulent Reynolds stresses.

In 3D, the radial (shown with the thick solid line in Figure
\ref{f:reynoldsstresses}) and lateral stress components (shown with thick
dash-dotted and dotted lines; their sum is shown with the thick dashed line),
increase from the inner boundary and across the gain radius (marked with a
vertical, dashed line at $r\approx 115$ km). The radial component reaches the
maximum around $r\approx 300$ km, roughly in the middle of the gain region,
while the overall contribution of lateral stresses peaks slightly deeper at
$r\approx 255$ km. The radial stress is about a factor of $2$ greater than the
sum of lateral stresses. Farther out, the stresses decrease and reach
approximately similar magnitude just below the shock. In the figure, the three
dashed, vertical lines located between 400 km and 600 km mark the innermost,
average, and outermost shock radii. The stresses show a mild increase across
the region occupied by the shock and eventually vanish upstream of the shock
inside the accretion flow. The sum of lateral stresses exceeds the radial
stress by $\approx 30$\% in a relatively narrow region behind the shock (at
$r\approx 430$ km). Before we begin discussing the qualitative and
quantitative between the average turbulent Reynolds stresses in 2D and 3D, we
note that the radial dependence of stresses in 2D (shown with thin solid and
dash-dotted lines for the radial and lateral stresses, respectively, in Figure
\ref{f:reynoldsstresses}) is quite similar to that in 3D. Moreover, flow
appears roughly isotropic below the gain radius, with radial and lateral
stresses in equipartition.

We note that the stresses in 2D are much larger in the region near the gain
radius than in 3D. This can be understood by recalling that the convection is
most vigorous in that region in our two-dimensional models (see Section
\ref{s:convenergytrans}). Farther out, the dominant, buoyantly-generated
radial stresses undergo redistribution in lateral directions (see, e.g.,
\cite{murphy+11}). This process of redistribution of the radial stress is
expected to be isotropic in the lateral directions, as observed in our 3D
simulations (thick dash-dotted and dotted lines in Figure
\ref{f:reynoldsstresses}). By volume-integrating the radial averages of the
stress components inside the gain region, we find that in both 2D and 3D the
integrated radial stress is larger by about $60$\% than the lateral stresses.
Furthermore, the integrated turbulent stresses are greater by about $30$\% in
3D than in 2D. The latter result may appear at first surprising, judging by
the run of turbulent stresses shown in Figure \ref{f:reynoldsstresses}.
However it is important to recognize that the curves shown in the figure
represent angular averages rather than volume-integrated contributions.
Therefore, even though the stresses are greater in 2D than in 3D near the gain
radius, the larger stresses in 3D in the upper layers of the gain region
result in their overall greater volume-integrated value.

The fact that the volume-integrated turbulent stresses are greater in 3D than
in 2D is consistent with the surface-to-volume difference between 2D and 3D
simulations (the argument we made in Section \ref{s:convenergytrans}). In this
scenario, 3D structures are more susceptible to perturbations and the flow
becomes disorganized over larger regions (in our case, a greater fraction of
the radial extent of the gain region). Specifically, full-width at half-maximum 
measure of the radial stress distribution is $25$\% greater in 3D than
in 2D. (The corresponding volume factor is still larger due to the 3D
distribution being centered at a higher radius.) Finally, we note that the
process of redistribution of the radial stress is appears more gradual in 3D
than in 2D (compare the two curves connected with the double arrow in Figure
\ref{f:reynoldsstresses}). In fact, this may only be a misimpression. Note
that the contribution of lateral stresses in 2D differs qualitatively from
that in 3D only in the lower half of the gain region. We find preliminary
evidence that this may be due to the difference in the work done by expansion
in that region, where the expansion work source term is large in 2D, while it
is nearly zero in 3D (see Figure \ref{f:fluxes}). We speculate that the
expansion increases the velocity fluctuations and thus contributes to the
increase of the turbulent stresses.
\subsection{Lagrangian analysis of the explosion mechanism}
\label{s:particle}

We believe that potentially important insights into the core-collapse
supernova explosion mechanism can be gained through analyzing the history of
fluid parcels as they enter the gain region and participate in the revival of
the shock. Our motivation partially follows from the realization that the
amount of evidence provided by the Eulerian analysis appears insufficient to
disentangle and clearly identify the role various physics plays in the
explosion process. To this end, we seeded our simulations with a large number
of tracer particles from the beginning of the simulations. Recall from Section
\ref{s:methods} than in our models, each tracer particle represents a fluid
element with a mass of $1\times10^{-6}$ \msun\ in two-dimensions and
$1\times10^{-7}$ \msun\ in three-dimensions.
\subsubsection{Residence times in the gain region}
\label{s:tres}
In our analysis we assume a particle resides inside the gain region if its
position is between the shock radius and the gain radius (both of which are
angle- and time-dependent). We define the residency time, $t_{res}$, as the
total time a tracer particle spends in the gain region. Thus, the maximum
residency time of any particle never exceeds the model time elapsed since the
beginning of the simulation. We construct a distribution of particle residency
times by binning the residency times with a resolution of $1$ ms.

\begin{figure}[t!]
\centering
\includegraphics[width=8cm]{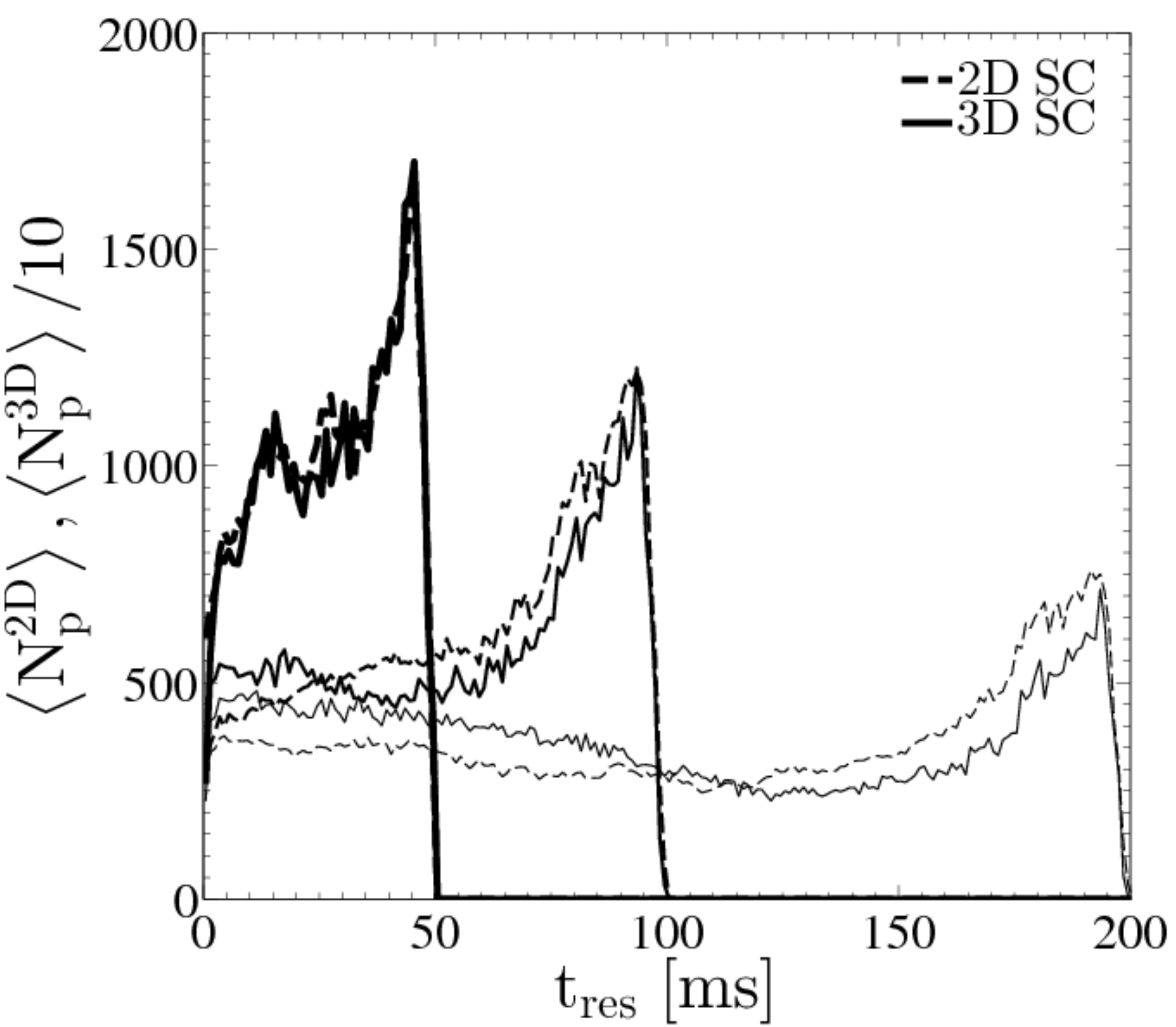}
\caption
{
	Distributions of tracer particles in the gain region as a function of
	particle residency time for SC models. This distributions are shown with
	dashed and solid lines for 2D and 3D families, respectively, at $t=50$ ms,
	$100$ ms, and $200$ ms. The data are averaged over all model realizations
	for a particular model dimension. For legibility, we smooth the curves
	with a boxcar method using a time window of 5 ms. Note that the scale
	differs for 3D models. See Section \ref{s:tres} for
	discussion.
\label{f:restime}}
\end{figure}

Figure \ref{f:restime}
shows the model realization-averaged distributions of particle numbers for our
SC models at $t=50$ ms, 100 ms, and 200 ms, as a function of particle
residency time. In the figure, the data for 2D and 3D models are shown with
dashed and solid lines, respectively. The total number of particles in the
gain region are approximately $5.2\times 10^4$, $6.3\times 10^4$, and
$7.3\times 10^4$ in 2D and $5.1\times 10^5$, $5.9\times 10^5$, and $7.3\times
10^5$ in 3D, for progressively later evolutionary times. We note that
particles occupying bins near $t_{res}=0$ should be interpreted as material
that has been just accreted (entered the gain region), while the farther to
the right the particle bin is the earlier it entered the gain region.

There is little difference in distributions of particle numbers between the 2D
and 3D models (shown with thick dashed and solid lines, respectively, in
Figure \ref{f:restime}) for particles that enter the gain region prior to the
onset of convection. This is expected, given that at those early times the
shock is nearly spherically symmetric and the post-shock flow is essentially
radial. These initial distributions are expected to evolve due to both
accretion of new material (particles passing through the shock), and loss of
particles already residing in the gain region through the gain radius and
their settling onto the proto-neutron star. Therefore, one may expect certain
differences in the distributions of particle numbers at later times. First,
notice that the number of particles with maximum residency times is always
smaller in 3D than in 2D, except at the early time ($t = 50$ ms). This means
that a portion of particles with the longest residency times that are accreted
through the gain radius is larger in 3D than in 2D. Second, this trend reverts
at $t\approx 75$, and at $t=200$ ms there are more particles with short
residency times present in 3D than in 2D. One possible explanation for this
behavior might be a systematic differences in accretion rates and shock radii
between 2D and 3D, but we found no evidence to support this possibility (cf.\
Section \ref{s:grchar}).

Additional information about the origin and evolution of particle number
distributions can be obtained by examining the spatial dependence of particle
residency times. These distributions reveal that, in 3D, there exists a
population of particles with low residency times located near the gain radius.
The systematic differences between particle number distributions in 2D and 3D
can therefore be explained by particles leaving and reentering the gain region
across the gain radius in 3D. This leads to the apparent deficit of particles
with long residency times and causes excess of particles with low residency
times in this case. We quantified the magnitude of this effect by examining
the contours of cumulative particle numbers at a given residency time. We
found that between $10$\% and $20$\% of particles located near the gain radius
participate in this process. This estimate is consistent with the difference
between the particle number distributions seen in Figure \ref{f:restime}. The
above effect signifies a difference in the flow dynamics near the gain radius
between 2D and 3D. In particular, we believe this behavior is due to a larger
amount of mass (by $10$--$20$\%) involved in convection at the bottom of the
gain region, where the convection is driven, and may be the reason for the
greater efficiency of convection in 3D (see Section \ref{s:unbinding} for
discussion).
\subsubsection{Thermodynamic evolution of the shocked material}
\label{s:eintgain}

One quantity of particular interest in the context of the ccSNe explosion
mechanism is the boost to the internal energy of the material residing in the
gain region provided by neutrino heating. This process can be analyzed by
considering the increase in the internal energy of shocked fluid parcels
(represented in our study by tracer particles) as a function of the particle
residency time. 

\begin{figure}[t!]
\centering
\includegraphics[width=8cm]{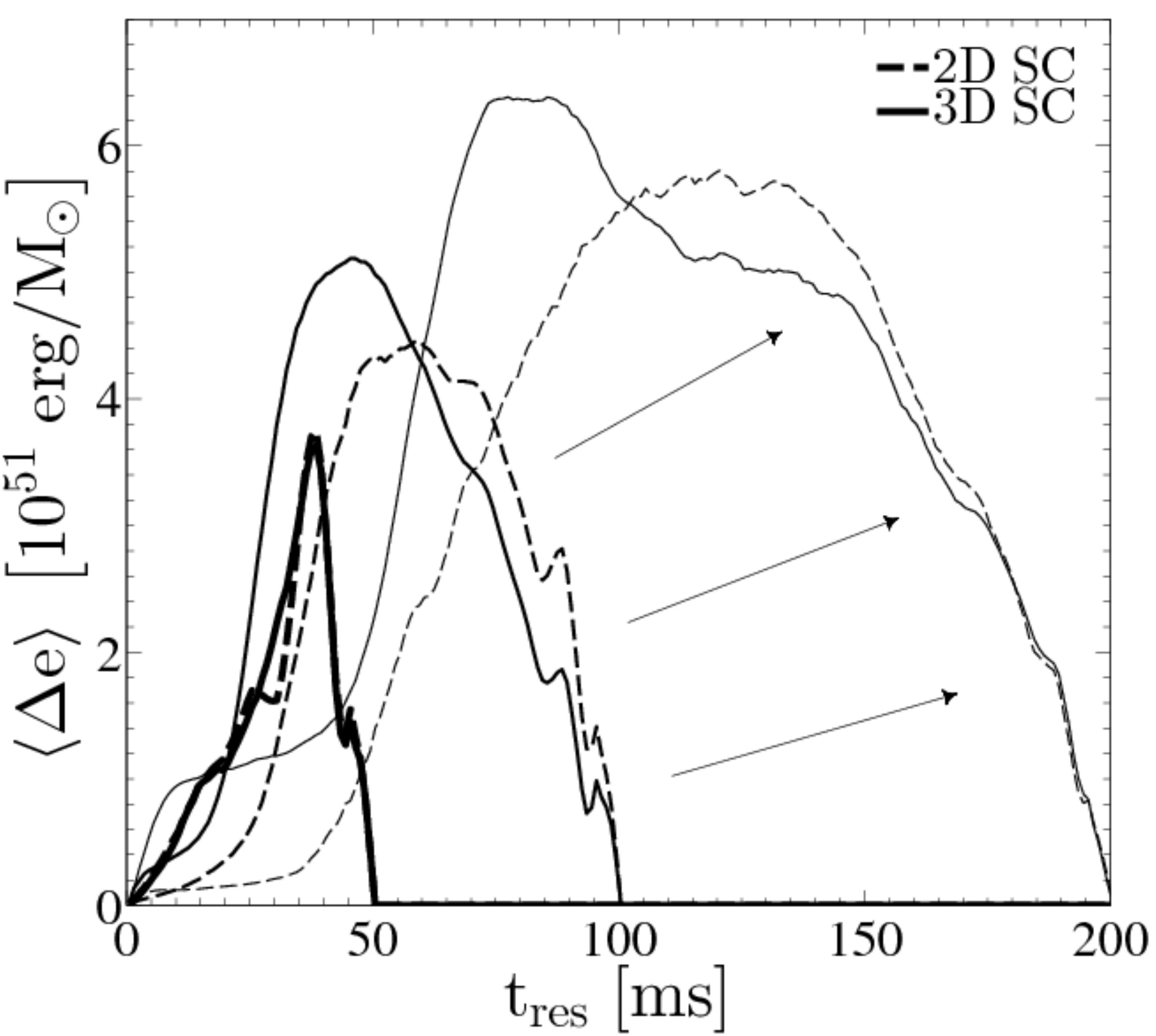}
\caption
{Distribution of the specific internal energy change for tracer particles
in the gain region as a function of particle residency time for SC models.
The distributions are shown at $t=50$, $100$, and $200$ ms with thick,
medium, and thin lines, respectively. The data are averaged over all model
realizations for a particular dimension, and are shown separately with
dashed and solid lines for 2D and 3D. For legibility, we smoothed the
curves with a boxcar method using a time window of 5 ms. See Section
\ref{s:eintgain} for discussion.
\label{f:restime_energy_avg}}
\end{figure}

Figure \ref{f:restime_energy_avg}
shows the model realization-averaged internal energy gain per unit mass,
$\langle\Delta e\rangle$, as a function of particle residency time for the 2D
and 3D SC models (shown with dashed and solid lines, respectively). We define
the internal energy gain of a tracer particle as the difference between the
particle internal energy at the evolutionary time shown and its internal
energy when it enters the gain region. We find that at early times ($t
\lesssim 50$ ms) there is essentially no difference in the way particles gain
energy in 2D and 3D (thick dashed and solid lines). At later times, we begin
to observe modest at first ($t=100$ ms) changes with the newly accreted
material progressively gaining energy faster in 3D than in 2D (the solid,
medium thick line representing 3D data lies always to the left of the dashed,
medium thick line representing the 2D data).

In addition to the apparent asymmetry in the 3D distribution at late times, we
also find qualitatively new unique features, in particular at $t=200$ ms.
First, we identify a region with almost constant increase in the energy gain
for particles that entered the gain region between $t=50$ ms and $t=100$ ms.
During this time, and as we discussed earlier in Section \ref{s:tres}, the
gain radius shrinks faster in 3D than in 2D. This allows the particles
residing close to the gain radius to gain comparatively more energy in 3D.
This process is responsible for the observed shape of the internal energy gain
distribution for residency times between $t_{res}\approx 100$ ms and
$t_{res}\approx 150$ ms. Second, there is a pronounced hump visible in Figure
\ref{f:restime_energy_avg} which broadly occupies a $35$ ms time interval for
particles that entered the gain region shortly after the vigorous convection
developed. This is marked by a rapid increase in the 3D energy gain curve
taken at $t=200$ ms (thin solid line) at $t_{res}\approx 100$ ms. We believe
this signifies the increase in the thermodynamic efficiency of the convective
engine in 3D. Third, there is a smaller hump for a population of particles
that entered the gain region shortly before ($t_{res} < 50$ ms) the shock was
launched.

Finally, we would like to point out that the similarity in the shapes of the
3D distributions at late times is quite striking. The arrows in the figure
connect specific features in the energy gain distributions that are well
preserved in 3D between $t=100$ ms and $t=200$ ms. This indicates that the
particles which entered the post-shock region during the first $100$ ms of the
simulation gained internal energy at a similar rate during the process of
shock revival.
\subsubsection{Unbinding of the shocked material}
\label{s:unbinding}
The explosion process necessarily requires gravitationally unbinding a
substantial fraction of the collapsing core. Alas, very little information
about this process can be found in the literature, perhaps with the exception
of discussions of global quantities such as the explosion energy. However,
armed with the tracer particles we are now in a position to investigate this
process in more detail. In particular, the results discussed in the previous
section indicate that profound differences in the heating efficiency exist
between 2D and 3D models. Although we are able to identify populations of
particles responsible for those differences, our focus was more on the
interplay between the advection and neutrino heating of the shocked material.
In the following discussion of the process of gravitational unbinding of the
shocked, stellar core we also take into account the flow dynamics inside the
gain region.

\begin{figure}[t!]
\centering
\includegraphics[width=8cm]{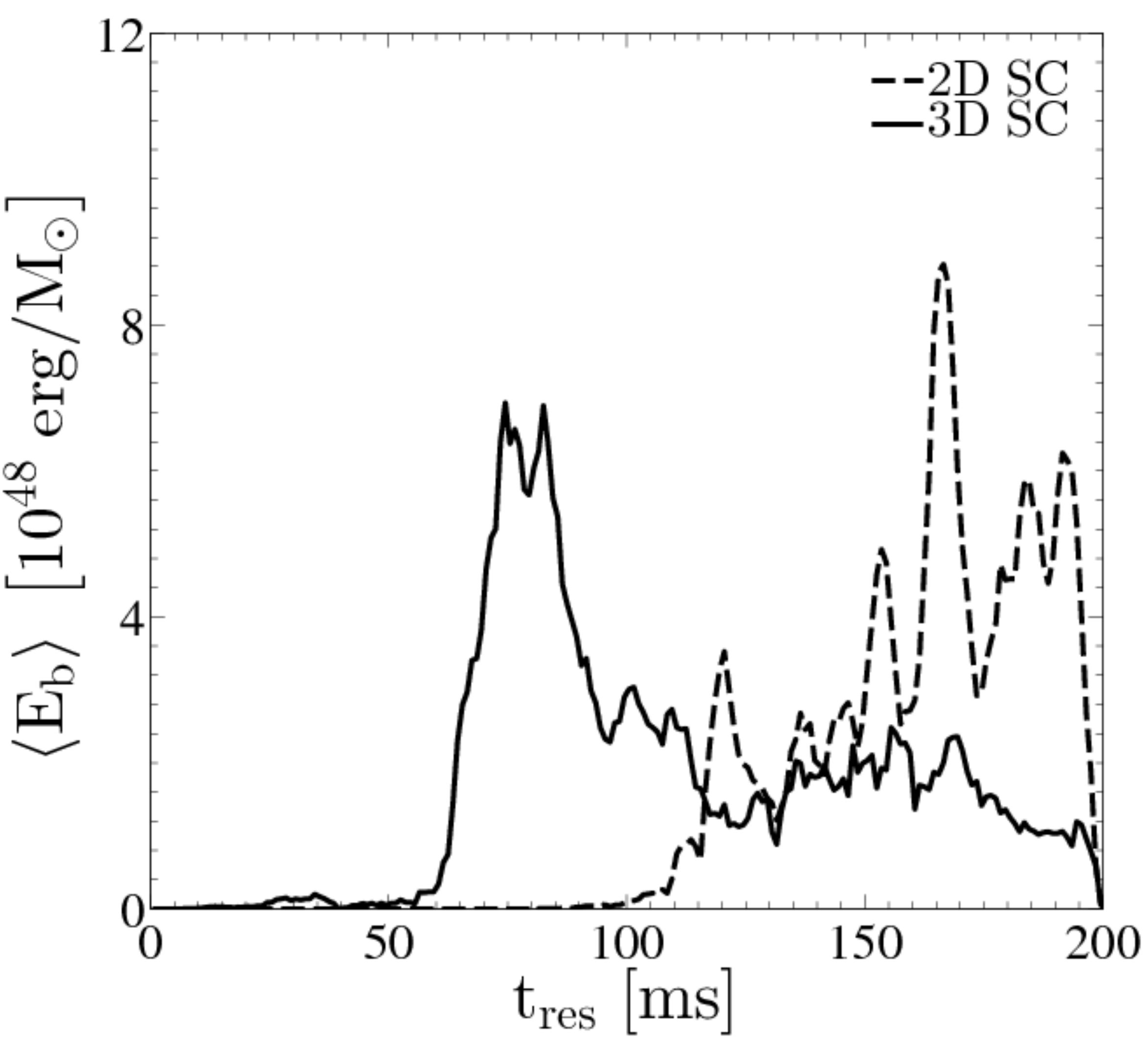}
\caption
{	
	Distribution of the positive specific gravitational binding energy for
	tracer particles in the gain region as a function of particle residency
	time for SC models. The data are taken at $t=200$ ms, and are averaged
	over all model realizations for a particular model dimension. The results
	are shown with dashed and solid lines for 2D and 3D, respectively. For
	legibility, we smooth the curves using a boxcar method with a time window
	of 5 ms. Note that in 2D, unbound material is composed primarily of matter
	which entered the gain region prior to the quasi-steady phase. However, in
	3D, there are substantial contributions from material which entered the
	gain region during the period of vigorous convection. See Section
	\ref{s:unbinding} for discussion.
\label{f:restime_binding}}
\end{figure}

Figure \ref{f:restime_binding}
shows the model realization-average of the positive part of the specific
gravitational binding energy, $E_{b}$, as a function of residency time for 2D
and 3D families of models at $t=200$ ms (marked by dashed and solid lines,
respectively). The data indicates dramatic differences in the energetics of
fluid elements in the gain region between 2D and 3D. In 2D, the unbound
portion of the gain region is almost exclusively made of material that entered
the gain region during the first $100$ ms of the evolution. However, in 3D one
can identify two different populations of particles. The first 3D population
consists of particles that entered the gain during the first $100$ ms.
Although this is the same material that is responsible for the explosion in
2D, in 3D this material is substantially less energetic. A careful examination
of the data shown in Figure \ref{f:restime_binding} reveals that, indeed,
around the explosion time the explosion energies in 2D are greater than in 3D.
This is consistent with the 2D models exploding, on average, faster than 3D
models. Also, as discussed in the previous section, the particles that enter
the gain region early on gain energy more efficiently in 2D than in 3D. It is
only at later times when the particles in 3D show a dramatic increase in their
internal energies. 

In the scenario presented above, the explosion is launched in 2D early on by a
relatively stronger central neutrino source. In 3D, however, the more
efficient convective engine is able to launch the shock at neutrino
luminosities lower by $\approx 4$\%. (Recall that we found evidence for up to
$20$\% more mass participating in convection across the gain radius in 3D, as
discussed in Section \ref{s:tres}). By integrating the particle energies, we
found that in 3D the same amount of positive binding energy is carried by
particles that entered the gain region during the first $100$ ms of evolution
as by those which entered the gain region during the following $50$ ms. Thus,
the process of unbinding material in the gain region appears more gradual in
three-dimensions, with significant energy gain during the quasi-steady phase.

The cause of the observed differences between energetics in 2D and 3D ccSNe
simulations, and so the question how the efficiency of the convective engine
depends on dimensionality, remains unknown at this time. However, we believe
the Lagrangian approach to analyzing the evolution of the gain region offers
significant advantages over Eulerian statistics for investigating the
efficiency of convection. This will problem will be the subject of a
forthcoming publication.
%
%
%
%
%
%
%
\section{Summary \& future work}\label{s:conc}

We have presented an analysis of two- and three-dimensional models of early
core-collapse supernova development in the collapsing core of a $15$ \msun\
progenitor. In our study, the simulations start at shock bounce, continue
through the onset of neutrino-driven convection, quasi-steady state of the
gain region, and early post-revival shock expansion until the energy of the
explosion had approximately saturated at 1.5 seconds. Our models were tuned to
match the estimated explosion energetics of SN 1987A by careful selection of
the parameterized neutrino luminosity. We considered the cases of slow and
fast contraction rates of the proto-neutron star to reflect on uncertainty in
the nuclear equation of state. We introduced and applied new diagnostic
methods for the explosion process, morphology and structure of convection, and
energetics of material inside the gain region.

Our conclusions are formulated based on a large database of explosion models.
This computationally expensive approach is necessary since the core-collapse
supernova problem involves highly nonlinear, strongly coupled physics which
results in the extreme sensitivity of realistic computational models to
perturbations. Therefore, conclusions of core-collapse supernova studies can
only be understood in a statistical sense. Moreover, such ensemble-based
conclusions are potentially affected by resolution effects, especially in
three-dimensions, due to the high computational cost of individual model
realizations.

The main findings of our work can be summarized as follows: 
\begin{itemize}
\item We found that the same explosion energy can be obtained in three-dimensions 
with less energy than in two-dimensions. This trend also holds
between two-dimensions and one-dimension. This result is in agreement with
some previous studies.

\item The stochastic character of the explosion process results in
non-uniqueness of the relation between neutrino luminosity and explosion
energy. In particular, in the process of tuning the neutrino luminosities, we
were able to obtain equally energetic 3D models with the neutrino luminosities
differing by as much as $3$\%. 

\item Dimensionality also contributes to the non-uniqueness of the above
relation. In particular, equally energetic explosions can be obtained in 3D
using neutrino luminosities smaller by $4$\% than in 2D. This implies the
efficiency of the convective engine is greater by that amount in 3D.

\item We proposed that the observed dependence of the efficiency of the
convective engine on the model dimensionality might be a consequence of the
difference in the surface-to-volume ratio of flow structures between 2D and
3D. In this scenario, the greater efficiency of the convective engine in 3D is
simply due to a larger surface-to-volume ratio in that case, which helps the
process of heat exchange between hot and cold flow structures. Quantifying
this possibly important effect is, however, beyond the scope of this paper.

\item We note that the relations between neutrino luminosities, accretion
rates, and explosion times show, in our models, less variation with the
model dimensionality than in other studies (see Figure \ref{f:lum}). We
believe this is purely due to the fact that the objective of our study was to
obtain models with similar energetics, rather than to determine the threshold
neutrino luminosity required for the explosion.

\item We found no evidence for the standing accretion shock instability (SASI)
in our models. Our results showed no persistent, oscillatory $l=1$ or $l=2$
mode behavior, considered to be a trademark of SASI. The absence of SASI in
our models is consistent with short advective times through the gain region.
Instead, the explosions are boosted in multidimensions by strong, neutrino-
driven  convection. At the same time, we would like to stress that SASI cannot
be excluded as an important, or even decisive, physics mechanism for different
parameters of collapsing core (e.g., progenitor structure, equation of state,
etc.).

\item We found that the shock front shows a rather modest degree of asymmetry
in our models, $r_{s}^{max}/r_{s}^{min}\approx 1.3$. We found tentative
evidence between the large-scale structure of convection and the shock
asymmetry. In particular, the shock is less deformed in the case when
convection has a richer structure (higher-order modes are present).

\item We found the neutron star recoil velocities reaching up to $\approx 475$
km/s, with typical values between $100$ km/s and $300$ km/s. These results are
in good agreement with the previous studies \citep[e.g.\ ][]{scheck+06}.

\item We introduced and applied new metrics to diagnose the dynamics and
structure of the gain region. In particular, we studied the characteristics of
the upflowing material (which we identified as the buoyant, convective
bubbles) and the solid angle it occupies. Using these measures, we found that
the structure of the gain region on large scales and its evolution did not
significantly differ between 2D and 3D. However, for the fast contracting
models, we found that the mass inside upflows always exceeded $50$\% of the
total gain region, hinting at prompt-like explosions. This situation is
qualitatively different in the case of the slow contracting models, where a
prolonged period of quasi-steady state in the gain region can be clearly
recognized. It appears that during that phase the mass in the upflows
fluctuates around the slowly increasing mean value with amplitude of several
percent, and on a timescale of 10 ms. We speculate that individual downflows
are responsible for the observed fluctuations.

\item Our method of identifying the convective bubbles allowed us to study the
statistical properties of the bubble distributions. We found that the solid
angle occupied by bubbles is initially small and progressively increases with
time (see Figure \ref{f:clustersize}). This indicates that the bubbles are
small at first and steadily grow as the evolution progresses. The process
appears to bear some similarity to the bubble merging process observed in
multi-mode Rayleigh-Taylor instabilities. Furthermore, we find a good
correlation between the solid angle occupied by rising material and the
upflowing mass. This highlights the convection takes place inside the whole
gain region, from the gain radius all the way up to the shock.

\item The development and evolution of convection inside the gain region can
also be studied by considering the temporal evolution of the number of
bubbles. We found that their number rapidly increased at $t\approx 50$ ms (see
Figure \ref{f:numclusters}), indicating the onset of convection. After the
convection set in, the number of bubbles progressively decreased with time. We
used the information about the number of bubbles combined with the solid angle
they occupy, and estimated that the convective bubbles are, on average, about
4 times smaller in 3D than in 2D at the time of the explosion.

\item We used the decomposition of the total energy flux to understand the
energy transport in the gain region. We found that the work done by expansion
and compression in that region differs between 2D and 3D, and we proposed that
this is due to the greater structural integrity of flow structures in 2D.

\item The analysis of the components of the total energy flux also provided
evidence that the region near the gain radius was not in thermal equilibrium
at the beginning of the quasi-steady phase. We found evidence for strong
cooling below and strong heating above the gain radius. Furthermore, we found
that the buoyant work was positive throughout the gain region in our models,
in accordance with the neutrino-driven convection scenario. 

\item We demonstrated that, for the same neutrino luminosities that produce
energetic explosions in well-resolved models, the explosion energies
dramatically decrease once the angular mesh resolution decreases below
6\degree. This implies there exists a minimal mesh resolution required for the
central engine to efficiently operate. This is expected. This also implies
that the neutrino-driven convection is the key ingredient of the explosion
mechanism in the energetic models considered here. We cannot exclude the
possibility that more physics (e.g.\ turbulence) may emerge at still higher
resolutions (i.e.\ less than about $0.05\times r$) to power the explosion more
efficiently than convection. 

\item We found that the the structure of turbulent energy spectra are steeper
in our 2D models ($E \sim l^{-2.68}$) than in our 3D models ($E \sim
l^{-1.86}$). These results should be taken with caution given the low
resolution of our models and an estimated Reynolds number on the order of
$100$.

\item Analysis of the turbulent Reynolds stresses revealed that the post-shock flow
is anisotropic in the radial direction due to buoyant driving. At the same
time, we found the flow in the gain region is isotropic in the lateral
directions.

\item Using a Lagrangian representation of the gain region, we found
substantial differences in the energetics of the shocked fluid elements
between 2D and 3D. We identified material near the base of the gain region
that experiences stronger heating during the period when the gain radius
shrinks. We also found that material accreted during the phase when the
convection is fully developed undergoes especially strong heating in 3D.

\item We found significant differences in the evolution of residency times and
binding energies of fluid elements inside the gain region between 2D and 3D.
In 3D, more mass is involved in the process of driving convection
in the region around the gain radius. At the same time, the same amount of
positive binding energy is carried by material that entered the gain region
during the first $100$ ms of evolution as by those which entered the gain
region during the following $50$ ms. However, in 2D, a smaller portion of the
gain region mass participates in driving convection, and practically all of
the positive binding energy is carried by material that entered the gain
region during the first $100$ ms.
\end{itemize}
Several possible directions for future research emerge from the current work.
It remains unclear why the convective engine is more efficient in three-dimensions. 
Although we were able to quantify the effect, we did not provide a
clear explanation for its origin. We obtained only rudimentary information
about the properties of turbulence in our largely underresolved and dominated
by numerical viscosity models. It is clear that to gain meaningful insight
into the role of turbulence in the process of core-collapse supernova
explosion a new generation of far better-resolved models exploiting more
efficient computational approaches is required. Also, one would like to evolve
those models until the neutron star recoil velocities saturate. Finally, our
current model does not include nuclear burning, and therefore we are unable to
discuss the nucleosynthetic yields and the compositional structure of our
models. The above issues will be the subject of a forthcoming publication.
%
%
%
\acknowledgements
We thank Konstantinos Kifonidis for his comments on the initial version of the
manuscript. The work of TH and TP has been supported by the NSF grant
AST-1109113, and TP has been partially supported by the DOE grant DE-FG52-09NA29548. 
This research used resources of the National Energy Research
Scientific Computing Center, which is supported by the Office of Science of
the U.S.\ Department of Energy under Contract No.\ DE-AC02-05CH11231.
Simulations were performed in part using the 3Leaf Systems SMP clusters at the
Florida State University High Performance Computing Center, and the Deszno SMP
cluster at Jagiellonian University, Cracow, Poland.
%
%
%
\bibliographystyle{apj}
\bibliography{ms}
%
%
%
\newpage
\appendix
\section{Estimating numerical viscosity and Reynolds number}
\label{s:reynoldsnumber}

In order to estimate the numerical viscosity and the numerical Reynolds
number, we use the information provided by the Lagrangian tracer particles.
Consequently, in our approach the velocity field is represented at discrete
points by tracer particles. We define the longitudinal Eulerian velocity
structure function of order-$p$ as
\begin{equation}
\label{e:LEVSF}
S^{L}_{p}\left(r_{sep}\right) = \langle{\left[\left(\vect{u}\left(\vect{x}\right) - \vect{u}\left(\vect{x}+r_{sep}\vect{\hat{r}}\right)\right)\vect{\cdot}\vect{\hat{r}}\right]^p}\rangle,
\end{equation}
where $\vect{x}$ is the particle position, $\vect{u}$ is the particle
velocity, $r_{sep}$ is the separation between particles, and $\vect{\hat{r}}$
is the particle displacement unit vector. The averaging operator,
$\langle\cdot\rangle$, is applied over a given $r_{sep}$ at a specific model
time.

Kolmogorov's turbulence theory \citep{monin+71,frisch95}, assuming negligible
intermittence, states that for $p=2$, the structure function, $S^{L}_{2}$,
is related to the average specific turbulent dissipation,
$\langle{\epsilon}\rangle$, via
\begin{equation}
S^{L}_{2}\left(r_{sep}\right) = C_{2}\langle{\epsilon}\rangle^{2/3} r_{sep}^{2/3},
\end{equation}
where $C_{2}$ is a constant on the order of 1. This result allows one to
estimate the turbulent dissipation (physical + numerical) present in a
numerical simulation. Within the same framework, the local, specific turbulent
dissipation is related to the average strain rate as
\begin{equation}
\epsilon = 2\nu\left\langle{\tens{S}\vect{:}\tens{S}}\right\rangle,
\end{equation}
where $\nu$ is the kinematic viscosity and $\tens{S}$ is the strain rate
tensor.

\begin{figure*}[t]
\centering
\includegraphics[width=8cm]{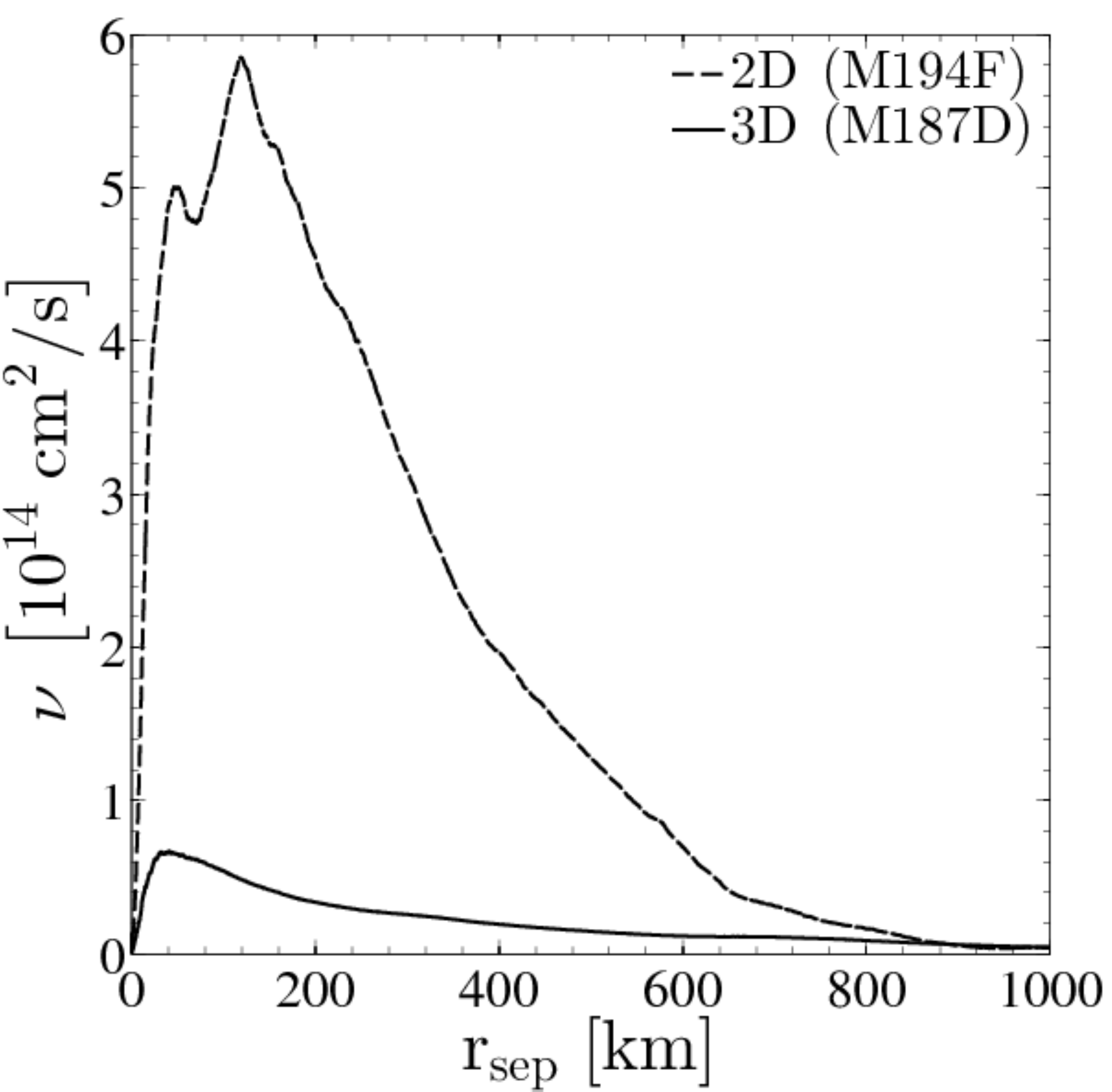}
\includegraphics[width=8cm]{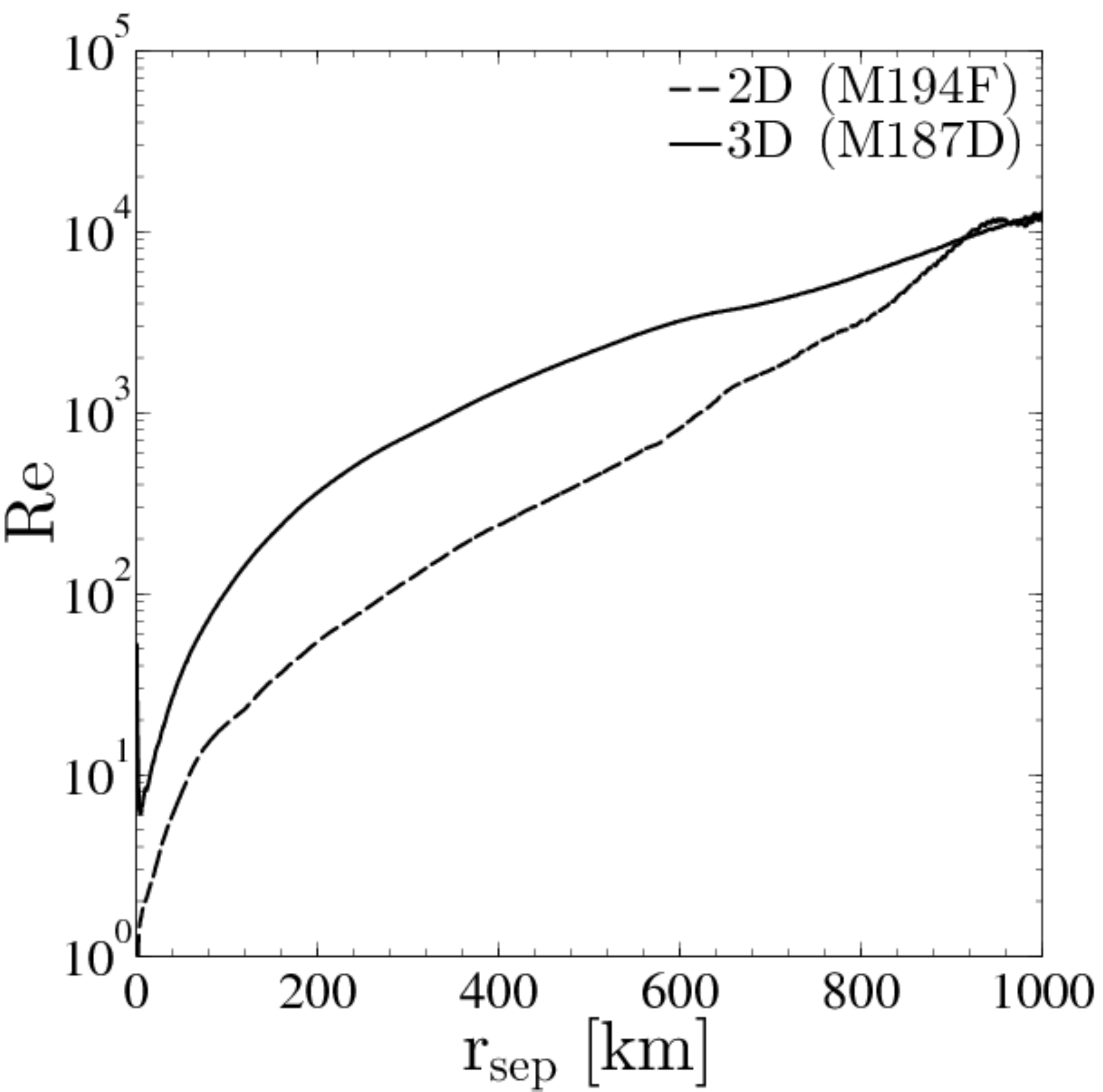}
\caption{Analysis results for models M194J and M187D. (left panel) Estimated numerical kinematic viscosity. (right panel) Estimated Reynolds number as a function of particle separation distance. We may expect our models to reflect flows with Reynolds numbers on the order of 10s or 100s.
\label{f:appendix}}
\end{figure*}

In order to estimate the numerical Reynolds number in our models, we assume
that the local turbulent dissipation is representative of the average
dissipation rate, $\epsilon\approx\left\langle{\epsilon}\right\rangle$. This
allows us to compute the kinematic viscosity as
\begin{equation}
\nu = \frac{{S_{2}^{L}}^{3/2}}{2C_{2}^{3/2}\left\langle{\tens{S}\vect{:}\tens{S}}\right\rangle r_{sep}}.
\end{equation}
Using $r_{sep}$ as our characteristic length scale and approximating the
characteristic velocity in the field by $\sqrt{S^{L}_{2}}$, we estimate the
Reynolds number as a function of separation distance via
\begin{equation}
Re = \frac{r_{sep} \sqrt{S_{2}^{L}}}{\nu}.
\end{equation}
Figure \ref{f:appendix}
shows the numerical kinematic viscosity and numerical Reynolds number
estimates for the 2D SC model M194J and the 3D SC model M187D at $t=150$ ms as
a function of particle separation distance. We find numerical viscosities on
the order of $1\times10^{14}$ cm$^2$s$^{-1}$ for the 2D model and
$1\times10^{13}$ cm$^2$s$^{-1}$ for the 3D model. At this time in the model
evolution, the average shock radius is about 500 km, providing an upper limit
on structure size. On that scale, the numerical Reynolds number is $\approx
200$ in 2D and $\approx 1000$ in 3D. However, as one can see, the numerical
viscosity, and so the numerical Reynolds number, vary with scale. Therefore,
our models cannot be represented by a single numerical Reynolds number.
Consequently, and given relatively low values of the numerical Reynolds number
present in our simulations, we tend to characterize the fluid flow inside the
gain region of our models as perturbed laminar, rather than turbulent.

Finally, we would like to remark that given the anisotropy of the flow field
obtained in our simulations (chiefly due to large-scale convection), the
assumption that the average dissipation rate can be represented by the local
dissipation rate may not hold true in all regions in our models. However, we
believe that the obtained estimates still provide valuable insights into the
nature of our simulations. In particular, given the low estimated numerical
Reynolds numbers, conclusions regarding the role of turbulence in the core-collapse 
supernova explosions should be considered, at most, tentative. Since
turbulence was not a particular focus of this study, we are not in a position
to estimate the resolution required to capture these effects.
\end{document}